\documentclass{article}
\usepackage{cite} 
\usepackage[font=small]{caption}
\usepackage[labelformat=empty, position=top]{subcaption}
\usepackage{authblk}
\usepackage{multicol}
\usepackage[centering,hmargin=2.5cm,vmargin=2.5cm]{geometry}

\usepackage{import}
\usepackage{my_style}
\numberwithin{equation}{section}

\makeatletter
\patchcmd{\@maketitle}{\LARGE \@title}{\fontsize{15}{19.2}\selectfont\@title}{}{}
\makeatother

\begin{document}

\title{Classical correspondence beyond the Ehrenfest time for open quantum systems with general Lindbladians}

\renewcommand\Affilfont{\small}

\author[1,2]{Felipe Hernández\footnote{Email: felipehb@stanford.edu}}
\affil[1]{Department of Mathematics, Stanford University, Stanford, CA 94305 USA}
\affil[2]{NTT Research, Inc., Physics \& Informatics Laboratories, Sunnyvale, CA 94085, USA}
\author[3]{Daniel Ranard\footnote{Email: dranard@mit.edu}}
\affil[3]{Center for Theoretical Physics, Massachusetts Institute of Technology, Cambridge, MA 02139, USA}
\author[2]{C. Jess Riedel\footnote{Email: jessriedel@gmail.com}}
\date{}
\maketitle
\begin{abstract}

Quantum and classical systems evolving under the same formal Hamiltonian $H$ may exhibit dramatically different behavior after the Ehrenfest timescale  $t_E \sim \log(\hbar^{-1})$, even as $\hbar \to 0$. 
Coupling the system to a Markovian environment results in a Lindblad equation for the quantum evolution.
Its classical counterpart is given by the Fokker-Planck equation on phase space, which describes Hamiltonian flow with friction and diffusive noise. The quantum and classical evolutions may be compared via the Wigner-Weyl representation. Due to decoherence, they are conjectured to match closely for times far beyond the Ehrenfest timescale as $\hbar \to 0$. We prove a version of this correspondence, bounding the error between the quantum and classical evolutions for any sufficiently regular Hamiltonian $H(x,p)$ and Lindblad functions $\LCk(x,p)$.  The error is small when the strength of the diffusion $D$ associated to the Lindblad functions satisfies $D \gg \hbar^{4/3}$, in particular allowing vanishing noise in the classical limit. 
Our method uses a time-dependent semiclassical mixture of variably squeezed Gaussian states.  
The states evolve according to a local harmonic approximation to the Lindblad dynamics constructed from a second-order Taylor expansion of the Lindbladian.
Both the exact quantum trajectory and its classical counterpart can be expressed as perturbations of this semiclassical mixture, with the errors bounded using Duhamel's principle.
We present heuristic arguments suggesting the $4/3$ exponent is optimal and defines a boundary in the sense that asymptotically weaker diffusion permits a breakdown of quantum-classical correspondence at the Ehrenfest timescale.  Our presentation aims to be comprehensive and accessible to both mathematicians and physicists.
In a shorter companion paper, we treat the special case of Hamiltonians that decompose into kinetic and potential energy with linear Lindblad operators, with explicit bounds that can be applied directly to physical systems.
\end{abstract}

\newpage
\setcounter{tocdepth}{2} 
\tableofcontents
\phantom{text}

\section{Introduction}
\label{sec:incoherent}

In this paper we study the correspondence between classical and quantum mechanics in systems that interact with an external environment.  That is, we include effects such as dissipation, diffusion, and decoherence that arise from the environmental interaction.  Such systems are referred to in the physics literature as \textit{open} quantum systems\footnote{In the mathematics literature, the term ``open system'' often refers to a dynamical system on a non-compact space.  In this paper we instead use the physicist's meaning of the term ``open system''. In particular, the entropy of the open quantum state obeying the Lindblad equation \ref{eq:lindblad-simple} and the entropy of the open classical state obeying the Fokker-Planck equation \ref{eq:fp-simple} can both increase with time.} and are important for understanding the emergence of classical behavior from quantum mechanics. \textit{Closed} quantum systems by definition have no such interactions with an environment, and the correspondence between classical and quantum mechanics provided by Egorov's theorem \cite{egorov1969canonical, zworski2022semiclassical, combescure1997semiclassical, hagedorn2000exponentially, silvestrov2002ehrenfest, robert2021coherent} is limited to the Ehrenfest time, which is logarithmic in Planck's constant, the semiclassical parameter $\hbar$. Beyond this timescale, the quantum wavefunction for a closed quantum system can develop coherence over long distances, which do not correspond to any classical state and are not readily observed in real-world macroscopic systems.  It has been argued in the physics literature that decoherence from the environment is responsible for the appearance of classical behavior \cite{zeh1973quantum, kubler1973dynamics, zurek1981pointer, zurek1994decoherence, shiokawa1995decoherence, zurek1998decoherence, zurek1999why, perthame2004quantum} (but cf.~\cite{ballentine1994inadequacy,fox1994chaos1,fox1994chaos2,schlautmann1995measurement,casati1995comment,zurek1995zurek,brun1996quantum,emerson2002quantum,wiebe2005quantum, kofler2007classical, kofler2008conditions}). Numerical simulations and analytical arguments \cite{spiller1994emergence, kolovsky1994remark,kolovsky1996condition, habib1998decoherence, bhattacharya2000continuous, bhattacharya2003continuous, zurek2003decoherence, toscano2005decoherence, pattanayak2003parameter, gammal2007quantum} suggest that the Wigner function of the quantum state and the corresponding classical state will become indistinguishable in the classical limit in the presence of sufficient decoherence.

The state of an open quantum system for $d$ variables is given by $\rho$, a positive semidefinite trace-class operator on $L^2(\bbR^d)$.  The strong physical assumption enabling our analysis is the Markov condition, which implies that the dynamics generate a quantum dynamical semigroup, governed by the Lindblad equation \cite{lindblad1976generators, davies1974markovian, davies1976markovian, davies1976quantum, davies1976classical, gorini1976completely, gorini1978properties, alicki2007quantum, chruscinski2017brief}.  Thus we take the Linblad equation as our starting point:
 the quantum state evolves 
according to $\partial_t\rho(t) = \LLQ[\rho(t)]$, with Lindbladian $\LLQ$ given by\footnote{We use $[\hat{A},\hat{B}]:=\hat{A}\hat{B}-\hat{B}\hat{A}$ and  $\{\hat{A},\hat{B}\}:=\hat{A}\hat{B} +\hat{B}\hat{A}$ for the commutator and anti-commutator of operators.  In particular, the latter should not be confused with the Poisson bracket, which we denote $\Poissonbracket{\cdot,\cdot}$.} 
\begin{align} \label{eq:lindblad-simple}
    \LLQ [\rho] &= -\frac{i}{\hbar}\left[\HQ,\rho\right]+
    \frac{\notdiffusionstrength}{\hbar} \sum_{k} \left(\LQk \rho \LQk{}^\dagger-\frac{1}{2}\left\{\LQk{}^\dagger\LQk{},\rho\right\}\right).
\end{align}
 The well-posedness of the Lindblad evolution in the present case of unbounded operators is addressed immediately after Definition~\ref{def:correspondingDynamics}, using the discussion of Galkowski and Zworski \cite{galkowski2024classical}. 
The first term corresponds to the Schrodinger evolution with self-adjoint Hamiltonian $\HQ$ and the second term incorporates the effect of the environment, as described by the Lindblad operators $\LQk$.  
Within this introduction we use a \introduce{coupling strength}
$\diffstrength >0$ to more transparently control the overall strength of the coupling with the environment, and in particular we will allow $\diffstrength$ to depend on $\hbar$ as we take $\hbar\to 0$.  
The Lindblad equation is traditionally written with $\diffstrength=1$ (i.e., $\sqrt{\diffstrength}$ absorbed into the definition of $\LQk$), as we will in fact do 
after the present introduction.

As we review in Section~\ref{sec:corresponding-dynamics}, the corresponding
classical dynamics for the classical distribution $\rhoc(t)$ 
are given by the Fokker-Planck equation $\partial_t \cstate = \LLC[\cstate]$ 
using the Liovillian \cite{risken1984fokkerplanck, gardiner2009stochastic}
\begin{equation}
    \label{eq:fp-simple}
    \LLC[\cstate] = -\partial_a [\cstate (\partial^a \HC + \GC^a)]  +	 \frac12 \partial_a (D^{ab} \partial_b \cstate).
\end{equation}
where $\HC=\Op^{-1}[\HQ]$ is the Wigner transform\footnote{The Wigner transformation is the inverse of Weyl quantization, $\Op$. This and other aspects of the Wigner-Weyl representation are reviewed in Section~\ref{sec:wigner-weyl}.} of the Hamiltonian, and where the friction vector $\GC^a$ and diffusion matrix $D^{ab}$ are given by 
\begin{align}
\GC^a &:=  \diffstrength \Imag \sum_k \LCk \partial^a \LCk^* \\
\label{Dab-first-def}
D^{ab} &:=  \diffstrength \hbar \Real \sum_k (\partial^a \LCk )( \partial^b \LCk^*)
\end{align}
using the ``Lindblad functions''  $\LCk=\Op^{-1}[\LQk]$.
In the mathematics literature, classical variables on phase space like $\HC$ and $\LCk$ are known as \introduce{symbols}.
We use phase space coordinate indices  $a,b \in \{1,\ldots,2d\}$ 
where the first $d$ indices are spatial and the second $d$ indices are momentum variables.  Indices are raised and lowered with the standard symplectic form $\sf=\left(\begin{smallmatrix}0 & \IdM_d \\ -\IdM_d & 0\end{smallmatrix}\right)$ and repeated indices are summed, so that for example $(\partial_a f)( \partial^a H) = (\partial_\x f)( \partial_\p H)-(\partial_\p f)( \partial_\x H) =: \Poissonbracket{f,H}$ is the Poisson bracket and $\partial_a\partial^a = 0$ vanishes by antisymmetry.  We will discuss varying $\diffstrength$ with $\hbar$ further in Section~\ref{sec:corresponedence-time}, but for now just note that with $\diffstrength=1$ the diffusion $\D$ vanishes in the classical limit $\hbar\to 0$ while the friction $\GC$ is fixed.
\footnote{Although it might initially seem strange that the classical dynamics ``depend'' on $\hbar$ (via $\D$), the interpretation is clear: making a choice of $\hbar$ relative to a fixed macroscopic scale sets the strength of the noise in the open quantum system, and hence the strength of the noise in the classical system to which it corresponds.} 
We sometimes refer to the diffusion in the classical dynamics as ``noise,'' in the sense of
Brownian motion arising from a Langevin stochastic differential equation.

We loosely refer to a ``quantum-classical correspondence'' when the quantum trajectory $\rho(t)$ resembles the classical trajectory $\rhoc(t)$.  For closed systems ($\diffstrength=0$), such a correspondence only lasts until the Ehrenfest time $\teh \sim \log(\hbar^{-1})$, while for open systems with $\diffstrength$ sufficiently large it is conjectured to last much longer. Our primary contribution in this paper is to prove such a correspondence for times that are a negative power of $\hbar$, hence exponentially larger than the Ehrenfest time, and for a general class of Lindbladians.  (An important special case is addressed in a short companion paper \cite{hernandez2023decoherence1}.)

We will now state a simplified version of our main result,
which demonstrates how our error bound scales with $\hbar$, $\diffstrength$, and $t$. It refers to coherent states, which are pure quantum states (i.e., rank-1 normalized operators) that are Gaussian with covariance matrix proportional to the identity, as reviewed in Section~\ref{sec:harmonic-gaussian}. We assume a fixed Hamiltonian function $H$ and Lindblad functions $\LCk$ that satisfy the following regularity conditions.
\begin{restatable}[Simplified admissible class of Lindbladians]{assum}{assumSimpleSuitableLindblad}
    \label{assum:simpleSuitableLindblad}
    For our simplified result, we assume
    \begin{itemize}
    \item \textbf{Symbol bounds}
    For multi-indices $n:=(n_1,n_2,\ldots,n_{2d}) \in (\mathbb{Z}_{\ge 0})^{\times 2d}$,
    \begin{align}
    |\partial^n H(\alpha)| &\leq C_n, & |n| &\geq 2,\\
    |\partial^n L_k(\alpha)| &\leq C_n, & |n|&\geq 1,\\
    |\partial^n L_k(\alpha)| &\leq C_n (1+|\alpha|)^{-1}, &  |n|&\geq 3,
    \end{align}
    where $|n|=\sum_a n_a$.
    \item \textbf{Nondegenerate diffusion}  The scaled diffusion matrix,
    \[
    \Omega^{ab}(\alpha) := \Real \sum_k \partial^a L_k(\alpha) \partial^bL_k^*(\alpha),
    \]
    is uniformly bounded from below, that is $\Omega\geq c\IdM$ for some $c>0$. 
    \end{itemize}
\end{restatable}
\noindent In particular, $H$ may grow at most quadratically at infinity and $L_k$ may grow only linearly at infinity. 
(For the more permissive --- but also more technically involved --- conditions under which our main result applies, see Assumption~\ref{assum:suitableLindblad} in Section~\ref{sec:definitions-and-assumptions}. 
We point out that under Assumption~\ref{assum:simpleSuitableLindblad} the Lindbladian $\sin(x)$ is \textit{not} admissible, but it is under Assumption~\ref{assum:suitableLindblad}.) 
We then have the following.
\begin{restatable}[Main result, simplified]{thm}{thmInformal}
\label{thm:informal}
Let $H\in C^\infty(\bbR^d\times\bbR^d)$ and $L_k\in C^\infty(\bbR^d\times\bbR^d)$ be Hamiltonian and Lindblad symbols satisfying Assumption~\ref{assum:simpleSuitableLindblad}.
Also let $\rho_0$ be a coherent state (i.e., a rank-1 normalized Gaussian operator with covariance matrix $\sigma\propto I$), or a probabilistic mixture (i.e., convex combination) of such states.   If $\rho(t)$ solves the Lindblad equation~\eqref{eq:lindblad-simple} with 
initial data $\rho_0$ and $f(t)$ solves the corresponding Fokker-Planck equation~\eqref{eq:fp-simple} with $f(t\liq 0) = \WW[\rho(t\liq 0)]$, 
then for any classical observable $A\in L^\infty(\bbR^d\times\bbR^d)$ corresponding to a quantum observable
$\hat{A} = \Op[A] \in \mathcal{B}(L^2(\bbR^d))$ we have:
\begin{equation} \label{eq:smooth-observable-bound}
\left|\Tr[\rho(t) \hat{A}] - \int f(t) A \, \dd \alpha\right| \leq  
(\|A\|_{L^\infty} +
\|\hat{A}\|_{\mathrm{op}})
\erate t
\end{equation}
with error rate
\begin{align}\label{eq:main-informal-r}
   \erate = C(H, \LCk)\, \hbar^{1/2} 
   \max\{ \diffstrength^{-3/2}, \diffstrength\}.
\end{align}
\end{restatable}

In the above theorem the constant $C(\HC, \LCk)$ depends only on the functions $\HC$ and $\LCk$, and is finite\footnote{More precisely, this constant only depends on $\|\HC\|_{C^{2d+4}}, \|\LCk\|_{C^{4d+6}}$, the ellipticity constants $\lambda$ and $\Lambda$ appearing in Assumption~\ref{assum:suitableLindblad}, and the nonlocal quantity in~\eqref{eq:nonlocal-Lk-norm}.} so long as $(\HC,\{\LCk\}_{k=1}^K)$ satisfies Assumption~\ref{assum:simpleSuitableLindblad}. 
For fixed coupling strength $\diffstrength$, the error accumulates in time as $t\sqrt{h}$, guaranteeing small error for times $t \ll \hbar^{-\frac{1}{2}}$. If we take $\diffstrength\to 0$ as $\hbar\to 0$, the error is dominated by the term $t\hbar^{1/2}\diffstrength^{-3/2}$.  So in general, if $\diffstrength \gtrsim \hbar^{1/3 - p}$ for some $p>0$,
or equivalently $D \gtrsim \hbar^{4/3 - p}$, the error is small for times $t \lesssim \hbar^{-q}$ for $q=\min\{\frac12,\frac{3p}{2}\}$. The correspondence time for different regimes is illustrated in Figure~\ref{fig:regimes}.

Theorem~\ref{thm:informal} above is a corollary of Theorem~\ref{thm:mainResult} below, which is stronger both quantitatively (specifying how $C(\HC,\LCk)$ scales with the derivatives of $\HC$ and $\LCk$ more precisely\footnote{Indeed, the purpose of allowing general coupling strength $\diffstrength$ in this introduction is to let us describe how our error rate in Theorem~\ref{thm:informal} scales with the overall amplitude of $\HC$ and $\LCk$ without tracking the dependence on other features of these functions. Theorem~\ref{thm:mainResult} contains strictly more information about the dependence of the error rate on the features of $\HC$ and $\LCk$, making $\diffstrength$ redundant.}) and qualitatively (controlling the correspondence between $\rho(t)$ and $f(t)$ without reference to any observable).
In a short companion paper \cite{hernandez2023decoherence1}, we apply the same techniques to the special case of Hamiltonians of the form $\HQ = \PQ/2m+V(\XQ)$ with linear and Hermitian Lindblad functions (and thus frictionless dynamics).
The special case there allows more explicit bounds and physical discussion.

In contrast to our shorter paper, Theorem \ref{thm:mainResult} also has the benefit of applying to any sufficiently smooth Hamiltonian and Lindblad operators. 
Some assorted examples of Hamiltonians that do not take the special form include: (1) non-linear optical systems (expressed in quadratures), like Kerr oscilators, (2) the beyond-leading-order terms in the non-relativistic expansion for a particle in an inhomogeneous gravitational field with kinetic term $p_\mu p_\nu g^{\mu\nu}(x)$, and (3) quasiparticles with an effective position-dependent dispersion relation.  Moreover, although linear Lindblad operators are widely deployed and convenient approximations, in many cases non-linear Lindblad operators are necessary to avoid unphysical effects \cite{gallis1990environmental}.

The strategy for proving Theorem~\ref{thm:mainResult} is to construct an auxilliary density matrix $\rhot(t)$ given by a time-dependent mixture of Gaussian states,
such that (1) $\rho(t)$ approximates $\rho(t)$ in the trace norm and (2)  $\WW[\rhot(t)]$ approximates $\rhoc(t)$ in the total variation distance (the $L^1$ norm).  
To this end, we introduce a new strategy for representing quantum states as a mixture of Gaussians with covariance matrices that are allowed to dynamically evolve but never get too strongly squeezed. This can be seen as a generalization of both the Glauber-Sudarshan P-function \cite{glauber1963coherent, cahill1969density, sudarshan1963equivalence, agarwal1970calculus} and the ``thawed Gaussian'' techniques of Heller and Graefe et al.~\cite{heller1975time,heller1981frozen,graefe2018lindblad}.  Our technique contrasts the traditional semiclassical analysis strategy of defining an appropriate symbol class and working strictly within it, since $\rhot(t)$ is generally a convex combination of states squeezed in different directions, thus a combination of symbols belonging to different (incompatible) symbol classes.  This gives us the flexibility to allow the Gaussian states to squeeze and stretch, granting us the full expressiveness of Heller's ``thawed'' approximation.  
Because $\rhot(t)$ is a good approximation to $\rho(t)$ in trace norm, this also suggests that approaches based on analysis within a single symbol class (for example, methods involving the FBI transform) are unable to obtain error estimates in trace norm with the optimal scaling in $\hbar$.

One might wonder how our bound depends on our choice of convention for the Lindblad equation in Eq.~\eqref{eq:lindblad-simple}, where the Lindblad operators $\LQk$ have a $\hbar^{-1}$ pre-factor just like the Hamiltonian.  For instance, this equation is sometimes written with an $\hbar^0$ or $\hbar^{-2}$ pre-factor instead on the Lindblad terms.
\footnote{The $\hbar^0$ and $\hbar^{-2}$ factors are natural boundaries: Suppose one uses a $\hbar^{-n}$ prefactor and takes $\hbar\to 0$ while holding the Lindblad functions and $\diffstrength=1$ fixed. For $n<0$, the physical diffusion $\D$ on phase space diverges (i.e., classical dynamics are swamped by environment-induced noise). For $n>2$, superpositions over macroscopic intervals $\alpha$, which decohere at a rate $\hbar^{-2} \alpha^a \D_{ab}\alpha^b$, become stable (never decohering) as $\hbar\to 0$. See Fig.~\ref{fig:regimes} and the discussion in Section~\ref{sec:corresponedence-time}. Our choice of $n=1$ lies in the middle of these two boundaries, yields finite friction $\GC^a$ as $\hbar\to 0$, and ensures $\LCk^*\LCk$ has the same physical units as $\HC$.}
These alternative conventions for $\hbar$ factors can be accommodated by taking $\diffstrength$ to depend differently on $\hbar$.  Regardless, we can also frame result our result in terms of the strength of the diffusion $D$ given by Eq.~\eqref{eq:diffusion-matrix}.  For instance, Theorem \ref{thm:informal} implies $D \gg \hbar^{4/3}$ suffices for an accurate quantum-classical correspondence. Such statements are independent of any conventions about the $\hbar$ factors appearing in the Lindblad equation.
\footnote{\label{fn:no-correct-scaling}It might seem that when deriving the Lindblad equation for a system coupled to an abstract bath (see the the heuristic argument in \cite{manzano2020short} or the more detailed \cite{nathan2020universal}), there should be a definitive answer about which power of $\hbar$ precedes the Lindblad operators (when $\diffstrength$ is fixed), or equivalently how $\diffstrength$ should depend on $\hbar$. Indeed, naively these derivations suggest $\diffstrength \sim \hbar^{-1}$, or an overall factor of $\hbar^{-2}$ on the Lindblad operators.  However, the Lindblad operators depend on the bath correlation function, which may actually depend on $\hbar$.  There is perhaps no canonical answer as to how one should choose these $\hbar$ factors in the abstract: different physical mechanisms for different system-bath couplings may have different $\hbar$ dependencies; see \cite{joos2013decoherence} for some examples of decoherence mechanisms and their associated $\hbar$-dependence.}

While we have touted that our bound is useful beyond the Ehrenfest time, one might ask: how interesting are the quantum and classical distributions beyond this time?  For simple chaotic systems with bounded accessible phase space, one expects that these systems ``thermalize'' after several multiples of the Ehrenfest time, i.e.\ spread somewhat uniformly over the allowed phase space, in which case our bound would be comparing two thermalizing distributions (which is non-trivial regardless).  However, in chaotic systems with large accessible phase spaces, or with both chaotic and non-chaotic degrees of freedom, or regardless with degrees of freedom that thermalize at very different speeds, this simple picture breaks down, and the dynamics beyond the Ehrenfest time may be much  more interesting.

\subsection{Structure of the paper}
In the rest of Section~\ref{sec:incoherent}, we discuss quantum-classical correspondence times, give a heuristic justification for the asymptotic scaling we see, and summarize previous and future work.
In Section~\ref{sec:pf-overview} we present a 
heuristic overview of the proof for Theorem~\ref{thm:informal} and~\ref{thm:mainResult}, including an explanation of the appearance of the factor $\diffstrength^{-\frac{3}{2}}$. In Section~\ref{sec:main-result} we make some definitions and formally state Theorem~\ref{thm:mainResult}.  
We prove Theorem~\ref{thm:mainResult} (which implies Theorem~\ref{thm:informal}) in Section~\ref{sec:main_proof}, but before this
we first review notation in Section~\ref{sec:notation} and present some preliminary facts about harmonic approximations for the Lindblad and Fokker-Planck equations in Section~\ref{sec:tech-prelim}.
In Sections~\ref{sec:linear-algebra-facts}, \ref{sec:harmonic-error}, and~\ref{sec:moyal} we prove some lemmas
needed in the main proof.  Appendix~\ref{sec:units-symplectic-covariance} discusses physical units and symplectic covariance, and illustrates them with Corollary~\ref{cor:MinDiff}.

Readers interested in understanding the argument in a simpler setting may prefer to review the companion paper \cite{hernandez2023decoherence1} which treats the special case of Hamiltonians of the form $H=p^2+V(x)$ with linear Lindblad operators. 

\subsection{Quantum-classical correspondence times for different coupling strengths} \label{sec:corresponedence-time}

We summarize what we know about the quantum-classical correspondence, or how well the quantum and classical trajectories match, for different regimes of coupling strength $\diffstrength$.  In each regime we ask about the loosely defined \introduce{correspondence time}, also called the ``(quantum) breaking time'': the timescale before which the trajectories are guaranteed to approximately match, and after which they may differ appreciably in some systems.  

The notion of a correspondence time depends on the metric by which we measure the distance between the quantum state $\rho(t)$ and classical distribution $f(t)$.  One possibility, and the route we take in this work, is to show the existence of a quantum trajectory $\rhot(t)$ such that both $\Lonenorm{\WW[\rhot(t)]-f(t)} =o(\hbar)$ and $\Trnorm{\rhot(t)-\rho}= o(\hbar)$.  That is, we find a trajectory that both (1) matches the quantum trajectory for all quantum observables and (2) matches the classical trajectory for all classical observables. Another possibility would be to demand both $\Lonenorm{\WW[\rho(t)]-f(t)} = o(\hbar)$ and $\Trnorm{\rho(t)-\Op[f(t)]} =o(\hbar)$, which we were not yet able to show using our method, though which we speculate may be possible as a corollary. Finally, there is a weaker notion of correspondence: one might only require that the trajectories match for ``macroscopic observables,'' e.g.\ requiring only that $|\Tr[\rho(t) \hat{A}]-\int d\alpha f(\alpha) A(\alpha)| = o(\hbar)$ for smooth symbols $A=\WW[\hat{A}]$ that do not depend on $\hbar$.  In fact, there has been some speculation that such a weaker notion of correspondence may hold for \textit{all} times even in \textit{closed} systems \cite{emerson2002quantum} except perhaps in certain fine-tuned situations, but there may also be numerical evidence to the contrary \cite{karkuszewski2002breakdown, carvalho2004environmental}.  Regardless, we do not explore this weaker notion of correspondence.

 \begin{figure}[tb]
    \centering
    \includegraphics[width=\linewidth, valign=t,scale=1.]{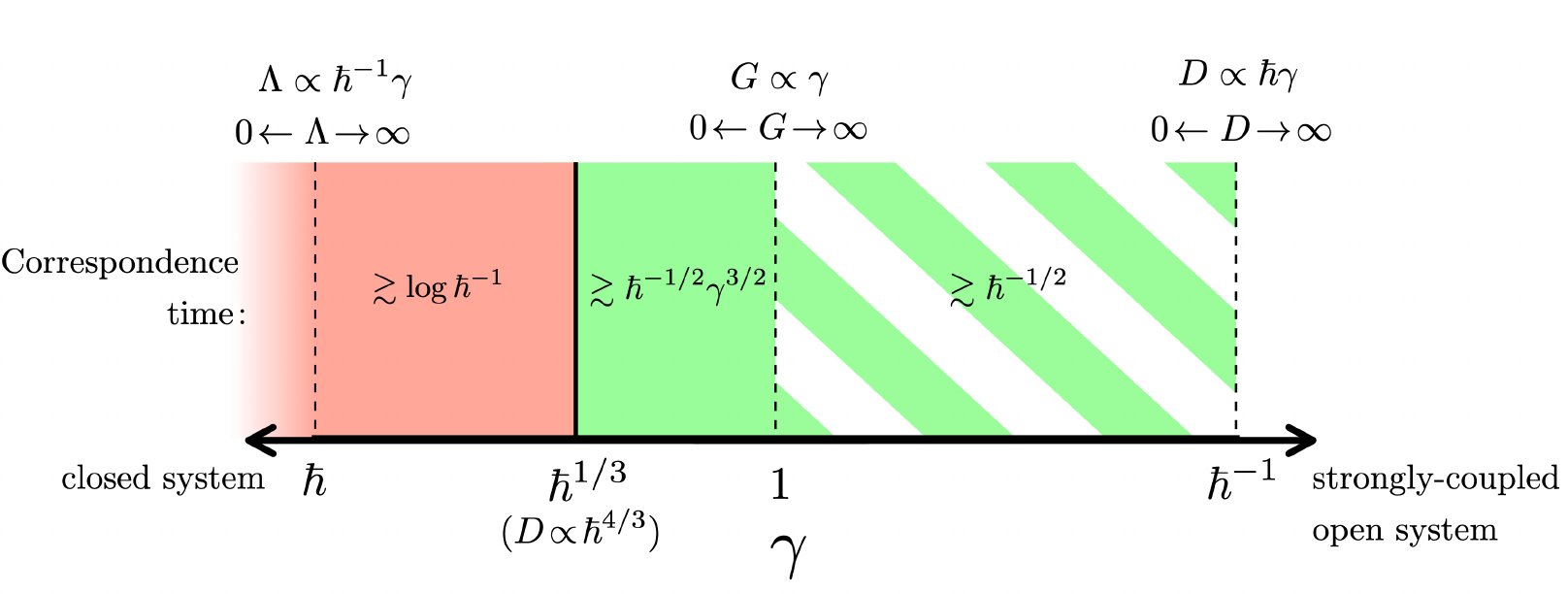}
    \caption{We illustrate the quantum-classical correspondence time  (also know as the quantum breaking time) in different regimes of the coupling strength $\diffstrength$.
     Unless they are set to exactly zero (e.g.\ $\LCk=0$), the classical diffusion $\D$ and classical friction $\GC$ scale like 
     $\diffstrength\hbar$ and $\diffstrength$ 
     respectively. 
     When $\hbar^0 \ll \diffstrength$, friction must be assumed exactly zero or else the classical dynamics will become will become singular (and likewise for diffusion when $\hbar^{-1} \ll \diffstrength$).}
    \label{fig:regimes}
\end{figure}

In Fig.~\ref{fig:regimes} we illustrate our conclusions about the quantum-classical correspondence time from Theorem~\ref{thm:informal}, when using the notion of correspondence and initial state specified there.  We take $\diffstrength$ to depend on $\hbar$, plotted along the horizontal axis, and we consider the correspondence time as well as the strength of the diffusion $D \propto \hbar \diffstrength$ and friction  $G \propto \diffstrength$. We also consider the \introduce{localization matrix} or ``decoherence matrix'' \cite{joos1985emergence}
\begin{align}\label{eq:localization-matrix-def}
    \locD : = \hbar^{-2} \D
\end{align} 
which characterizes the inverse timescale on which a Schr\"odinger cat state (two wavepackets initially superposed over an arbitrary fixed macroscopic distance) will decohere;\footnote{More precisely, for linear Lindblad operators with constant diffusion matrix $\D$, the matrix $\locD_{ab}=\hbar^{-2}\sf_{ac}\sf_{bd}\D^{cd}$ characterizes how a superposition of two wavepackets with separation $\alpha$ decoheres: the interference terms are suppressed by a factor $\exp(-t \alpha^a \locD_{ab}\alpha^a)$ \cite{joos1985emergence}.}
it scales as $\locD \propto \diffstrength\hbar^{-1}$.

The regime $\hbar^{1/3} \ll \diffstrength \ll 1$, or equivalently $\hbar^{4/3} \ll D \ll \hbar$, is shaded green, because there our main result shows the correspondence time is at least a negative power of $\hbar$ (and the true correspondence time may indeed be much longer).
\footnote{For $D \gg \hbar^{4/3}$ in the chaotic system studied in Ref.~\cite{toscano2005decoherence}, it appears the correspondence holds as the distributions approach their steadystate (after which the correspondence continues to hold trivially), meaning the correspondence time is in fact infinite. In contrast, for $D  \ll \hbar^{4/3}$, the trajectories diverge at the Ehrenfest time. In this sense, the border $D  \sim \hbar^{1/3}$ may be a sharp threshold.}
In this regime we also have that  friction $\GC$ and diffusion $\D$ vanish as $\hbar \to 0$, approaching closed Hamiltonian mechanics.  The regime $1 \ll \diffstrength \ll \hbar^{-1}$, or equivalently $\hbar^{4/3} \ll D \ll \hbar$, is partially shaded green, indicating the fact that the correspondence time $\hbar^{-1/2}$ is long, but that the friction $\GC$ will diverge as $\hbar\to 0$ --- making the Fokker-Planck equation singular --- unless the Lindblad functions are specifically taken to satisfy $0 = \Imag \sum_k \LCk \partial^a \LCk^*$ (i.e., unless the friction vanishes regardless of $\diffstrength$). At the border between these two regimes, $\diffstrength \sim 1$, the corresponding classical dynamics generically exhibit finite friction.\footnote{Instead of using the coupling strength $\diffstrength$, one could consider a family of quantum and classical systems where the Lindblad functions are taken to depend on $\hbar$ in a more complicated way, e.g., so that the friction and diffusion are both finite as $\hbar\to 0$. As briefly discussed in footnote~\ref{fn:no-correct-scaling}, it is not clear that there is a single ``correct'' scaling.}

For $\diffstrength \ll \hbar^{1}$ (including the  case $\diffstrength=0$ of exactly closed systems), decoherence is too weak to prevent Schr\"odinger cat states from being generated in chaotic systems, leading to a breakdown of correspondence at the Ehrenfest time $\teh\sim\log \hbar^{-1}$. Based on the numerical results of Toscano et al.~\cite{toscano2005decoherence,toscano2006quantum,wisniacki2009scaling} and unpublished work with Y. Borns-Weil, we conjecture that this lack of correspondence extends through the regime $\diffstrength \ll \hbar^{1/3}$ 
(marked by  ``$\log \hbar^{-1}$'' in Fig.~\ref{fig:regimes}). 

To summarize, if our conjecture is true, then 
\begin{enumerate}
    \item In the regime 
    $\D \ll \hbar^{4/3}$ (i.e., $\diffstrength \ll \hbar^{1/3}$), there is a loss of correspondence after the Ehrenfest time for at least some observables.
    \item The regime 
    $\hbar^{4/3} \ll \D \ll \hbar^{1}$ (i.e., $\hbar^{1/3} \ll \diffstrength \ll \hbar^{0}$) achieves correspondence beyond the Ehrenfest time; this regime characterizes the seemingly reversible macroscopic classical systems of everyday life. 
    \item The regime 
    $\hbar^{1} \ll \D \ll \hbar^{0}$ (i.e., $\hbar^{0} \ll \diffstrength \ll \hbar^{-1}$)
    also exhibits the quantum-correspondence, but the classical dynamics are singular (due to divergent friction) unless the Lindblad functions induce precisely zero friction.
    \item In the regime $\hbar^{0} \ll \D$ (i.e., $\hbar^{-1} \ll \diffstrength$), the diffusion diverges, giving singular classical dynamics.
\end{enumerate}

While we describe the regime with $D \to 0$ as vanishing diffusion, or vanishing noise, we must take some care with timescales.  The formal limit $D \to 0$ in the Fokker-Planck equation Eq.~\eqref{eq:fp-simple} indeed results in deterministic flow (in particular the classical Hamiltonian flow, if friction also vanishes). Fixing a timescale and taking $D$ sufficiently small, the evolution of smooth observables should be well-approximated by the $D=0$ classical flow.  Thus we say the classical evolution gives the appearance of zero noise over fixed timescales.  However, for any fixed $D>0$, at sufficiently large times $t \gtrsim \log(D^{-1})$ the diffusion may have dramatic effect, due to the exponential amplification of the noise by chaotic dynamics.

\subsection{Heuristic justification of the \ToP{$\hbar^{4/3}$}{ħ⁴ᐟ³} threshold from the Moyal bracket}
\label{sec:intuition-moyal}
While in Section~\ref{sec:pf-overview} we outline the reasoning that we ultimately make precise, here we offer an alternative heuristic argument below, via the Moyal bracket.  This argument does not rely on any harmonic approximation, but it suggests the same scaling for the error as given in Eq.~\eqref{eq:main-informal-r}.  The agreement with \eqref{eq:main-informal-r} suggests the dependence on $\hbar,\diffstrength$ may be optimal, or at least not an artifact of the harmonic approximation.

In a closed quantum system, the Wigner function $f$ evolves under Hamiltonian $\HC$ by \cite{curtright2014concise}
\begin{align}
    \partial_t f & = \Moyalbracket{\HC,f} \\
    &= \frac{2}{\hbar} \HC \sin \left( \frac{\hbar}{2}\, \cev{\partial}\!{}_a\vec{\partial}{}^a \right) f \\ 
    &= \sum_{n=0}^\infty \frac{(-\hbar^2/4)^n}{(2n+1)!} \left(\partial_{a_1}\cdots \partial_{a_{2n+1}} \HC\right)\left(\partial^{a_1}\cdots \partial^{a_{2n+1}} f\right) \\ 
    \label{eq:moyal-bracket-expansion}
    &= (\partial_a \HC)(\partial^a f) - \frac{\hbar^2}{24}(\partial_a\partial_b\partial_c \HC)(\partial^a \partial^b \partial^c f) + \ldots
\end{align}
where $\Moyalbracket{\cdot,\cdot}$ is the Moyal bracket, and 
$\cevpartial$ and $\vecpartial$ denote partial derivatives that are understood to act on everything left and right (extending beyond the parentheses), as illustrated by the subsequent line.
(The power series is a formal expansion, and we do not discuss its convergence, but it is useful for the intuition below.)

Say $\HC$ only varies over order-unity scales (i.e., independent of $\hbar$), and say the Wigner function $f$ has minimum length scale $w$ that may depend on $\hbar$, e.g.\ maybe $f$ has long tendrils, with minimum width $w$. 
Then $\partial^3 f \lesssim w^{-3} f$, so the leading $\hbar$-dependent term above is roughly $\hbar^2 w^{-3} f$, or  
\begin{align}\label{eq:wigner-expand}
    \partial_t f \sim (\partial_a \HC)(\partial^a f) + [\hbar^2 w^{-3} f] + \ldots
\end{align}
So given a classical solution $f(t)$, the error between the quantum and classical evolution generators acting on $f$ is like
\begin{align} \label{eq:wigner-expand-norm}
    \norm{\partial_t f - (\partial_a \HC)(\partial^a f)}_{L_1} \lesssim \hbar^2 w^{-3}.
\end{align}
We can ignore the higher-order terms $\hbar^{2n}w^{-(2n+1)}$ because they are small when the leading term $\hbar^2 w^{-3}$ is small, i.e.\ when $w \gg \hbar^{\frac23}$.

Now consider an open system with diffusion $D$. The classical evolution under the Fokker-Planck equation \eqref{eq:fp-simple} will produce a distribution $f$ with minimum length scale 
\begin{align} \label{eq:min-width}
    w \sim \sqrt{D/\lambda_L},
\end{align}
for maximal local Lyapunov exponent $\lambda_L$.  (This is the scale at which the diffusion balances the squeezing; see Fig.~\ref{fig:ellipses}.) If we assume linear Lindblad operators for simplicity, i.e.\ constant diffusion $D$, there is no quantum correction associated to this term (see Section~\ref{sec:harmonic-approx}). Therefore Eq.~\eqref{eq:wigner-expand-norm} again holds, and
so
\begin{align}
    \norm{\partial_t f - (\partial_a \HC)(\partial^a f)}_{L_1} \lesssim \hbar^2 D^{-\frac32}.
\end{align}
Note this quantifies the rate at which the quantum and classical evolution can diverge.  Using a Duhamel-type argument as in Sections~\ref{sec:quantum-duhamel} and \ref{sec:classical-duhamel}, the cumulative error after time $t$ is then at most
\begin{align}
    \norm{f(t) - \WW[\rho(t)]}_{L_1} \lesssim t \hbar^2 D^{-\frac32}.
\end{align}
which matches the $\erate \sim \hbar^{1/2}\diffstrength^{-3/2}$ scaling for the error rate in Eq.~\eqref{eq:main-informal-r}.  We again conclude the quantum and classical evolutions match (for times at least $t \ll \hbar^{1/2}$) when $D \gg \hbar^{4/3}$.

Some previous literature \cite{kolovsky1994remark, zurek2003decoherence}, in accords with some numerical studies \cite{pattanayak2003parameter, gammal2007quantum}, has used a different heuristic to conclude that the weaker condition $D \gg \hbar^2$, rather than $D \gg \hbar^{4/3}$, is sufficient for matching quantum and classical evolutions as $\hbar \to 0$.
Here is one attempt to paraphrase these arguments in the context of the calculation above, although this paraphrase may be incorrect: 
The first two terms in Eq.~\eqref{eq:wigner-expand} are schematically size $w^{-1} f$ and $\hbar^{-2} w^{-3} f$ respectively, and one might claim the second and higher terms in Eq.~\eqref{eq:wigner-expand} can be dropped when the second term is small compared to the first term, or $w \gg \hbar$, which by \eqref{eq:min-width} requires only $D \gg \hbar^2$.

However, we suggest that the second term being small relative to the first does not justify dropping it since, in fact, both terms may be large.
To emphasize with a related example, consider a Gaussian coherent state in phase space with minor axis of thickness $w \sim \hbar$, traveling at unit speed parallel to this short axis. Then both $\Lonenorm{\partial_t f}$ and $\Lonenorm{\partial_t \WW[\rho]}$ are diverging like $\hbar^{-1}$ as $\hbar \to 0$, because although the wavepacket travels at unit speed, the small support of the wavepacket quickly becomes disjoint from its previous location. For $f(t)$ and $\WW[\rho(t)]$ to match after time $t$, it is not sufficient for them to diverge at a rate slow compared to the large rate $\Lonenorm{\partial_t f}$.  Instead, they must diverge at a rate small compared to $t$.

\subsection{Previous work}
\label{sec:prev-work}

In the introduction, we briefly cited some of the large literature on the quantum-classical correspondence that motivated the present paper.  Here, we will discuss in a bit more detail some  earlier approximation techniques and how they relate to our results.

Ehrenfest's theorem \cite{ehrenfest1927bemerkung} from 1927 states that for Hamiltonians of the form 
$\HQ = \frac{\PQ^2}{2m} + V(\XQ)$ and any wavefunction $\psi$
solving the Schr\"{o}dinger equation (and hence for a closed system), the observables $\XQ$ 
and $\PQ$ instantaneously satisfy
\begin{equation}
\begin{split} \label{eq:ehrenfest}
     \frac{\dd}{\dd t}\langle \psi| \XQ |\psi\rangle
    = m^{-1}\langle \psi| \PQ|\psi\rangle  \qquad\qquad
    \frac{\dd}{\dd t}\langle \psi| \PQ |\psi\rangle
    = -\langle \psi| \nabla V(\XQ)|\psi\rangle.
\end{split}
\end{equation}
As Ehrenfest remarked, when a state $\psi$ is localized 
in position,
one can approximate $\langle \psi| \nabla V(\XQ)|\psi\rangle \approx \nabla V(\langle \psi |\XQ|\psi\rangle)$ in Eq.~\eqref{eq:ehrenfest} to obtain an ODE for for the time evolution of the expectation values $\langle \psi |\XQ|\psi\rangle$ and $\langle \psi |\PQ|\psi\rangle$, yielding Hamilton's classical equation of motion. This provides heuristic justification for the correspondence between classical and quantum mechanics.  
More rigorously, when paired with a bound for the rate of stretching in phase space of the function $\psi(t)$, one can use Ehrenfest's theorem to prove a comparison between the quantum and classical evolutions at some finite time.  
In contrast, Egorov's theorem \cite{egorov1969canonical} (see Zworski \cite{zworski2022semiclassical} for a modern introduction) 
is a finite-time comparison of 
a Heisenberg-picture operator
$\hat{A}(t) =  e^{it\HQ/\hbar} \Op[A_0] e^{-it\HQ/\hbar}$ (evolved with the Schr\"{o}dinger equation) and the quantization of the corresponding classical variable $A_\cl(t) = e^{it\LLCq} A_0 e^{-it\LLCq}$ (evolved with Liouville's equation).

Heller \cite{heller1975time} first approximated the evolution of a Gaussian state in a non-harmonic potential of a closed quantum system by making a local harmonic approximation, leading to a Gaussian whose center follows the classical trajectory and whose shape distorts over time.  This method is sometimes called the ``thawed Gaussian approximation.'' (In contrast, the ``frozen Gaussian approximation'' \cite{heller1981frozen} uses a covariance matrix fixed in time.) Much more recently, Graefe et al.~\cite{graefe2018lindblad} present an analogous approximation for open systems.  The Gaussian approximation method has been used to simulate a variety of quantum-mechanical phenomena (see Refs.~\cite{davis1979semiclassical,lee1982exact,drolshagen1983time}), in addition to sampling-based methods for the Fokker-Planck equation \cite{scharpenberg2012adaptive}.  

In terms of analytical results for bounding the error introduced by the Gaussian approximation, an error bound for the thawed Gaussian approximation was first calculated by Hagedorn~\cite{hagedorn1998raising} (see Theorem 2.9) for closed systems of the form $\HQ = \PQ^2 + V(\XQ)$.
For a more recent treatment with an emphasis on numerical methods see Lemma 5 of Bergold \& Lasser~\cite{bergold2022error}.
We are not aware of any analogous results for open systems. We present such a result in Lemma \ref{lem:HarmErrorQ}, for a general class of Hamiltonian and Lindblad operators.  
Note that even within the setting of closed systems, one can reach longer timescales by generalizing the set of states one is willing to consider from Gaussian coherent states to more general WKB states.  The degeneration of wavepackets into delocalized states was studied using local harmonic approximations by Schubert, Vallejos, \& Toscano \cite{schubert2012wave}.

The formal correspondence between the quantum Lindblad equation for the Wigner function and the classical Fokker-Planck equation has frequently been discussed for the case of linear Lindblad operators. For more general Lindblad operators, the formal limit of the Lindblad equation (i.e.\ dropping terms subleading in $\hbar$) has been shown to yield a Fokker-Planck equation in Refs.~\cite{frigerio1984diffusion,tzanakis1998generalized,strunz1998classical,dubois2021semiclassical},\footnote{In particular, Dubois et al.~\cite{dubois2021semiclassical} consider the case of a curved phase space, necessitating modified Poisson brackets.} 
similar to our development in Section~\ref{sec:corresponding-dynamics}.

The question of how long the quantum-classical correspondence holds in open quantum systems and how much diffusion is necessary has been discussed extensively \cite{zurek1994decoherence,zurek1995zurek, habib1998decoherence,bhattacharya2000continuous,bhattacharya2003continuous,toscano2005decoherence,toscano2006quantum,wisniacki2009scaling}, though without rigorous general results. It has been suggested that the condition $D \gg \hbar^2$ is sufficient to ensure a lack of coherent superposition over order-unity scales \cite{zurek1994decoherence}, which is one component of a quantum-classical correspondence.  More strongly, some arguments suggest that $D \gg \hbar^2$ is sufficient \cite{zurek2003decoherence,kolovsky1994remark,kolovsky1996condition, pattanayak2003parameter, gammal2007quantum} to ensure closely matching quantum and classical evolutions, though see the comments at the end of Section~\ref{sec:intuition-moyal}. In contrast, numerical evidence and heuristic arguments for specific systems (kicked harmonic oscillators) in \cite{toscano2005decoherence,toscano2006quantum,wisniacki2009scaling} suggest the error between the quantum and classical trajectories is genuinely proportional to $\hbar^2 D^{-3/2}$, and in particular the error may be large when $\hbar^2 D^{-3/2}$ is large, even as $\hbar \to 0$. The numerical evidence thus suggests $D \gtrsim \hbar^{4/3}$ is actually necessary for quantum-classical correspondence in some systems.  The heuristic in Section~\ref{sec:intuition-moyal} is consistent with this conclusion.  If that were true, our bound in Theorem \ref{thm:informal} would have optimal dependence on $\diffstrength$ and  $\hbar$, and $D \sim \hbar^{4/3}$ would be a genuine threshold.

\subsection{Future work}\label{sec:future-work}

We list several questions left open, roughly ordered from more significant questions at the top to more minor questions at the bottom which may only require small improvements to our argument.
\begin{enumerate}
    \item  Does a similar bound apply in the case of an arbitrary initial state, rather than a mixture of Gaussian wavepackets? We expect that arbitrary initial states will decohere into an approximate mixture of Gaussian wavepackets, without substantially changing the expectation of classical smooth variables on phase space, on a timescale that vanishes as $\hbar\to 0$.  (Indeed, there is reason to think this may happen exactly in finite time~\cite{diosi2002exact, brodier2004symplectic, eisert2004exact, riedel2016quantum}.) 
    \item Does a similar bound apply in the case of a degenerate diffusion matrix, such as when position but not momentum is decohered? Degenerate diffusion matrices arise naturally, e.g., in the case of collisional decoherence \cite{joos1985emergence, gallis1990environmental, schlosshauer2008decoherence}. 
    \item Do similar results hold for different phase spaces, e.g.\ for the correspondence between classical spins and large quantum spins? There generalizations of the Moyal product may be used. 
    \item Do similar error bounds apply uniformly in time for some systems?  One might expect that even though the errors accumulate, they may be continuously washed away as the system thermalizes.  Then the ``correspondence time'' discussed in Section~\ref{sec:corresponedence-time} would be infinite in the appropriate regime, consistent with the numerical simulations in \cite{toscano2005decoherence}. The Duhamel-based bound presented here, which simply adds together the errors that accumulate at each time step without allowing them to cancel, would have to be modified.
    \item Can the scaling exhibited in Theorem \ref{thm:informal} in terms of $\diffstrength$ and $\hbar$ be shown to be optimal? As discussed in Section~\ref{sec:prev-work}, evidence from \cite{toscano2005decoherence,toscano2006quantum,wisniacki2009scaling} suggests this may be the case.
    \item Can the results be generalized to handle $\HC$ and $\LCk$ that are irregular in ways that violate Assumption~\ref{assum:suitableLindblad} but only in regions of phase space that are essentially inaccessible to the quantum state? For instance, currently we must assume the Hamiltonian grows at most quadratically at infinity so that the local harmonic dynamics associated with $\nabla^2 \HC$ have strength that is bounded over phase space, but this shouldn't be necessary if the Hamiltonian diverges positively in all directions and the state has bounded energy since this means it is confined to a bounded region that never sees this growth.
    \item  Can one more directly relate the quantum evolution $\rho(t)$ and classical evolution $f(t)$, without the intermediary $\rhot$? Perhaps one can bound $\Trnorm{\rho(t)-\Op[f(t)]}$ and/or $\Lonenorm{\WW[\rho(t)]-f(t)}$.
    \item Can the heuristic in Section~\ref{sec:intuition-moyal} using the Moyal bracket be made rigorous?  
    \item  Can the length and complexity of the argument be reduced?  In particular we expect the size and especially $d$-dependence of the constants can certainly be improved. More fully exploiting symplectic symmetry may help.  See Appendix~\ref{sec:units-symplectic-covariance} for more discussion of this point.
\end{enumerate}

\subsection{Acknowledgements}
We thank Yonah Borns-Weil for valuable discussion and collaboration on related work and Wojciech Zurek for foundational insights that inspired this work. We thank Joseph Emerson, Fabricio Toscano, and Diego Wisniacki for helpful discussion of numerical evidence for the quantum-classical correspondence in closed and open systems.  
We also thank Jeff Galkowski and Maciej Zworski for pointing out to us how to derive the contraction property of the Lindblad evolution.
DR acknowledges funding from NTT (Grant AGMT DTD 9/24/20).  FH was supported by the Fannie and John Hertz Foundation.

\section{Overview of the proof}
\label{sec:pf-overview}

\begin{figure}[tb]
    \centering
    \includegraphics[width=\linewidth, valign=t,scale=1.]{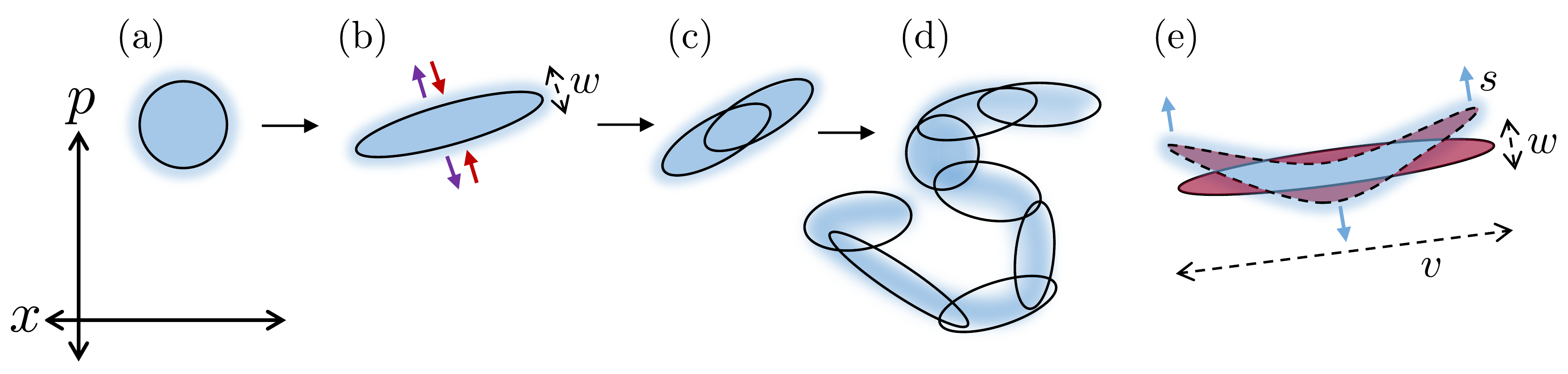}
    \caption{
    (a) An initial pure quantum Gaussian state  $\rho(t \liq 0)$ evolves in phase space.
    (b) At short times the dynamics admit a local harmonic (quadratic) approximation, broadening the distribution via diffusion (purple arrows) and possibly squeezing it via classical flow (red arrows). For diffusion strength $D$ and local Lyapunov exponent $\lambdaL$ of the flow, the Gaussian state (ellipse) has a minimum thickness: the diffusion broadens the ellipse at speed $\dot{w} \sim D/w$, while the the Hamiltonian flow can shrink the width by at most $\dot{w} \sim -w /\lambdaL$, with the competing effects balanced at $w \sim (D/\lambdaL)^{1/2}$. 
    (c) After $\rho(t)$ becomes mixed due to diffusive broadening, it can be approximated by a mixture $\rhot(t)$ of pure Gaussian states (ellipses) that are individually less squeezed. Each evolves by its own local harmonic dynamics while continuously being further decomposed.
    (d)  As $\rho(t)$ spreads in phase space, our approximation $\rhot(t)$ uses ellipses of fixed area $\hbar$ but varying amounts of squeezing.
     (e) 
     The minimum thickness $w$ controls the error of the harmonic approximation: 
     the dynamics are perturbed by the leading-order anharmonicity $\nabla^3 \HC$, which is strongest (relative to the center) at the tips of the ellipse lying on either end of the long axis $v \sim \hbar/w$.  This changes the speed of the local flow by $s \lesssim v^2 \|\nabla^3 \HC\|$, 
     so the discrepancy (red shaded area) between the true distribution (curved boomerang) and the ellipse grows at rate $\lesssim sv$.  
     Compared to the ellipse's area $\hbar$, this gives an error rate $sv/\hbar \lesssim  (\hbar^{4/3}/\D)^{3/2} \lambdaL^{3/2} \|\nabla^3 \HC\| $, which is small when $\D \gg \hbar^{4/3}$.}
    \label{fig:ellipses}
\end{figure}

We sketch the ideas behind the proof of Theorem~\ref{thm:informal}.  We offer a synopsis before elaborating, perhaps initially opaque: we approximate $\rho(t)$ with a mixture $\rhot(t)$ of pure Gaussian states, each of which evolves according to a local quadratic expansion of the Lindbladian, while being continuously decomposed into a further mixture of Gaussian states, which never become overly stretched or squeezed due to the diffusion induced by the Lindblad operators.  See Fig.~\ref{fig:ellipses}.

A key tool is the use of Gaussian quantum states $\tauQas$, which are  precisely the states that have Gaussian Wigner functions, each specified by its mean $\alpha \in \mathbb{R}^{2d}$ and covariance matrix $\sigma$. We review intuition here. (See Section~\ref{sec:harmonic-gaussian} for details.) We often visualize Gaussian states $\tauQas$ in phase space as ellipses centered at $\alpha$, with principal axes and (squared) lengths given by the eigenvectors and eigenvalues of $\sigma$.  
These ellipses\footnote{In more than one spatial dimension (two dimensions of phase space), one can imagine Gaussian states as ellipsoids.} must have volume at least $(\hbar/2)^d$, achieving this minimum when the states are pure, i.e., when $\text{rank}(\tauQas)=1$.  By a generalization of Heisenberg's uncertainty principle,  $\sigma$ then has eigenvalues that come in pairs $(\lambda_1,\lambda_2)$ with product $\lambda_1 \lambda_2 = \hbar^2/4$. 
In the isotropic case $\sigma = (\hbar/2)\IdM_{2d}$, we call these pure Gaussian states ``coherent states,'' otherwise we refer to them as ``squeezed,'' imagining squeezed ellipses.

We approximate the quantum evolution $\rho(t)$ by  $\rhot(t)$, a positive mixture of pure Gaussian states: 
\begin{equation} \label{eq:overview-mixture}
\rho(t) \approx \rhot(t) := 
\iint \tauQas \dd\pmsrt(\alpha,\sigma)
\end{equation}
for some time-dependent probability measure $\pmsrt$ 
supported on pairs $(\alpha,\sigma)$ of points $\alpha$ in phase space and allowed covariance matrices $\sigma$ (i.e., both positive-definite, $\sigma > 0$, and (scaled) symplectic, $2\hbar^{-1}\sigma \in \mathrm{Sp}(2d,\bbR)$).
We assume the initial state $\rho(t\liq0)$ is a mixture of such Gaussian states, so that at time $t=0$ we can take $\rhot=\rho$ and the approximation is exact. In general, $\rho(t>0)$ is \textit{not} precisely a positive mixture of Gaussian states, so our task is to choose a suitable $\pmsrt$ and control the error $(\rho-\tilde{\rho})$.

To this end, we consider how a single $\tauQas$ evolves using the second-order expansion of the Lindbladian with respect to $\alpha$. We call this second-order expansion a ``harmonic approximation,'' because it approximates the true Hamiltonian by a generalized harmonic oscillator.\footnote{See Section~\ref{sec:harmonic-approx} for a precise definition of the harmonic approximation. We say ``generalized harmonic oscillator'' because, in addition to being skewed in phase space, the oscillator may be unstable in any number of directions.} 
In our harmonic approximation, pairs of Lindblad functions $\LCk$ are also expanded to quadratic order (roughly corresponding to a linear expansion of each $\LCk$), 
so that the dynamics are given by a damped harmonic oscillator with constant diffusion, or Brownian noise.  
Two key features of the harmonic approximation are that (1) it exactly preserves Gaussian states, and (2) the harmonic approximation of the quantum and classical dynamics agree.\footnote{For quadratic Hamiltonian and linear Lindblad operators, the agreement of the Lindblad equation and Fokker-Planck equation can be confirmed readily from the Moyal product expansion~\eqref{eq:moyal-bracket-expansion}. The exact preservation of Gaussian states follows from the observation that harmonic oscillators merely induce linear dynamics on phase space. A complete demonstration is found in Section~\ref{sec:harmonic-gaussian}.}
So under this approximation, $\tauQas$ remains a Gaussian state, with the center $\alpha$ following the classical flow 
while the covariance $\sigma$ evolves as
\begin{equation}
    \label{eq:cov-growth-intro}
    \partial_t \sigma = 
        (\hh + \GG) \sigma + \sigma (\hh + \GG)^\tp + \D,
\end{equation}
where $\hh = \sf \nabla^2 \HC$ consists of second derivatives of the Hamiltonian, and where $\D$ and $\GG$ are determined by the Lindblad operators, with $\D$ describing diffusion and $\GG$ related to friction.  (See Lemma \ref{lem:gaussian_evolution}.\footnote{Note we have set $\gamma=1$ in Lemma \ref{lem:gaussian_evolution}, i.e.\ we absorb $\sqrt{\gamma}$ into $L$, as we do for cleanliness beginning in Section~\ref{sec:main-result}.})  The effect of the Hamiltonian, through $\hh$, is to symplectically squeeze and stretch the ellipse associated to $\sigma$ without changing its volume.  In contrast, the diffusion term $\D$ implements diffusive broadening in phase space, increasing the volume of the ellipse and hence the entropy of the state $\tauQas$.

Crucially, because the quantum and classical evolutions on phase space are identical for harmonic dynamics, the quantum evolution is well-approximated by the classical evolution whenever the local harmonic approximation is good.  The error introduced by the harmonic approximation increases as the covariance
matrix becomes squeezed and $\tauQas$ extends over a larger distance in phase space.  In particular, because the error in the harmonic approximation appears at third order, we loosely expect a bound of the form
\begin{align}
\label{eq:harm-error-intro}
    \textrm{harmonic approximation error } \propto \frac{1}{\hbar} \|\sigma\|^{3/2}
\end{align}
since $\|\sigma\|^{1/2}$ is the the diameter of the effective support of the 
Gaussian packet (the ``length of the ellipse''), and the factor of $\hbar^{-1}$ appears in the Schrodinger
equation. See Figure ~\ref{fig:ellipses} (e).

In closed chaotic systems, a pure Gaussian state stretches exponentially quickly so that $\|\sigma(t)\|\sim \|\sigma(0)\| e^{\lambdaL t}$ where $\lambdaL$ is the largest local Lyapunov exponent of the system, which summarizes the maximum amount of stretching in the relevant region of phase space on the relevant timescale. Thus by Ehrenfest time we can already have $\|\sigma\|^{3/2}\gg \hbar$, so that the harmonic approximation error is large in closed systems.  If one tried to decompose the corresponding over-stretched ellipse into a mixture of less-stretched ellipses, these would have volume less than $(\hbar/2)^d$, violating the uncertainty principle and hence not corresponding to admissable quantum states. However, in open systems, the diffusion 
prevents the Gaussian states from becoming squeezed too thin.  In particular, the strength of the diffusion $\D$ becomes stronger, relative to Hamiltonian squeezing associated with $\lambdaL$, as the ellipse gets narrower, resulting in a minimum thickness $w \sim \sqrt{\D/\lambdaL}$ (see Figure ~\ref{fig:ellipses}(b)).  This means that the mixed Gaussian can be continuously decomposed into pure Gaussians of maximum length $v \sim \hbar/w \sim \sqrt{\lambdaL\hbar/\D}$, and these new states can be separately evolved with the harmonic approximation about their respective centroids, thus controlling the error of the harmonic approximation.

More precisely, for a given Gaussian $\tauQas$ consider the time derivative of the smallest eigenvalue $\sigma$, denoted $\lambdamin[\sigma]$.
By first order variation
of the eigenvalue $\lambdamin[\sigma]$, with unit eigenvector denoted $v$, and using the evolution equation for the covariance matrix~\eqref{eq:cov-growth-intro},
we have\footnote{Although the unit eigenvector $v$ is changing with time, its derivative is necessarily orthogonal to itself, $v^\tp (\partial_t v)=0$, ensuring that $\partial_t (v^\tp \sigma v) = (\partial_t v^\tp) \sigma v + v^\tp (\partial_t \sigma) v + v^\tp \sigma(\partial_t v) = \lambdamin[\sigma](\partial_t v^\tp) v + v^\tp (\partial_t \sigma) v + \lambdamin[\sigma] v^\tp (\partial_t v) = v^\tp (\partial_t \sigma) v$.}
\begin{equation} \label{eq:overview-lambdamin}
    \begin{split}
        \partial_t \lambdamin[\sigma]
        &= v^\tp(\partial_t \sigma) v \\
        &= v^\tp (\hh+\GG) \sigma v + v^\tp \sigma (\hh+\GG)^\tp v
        + v^\tp \D\\
        &\geq \lambdamin[\D] - 2\lambdamin[\sigma] \|\hh+\GG\| ,
    \end{split}
\end{equation}
where $\lambdamin[\D]$ denotes the minimum eigenvalue of $\D$. 
We see that $\lambdamin[\sigma]$
is growing so long as 
\begin{align}
    \lambdamin[\sigma] \lesssim \frac{\lambdamin[\D]}{\|\hh+\GG\|} \sim \hbar \diffstrength.
\end{align}
The second relation follows from treating $\HC$ and $\LCk$ as fixed classical functions (independent of $\hbar$ and $\diffstrength$) so that\footnote{Per the discussion in Section~\ref{sec:corresponedence-time}, we are here assuming $\GG = 0$ or $\diffstrength \lesssim \hbar^{0}$ so that $\GG \lesssim \hbar^0$.} $\hh\propto \hbar^0$, $\GG\lesssim \hbar^0$, and $\D \propto \hbar\diffstrength$.
(In Theorem~\ref{thm:mainResult} we drop $\diffstrength$ and work directly with $\D$, $\hh$, and $\GG$, but for this overview it will be simpler to use $\diffstrength$ as in Theorem~\ref{thm:informal}.)

Thus if $\lambdamin[\sigma]$ initially satisfies $\lambdamin[\sigma] \gtrsim \hbar\diffstrength$, it will never shrink below $\lambdamin[\sigma] \sim \hbar\diffstrength$.  Then the mixed state with covariance $\sigma$
can be decomposed into (pure) coherent states whose covariance matrix has minimum eigenvalue $\lambdamin[\sigma]\sim \hbar\min\{1,\diffstrength\}$ and
maximum eigenvalue $\lambdamax[\sigma]\sim\hbar\max\{1,\diffstrength^{-1}\}$ because\footnote{The max arises because when $\diffstrength \lesssim 1$ the mixed state can be decomposed into coherent states with $\sigma = \frac{\hbar}{2}\IdM_{2d}$, which are the Gaussian pure state that are least extended in phase space. In this case, additional diffusion --- larger $\diffstrength$ --- cannot help because the states are already fully unsqueezed.} the eigenvalues of pure-state covariance matrices come in pairs multiplying to $\hbar^2/4$. 
By Eq.~\eqref{eq:harm-error-intro}, the harmonic approximation error for such 
coherent states is $\hbar^{-1} \|\sigma\|^{3/2} \sim \hbar^{1/2} \max\{1,\diffstrength^{-3/2}\}$.  
This 
is the instantaneous error, which we integrate in time (using Duhamel's principle in the sense of Eq.~\eqref{eq:duhamel-quantum}) to yield the final 
error of $t\hbar^{1/2}\max\{1,\diffstrength^{-3/2}\}$ that appears in Theorem~\ref{thm:mainResult}. 

So far we have described a process of evolving $\tauQas$ according to a local harmonic approximation, which we then decompose into pure Gaussian states, which we then further evolve, and so on. While this picture is instructive and closely resembles the logic of the proof, there we more cleanly track the continuous decompositions by simply specifying a PDE for the probability measure $\pmsrt$ 
defining $\rhot$ in Eq.~\eqref{eq:overview-mixture}.  We define $\rhot(t)$ to evolve like
\begin{equation} 
\partial_t \rhot(t) = 
\int \LLQLa[\tauQas]\dd\pmsrt(\alpha,\sigma)
\end{equation}
where $\LLQLa$ is the harmonic approximation about the point $\alpha$ to the full Lindbladian $\LLQ$.  We re-express $\LLQLa[\tauQas]$ above as a change in the measure $\pmsrt$. Even for fixed $\rhot(t)$, we have freedom in how we choose $\pmsrt$, corresponding to our freedom to decompose mixed Gaussian states in multiple ways. The discussion below Eq.~\eqref{eq:overview-lambdamin} ensures we can choose the distribution $\pmsrt$ to be supported on pure states with $\lambdamin[\sigma] \gtrsim \gamma \hbar$ and $\norm{\sigma} \lesssim \hbar \gamma^{-1}$, which controls the error of the harmonic approximation as discussed above.

\section{Statement of the main result}
\label{sec:main-result}

For the rest of the paper we will drop the coupling strength $\diffstrength$ from the Lindblad equation \eqref{eq:lindblad-simple} by setting $\diffstrength=1$ (equivalently, absorbing it into the Lindblad operators).  

As discussed in depth in Appendix~\ref{sec:units-symplectic-covariance}, the theorem we present in this section  ``ignores physical units'': we imagine a fixed choice of length, time, and mass units has been made, so that physically dimensionful quantities are represented by dimensionless numbers, and in particular it makes sense to (1) require that $\hbar<1$, and (2) use the Euclidean norm of a vector $\alpha = (\alpha^\x,\alpha^\p)\in\bbR^{2d}$ in phase space: $|\alpha|^2 = |\alpha^\x|^2+|\alpha^\p|^2$.  
Indeed one could generally obtain a tighter bound by optimizing over the choice of units.  This is due to the fact that our results are not invariant under linear symplectic transformations, despite the Fokker-Planck equation enjoying this symmetry. See Appendix~\ref{sec:units-symplectic-covariance} for more on this.

To help navigate the notation in this paper, the reader may refer to the glossary in Table~\ref{tab:glossary}.

\subsection{Definitions and assumptions}\label{sec:definitions-and-assumptions}

We will use the Weyl quantization $\Op[\,\cdot\,]$ to map classical functions of phase space to operators as follows:\footnote{Other quantizations are also perfectly acceptable, the Weyl quantization simply has simplifying properties that we make use of.}
\begin{align}
\label{eq:weyl-op-action-def-intro}
(\Op[E]\psi)(y) = (\hat{E}\psi)(y) =
(2\pi \hbar)^{-d}
\int \! \dd x \dd p\, e^{\frac{i}{\hbar} (x-y)\cdot p}
E\Big(\frac{x+y}{2},p\Big) \psi(x).
\end{align}
The inverse map can be used to define the Wigner function $\WW[\rho]:=\Op^{-1}[\rho]/(2\pi\hbar)^d$ of a quantum state $\rho$.  In terms of the state's kernel $K_{\rho}$, the Wigner function can be written
\begin{align}
\label{eq:wigner-weyl-kernel-def-intro}
\WW[\rho](x,p) = 
(2\pi \hbar)^{-d} \int e^{\frac{i}{\hbar} y\cdot p} K_{\rho}(x+y/2,x-y/2) \dd y.
\end{align}
The oscillatory integral is a distributional Fourier 
 transform in the $y$ variable, so is well defined as a distribution in $(x,p)$. (For more details on $\Op$ and $\WW$, see Section~\ref{sec:wigner-weyl}.)

\begin{restatable}[Corresponding dynamics]{defin}{defCorrespondingDynamics}
\label{def:correspondingDynamics}
Let $H, \LCk \in C^\infty(\bbR^{2d})$ be smooth functions on phase space with $1\leq k\leq K\in\bbN$. The \introduce{Markovian open quantum system} corresponding to the data $(\HC,\{\LCk\}_{k=1}^K)$ at semiclassical parameter $\hbar$ is defined by the Lindblad equation $\partial_t \rho =\LLQ [\rho]$ with Lindbladian 
\begin{align} \label{eq:lindblad-simple-restate}
    \LLQ [\rho] &:= -\frac{i}{\hbar}\left[\HQ,\rho\right]+ \frac{1}{\hbar}\sum_{k} \left(\LQk \rho \LQk{}^\dagger-\frac{1}{2}\left\{\LQk{}^\dagger\LQk{},\rho\right\}\right)
\end{align}
where $\HQ=\Op[H]$ and $\LQk=\Op[\LCk]$, and
where $\rho(t)$ is a trace-class operators on the Hilbert space
$L^2(\bbR^d)$.
The \introduce{corresponding classical dynamics} are given by the 
Fokker-Planck equation $\partial_t \cstate = \LLC [\cstate]$ with Liovillian
\begin{equation}
\label{eq:fokker-planck-restate}
\LLC [\cstate] := 
 -\partial_a [\cstate (\partial^a \HC +  \GC^a)]  +
 \frac{1}{2} \partial_a (\D^{ab} \partial_b \cstate)
 \end{equation}
where 
\begin{align}
    \label{eq:diffusion-matrix}
    \D^{ab} :=&\hbar \Real \sum_k (\partial^a \LCk)(\partial^b \LCk^*)=: \hbar  \scD^{ab} & & (\textbf{diffusion matrix})\\
    \label{eq:friction-vector}
    \GC^a :=& 
     \Imag \sum_k \LCk \partial^a \LCk^* & & (\textbf{friction vector})
\end{align}  
When $\GC^a=0$, we say the dynamics are \introduce{frictionless}. 
Given a quantum trajectory $\rho(t)$ that evolves according to the Lindblad equation~\eqref{eq:lindblad-simple-restate} from an initial state $\rho(t\liq 0)$ with a non-negative Wigner function $\WW[\rho(t\liq 0)]$, the \introduce{corresponding classical trajectory} $\rhoc(t)$ is the solution to the Fokker-Planck equation~\eqref{eq:fokker-planck-restate} with initial distribution $\rhoc(t\liq 0) = \WW[\rho(t\liq 0)]$.
\end{restatable}
\noindent  As shown\footnote{This was pointed out to us by Jeff Galkowski and Maciej Zworski, who learned of this reference from Simon Becker.}by Davies in~\cite{davies1977neutron}, the semigroup $e^{t\LLQ}$ is a contraction on the space of density matrices so long as $i\HQ -\sum_k \LQk^\dagger\LQk$ is the generator of a strongly continuous contraction semigroup on $L^2(\bbR^d)$. In~\cite{galkowski2024classical}, Galkowski and Zworski derive the latter condition from the Hille-Yosida theorem (see the proof of Proposition 4.6 and also Proposition A.2) in the case that $H$ and $L_k$ are $C^\infty$ and have derivatives growing at most linearly
at infinity.

We review\cite{strunz1998classical,bondar2016wigner,dubois2021semiclassical} in Section~\ref{sec:corresponding-dynamics} why the Fokker-Planck equation \eqref{eq:fp-simple} describes the classical dynamics naturally corresponding to the Lindblad equation \eqref{eq:lindblad-simple-restate}, and in particular why $\D^{ab}(\alpha)$ is interpreted as the classical diffusion matrix.  For the purposes of stating our
assumptions and our bounds, it is will also be useful to refer to the \introduce{scaled diffusion matrix}
\begin{align}
\label{eq:delta-def}
    \scD^{ab}:=& \Real \sum_k (\partial^a \LCk)(\partial^b \LCk^*) = \frac{1}{\hbar} D^{ab}.
\end{align}
Note that $\scD$ is independent of $\hbar$ and only depends on the classical
functions $\LCk$.

Our results will apply to data $(H,\{\LCk\}_{k=1}^K)$ that satisfy some regularity and decay assumptions.   The first condition is 

\begin{restatable}[Admissible class of Lindbladians]{assum}{assumSuitableLindblad}
    \label{assum:suitableLindblad}
    We say that the tuple of functions 
    $(H,\{\LCk\}_{k=1}^K)$ is \introduce{admissible} if the following
    hold:
    \begin{enumerate}[a.]
        \item 
        For $2\leq j < \infty$, all $j$-th order mixed partial derivatives of the Hamiltonian are bounded over phase space: $\sup_\alpha |\partial_{a_1}\cdots \partial_{a_j} \HC(\alpha)|<\infty$. For $1\leq j < \infty$, the same is true for the Lindblad functions: $\sup_\alpha |\partial_{a_1}\cdots \partial_{a_j} \LCk(\alpha)|<\infty$.
        \item For $3\leq j\leq 2d+4$, 
        the $j$-th order mixed partial derivatives of the Lindblad functions, weighted by the functions themselves, grow sublinearly\footnote{This assumption can be relaxed to allow for any polynomial growth of the product $|\LCk(\alpha)||\partial_{a_1} \cdots \partial_{a_n} \LCk(\beta)|$ at the cost of requiring bounded higher-order derivatives.} at infinity:
        \begin{align}
        \label{eq:nonlocal-Lk-norm}
            \sup_{\alpha,\beta} \, \frac{|\LCk(\alpha)||\partial_{a_1} \cdots \partial_{a_j} \LCk(\beta)|}{1+|\alpha-\beta|} < \infty.
        \end{align}
        \item The matrix $\scD$ defined in~\eqref{eq:delta-def} is uniformly lower bounded,
        \begin{equation}
            \inf_{\alpha}
            \lambdamin[\scD(\alpha)] > 0.
        \end{equation}
    \end{enumerate}
\end{restatable}
The first assumption allows $H$ to be unbounded but requires it grows at most quadratically at infinity.   We stress that although
we require $C^\infty$ regularity of $H$ and $L_k$, this is 
only so that the argument of Galkowski and Zworski~\cite{galkowski2024classical} applies to prove that the
Lindblad evolution is positivity preserving.  In particular,
we only use quantitative estimates on $\partial^\alpha H$ 
and $\partial^\beta L_k$ for $|\alpha| \leq 2d+4$ and 
$|\beta| \leq 4d+6$.  The second assumption ensures that the friction
$G^a$ is bounded and, for example, is satisfied for Lindblad functions of the form $L(\alpha) = \alpha^a + g(\alpha)$ where $g$ is any Schwartz-class function.

To state our main result we introduce some quantities that we use to bound the  error between the classical and quantum evolutions.  The first measures the strength of the diffusion term in the evolution of the covariance matrix~\eqref{eq:cov-growth-intro} relative to the squeezing terms caused by the Hamiltonian flow and the friction.

\begin{restatable}[Relative diffusion strength]{defin}{defEffInvDiffStrength}\label{def:eff-inv-diff-strength}
Given an admissible tuple $(\HC,\{\LCk\}_{k=1}^K)$, we define the \introduce{relative diffusion strength} $\rds$ to be 
\begin{align}
\label{eq:zz-def}
\rds := \min\left\{  
         \frac12 \inf_\alpha 
        \frac{\lambdamin[\scD(\alpha)]}{\lambdamax[\nabla^2\HC(\alpha)]},
        \inf_\alpha 
        \left(\frac{\lambdamin[\scD(\alpha)]}{\lambdamax[\scD(\alpha)]}\right)^{1/2}
        \right\}
\end{align}
In the special case that the dynamics are frictionless (i.e., when $\GC$ define by~\eqref{eq:friction-vector} vanishes), we define $\rds$ to be the larger quantity
\begin{align}
\label{eq:zz-def-frictionless}
\rds := \min\left\{ 
        \inf_\alpha 
        \frac{\lambdamin[\scD(\alpha)]}{\lambdamax[\nabla^2\HC(\alpha)]},1
        \right\}.
\end{align}
\end{restatable}

The relative diffusion strength compares the diffusion term to the Hamiltonian and friction terms in the evolution equation for the covariance matrix~\eqref{eq:cov-growth-intro}.  The Hamiltonian term [represented by $\hh=\sf\nabla^2\HC$ in~\eqref{eq:cov-growth-intro}] is simply bounded with the largest eigenvalue\footnote{In Hamiltonian systems, the local flow generated by the Hamiltonian $\HC$ is $\partial^a \HC = \sf^{ab}\partial_b \HC$. The Jacobian of this vector field is $\hh^a_{\pha b} := \partial_b \partial^a \HC = \sf^{ac}(\nabla^2\HC)_{bc}$. The Hessian $\nabla^2\HC$ is necessarily symmetric, so the Jacobian $\hh^a_{\pha b}$ is a Hamiltonian matrix by construction. Because the symplectic form $\sf$ is an orthogonal matrix, $\|\hh\| = \|\nabla^2\HC\|$.} of the Hessian of the Hamiltonian $\lambdamax[\nabla^2H]$. On the other hand, the friction term [represented by $\GG = \nabla\GC$ in~\eqref{eq:cov-growth-intro}] is bounded indirectly with $\lambdamax[\scD]$ using the matrix inequality $\scD+i\GG\sf \ge 0$. 
The fact that $\rds$ depends on the condition number of $\scD$, and therefore is not monotone in the diffusion $D$, is an artifact of our proof that we believe to be suboptimal.\footnote{The friction term can squeeze the state, potentially increasing the discrepancy between the quantum and classical states, and hence must sometimes lower the relative diffusion strength $\rds$. However, we bound it with $\lambdamax[\scD]$, and pure (i.e., frictionless) diffusion can only \emph{reduce} the discrepancy, and would ideally only \emph{increase} $\rds$.  Since our argument in its current form cannot distinguish these, we have been forced to define the relative diffusion strength $\rds$ so that it has the undesirable property that adding pure diffusion to the dynamics can weaken our bound, which manifest as lowering the $\rds$ defined here.  
We attribute this deficiency to the crude operator norm estimates and use of the triangle inequality in Lemma~\ref{lem:covar-diff-decomp}.
} In the frictionless case we only need to compare the diffusion to the Hamiltonian squeezing term (without needing to bound $\GG$ in terms of $\scD$), and therefore recover the desired monotonicity in $D$.

Now we introduce a preferred set of pure (i.e., rank-1) quantum state. The \introduce{pure Gaussian states} are
$\tauQas = \ket{\alpha,\sigma}\bra{\alpha,\sigma}$ where $\bra{\alpha,\sigma}$ is a wavepacket with Gaussian envelope and a quadratic phase.  It is parameterized by the phase space mean $\alpha$ and the covariance matrix $\sigma = \frac{\hbar}{2}\bar\sigma$ where $\bar\sigma$ is positive definite \emph{and} (per the uncertainty principle) symplectic: $\bar\sigma > 0$, $\bar\sigma\in\mathrm{Sp}(2d,\bbR)$. The Wigner function of a Gaussian state is
\begin{equation}
    \WW[\tauQas](\alpha+\beta) = \tauCas(\alpha+\beta)  = (2\pi)^{-d} (\det \sigma)^{-1/2} \exp(-\beta^\tp \sigma^{-1} \beta/2).
\end{equation}
For more details about Gaussian states we refer to Section~\ref{sec:harmonic-gaussian}.

A special kind of pure Gaussian state are the \introduce{coherent states} $\tauQa:= \tauQ_{\alpha,\sigmaco}$ with covariance matrix $\sigmaco :=\frac{\hbar}{2}\IdM_{2d}$.  In this paper we will make use of the following class of states that are ``almost coherent'' in the sense that their condition number is controlled.

\begin{restatable}[Not-too-squeezed states]{defin}{defNTS}\label{def:NTS}
Given a squeezing ratio
$\NTSm \leq 1$,  
we say a pure\footnote{In this paper we will only work with pure NTS states, but there are reasons to consider generalizations to mixed states with appropriately bounded covariance matrices, e.g., when extending our main result to the case of degenerate diffusion. See Section~\ref{sec:future-work}.}
Gaussian state $\tauQas$ is \introduce{not too squeezed} (\introduce{NTS}) when its covariance matrix obeys $\sigma \geq \NTSm \sigmaco$. The set of such covariance matrices is 
\begin{align}
    \SNTS(\NTSm) := 
    \left\{\sigma \middle| \frac{\sigma}{\hbar/2} \in \mathrm{Sp}(2d,\bbR), \sigma \geq \NTSm\frac{\hbar}{2}\IdM_{2d}\right\}
\end{align}
\end{restatable}
When $\sigma = \frac{\hbar}{2}\bar{\sigma}$ is the covariance matrix of a pure state (so that $\bar{\sigma}$ is positive-definite and symplectic),
the minimum and maximum eigenvalues come in pairs $(\hbar/2)\lambdas^{-1}$ and $(\hbar/2)\lambdas$. 
Therefore we in fact have $\lambdas\sigmaco \le \sigma \le \lambdas^{-1}\sigmaco$ whenever $\sigma\in \SNTS(\NTSm)$. 
By the uncertainty principle, the phase-space standard deviations satisfy $\lambdas\sqrt{\hbar/2} \le \Delta x \le \lambdas^{-1}\sqrt{\hbar/2}$ and $\lambdas\sqrt{\hbar/2} \le \Delta p \le \lambdas^{-1}\sqrt{\hbar/2}$.
When $\lambdas=1$, the only states allowed are the coherent states, i.e., the unsqueezed pure Gaussian states for which $\Delta x = \Delta p = \sqrt{\hbar/2}$.

\begin{restatable}[Suitable class of initial states]{assum}{assumSuitableStates}
\label{assum:suitableStates}
We assume the initial state $\rho_0=\rho(t \liq 0)$ is a mixture of pure Gaussian states $\tauQas$ that are squeezed relative to the coherent states $\tauQa$ by no more than the effective inverse diffusion strength \eqref{eq:zz-def} of the dynamics, i.e.,
\begin{align}
    \rho_0 = \int_{\bbR^{2d}} \int_{\SNTS(\rds/2)} \tauQas \dd\pmsr_0(\alpha,\sigma).
\end{align}
for some probability measure $\pmsr_0$ supported on the set $\bbR^{2d}\times \SNTS(\rds/2)$ of covariance matrices that are not too squeezed,
where $\rds$ is the relative diffusion strength parameter
defined in Definition~\ref{def:eff-inv-diff-strength}.
\end{restatable}

The other important parameters that we introduce which quantify the divergence between the classical and quantum trajectories are the ``anharmonicity'' factors.  These measure the failure of $\HC$ to be a quadratic function and $\LCk$ to be a linear function.
The classical anharmonicity factor over phase space is
\begin{align} \label{eq:harmErrConstCtt-def}
	\harmErrConstCtt{\HC}{\LCk} := \left(\CkSN{H}{3} + 
     \CkSN{G}{2} + \CkSN{\scD}{1} \right),
\end{align}
where the $C^k$ seminorms are defined 
in~\eqref{eq:c-k-seminorm-def}.  
This factor goes into the error rate of the classical evolution. 
Note that $G$ is a linear function and $\Omega$ is a constant when the Lindblad operators are linear, and thus $B^{\text{anh}}_c[H,\LCk]$ vanishes for systems with a quadratic Hamiltonian and a linear Lindbladian.  Thus $B^{\text{anh}}$ is a very natural factor with which one may measure the growth of the error in the semiclassical correspondence.

On the quantum side we do not arrive at such a natural definition for the ``anharmonicity factors'' in particular because we need more than just three derivatives.  Nevertheless all higher order derivatives come with an additional factor of $\hbar^{1/2}$.
The quantum anharmonicity factors are defined
using the ``anharmonicity seminorms'' $\constQ^{q,r}_\hbar$ and 
its nonlocally weighted version $\constN^{q,r}_{\hbar;s,\nu}$
\begin{align}
\label{eq:constQ-def}
\constQ^{q,r}_{\hbar}[E]
&:= \sum_{j=q}^r \hbar^{(j-q)/2}\sup_\alpha \TsrNrm{\nabla^jE(\alpha)}
= \sum_{j=q}^r \hbar^{(j-q)/2}\CkSN{E}{j} \\
\label{eq:constN-def}
\constN^{q,r}_{\hbar;s,\nu}[E](\alpha)
&:= 
\sum_{j=q}^r \hbar^{(j-q)/2} 
\sup_\beta 
\frac{\TsrNrm{\nabla^jE(\alpha+\beta)}}{(1+\nu^{-1}|\beta|)^s}.
\end{align}
To compare with more standard semiclassical analysis notation we observe that the quantities $\constQ$ and $\constN$ above 
define seminorms on otherwise standard symbol classes.  In particular, note that $\constQ^{q,r}_\hbar[E]$ is bounded
for symbols in $S(1)$  satisfying
\[
\sup_\alpha\|\nabla^j E(\alpha)\| \leq C_j 
\]
Moreover $\constN^{q,r}_{\hbar;s,\nu}[E](\alpha)$
is bounded for symbols in $S((1+\nu^{-1}|\alpha|)^s)$, satisfying
\begin{equation}
\label{eq:stupid-S-def}
\|\nabla^j E(\alpha)\|
\leq C_j (1+\nu^{-1}|\alpha|)^s.
\end{equation}
However note that these seminorms do not include bounds on derivatives of order less than $q$.

Note that if $q=r$ there is no $\hbar$-dependence
in $\constQ^{q,r}_{\hbar}[E]$ so\footnote{Likewise, $\harmErrConstCtt{\HC}{\LCk} = \constQ^{3,3}[\HC]+\constQ^{2,2}[\GC]+\constQ^{1,1}[\scD]$.} we can drop the appearance
of $\hbar$.  
Then we define
\begin{align}
    \begin{split}\label{eq:harmErrConstQtt-def}
    \harmErrConstQtt{\HC}{\LCk} :=\, &
        \constQ^{3,2d+4}_{\hbar}[\HC]
         + \sum_k
         \constQ^{1,1}[\LCk]
         \constQ^{2,2d+3}_{\hbar}[\LCk]
    \end{split}
    \\
    \begin{split}\label{eq:harmErrConstQf-def}
    \harmErrConstQfBare[\LCk,\hbar,\vsc] :=\, &
    \sum_k
        \left[
        \sup_{\alpha} |\LCk(\alpha)| \constN_{\hbar;1,\vsc}^{3,2d+6}[\LCk](\alpha)
        +
         \vsc (\constQ^{2,4d+6}_{\hbar}[\LCk])^2 
        \right]
    \end{split}
\end{align}
Note that $\harmErrConstQtt{\HC}{\LCk}$ and 
$\harmErrConstQfBare[\LCk,\hbar,\vsc]$ are finite when 
$H$ and $\LCk$ satisfy the hypotheses of Assumption~\ref{assum:suitableLindblad}, and vanish when $H$ is quadratic and $\LCk$ are linear functions of $\alpha$.
The quantum anharmonicity factors are much more complicated than the classical ones essentially because they are needed to control the higher-order error terms in the Moyal product expansion for the symbol of products of operators.  Moreover it is quick to check 
(after unwrapping the perhaps cumbersome notation) 
that $\harmErrConstCtt{\HC}{\LCk} \leq \harmErrConstQtt{\HC}{\LCk} + \harmErrConstQfBare[\LCk,\hbar,\vsc]$.

\subsection{Statement of Theorem~\ref{thm:mainResult}}\label{sec:main-result-statement}

We are ready now to state the main result.
\begin{restatable}[Main result]{thm}{thmMainResult}
\label{thm:mainResult}
    Consider an open system with data $(\HC,\{\LCk\}_{k=1}^K)$ which is admissible in the sense of Assumption~\ref{assum:suitableLindblad} with quantum trajectory $\rho(t)$ solving the Lindblad equation~\eqref{eq:lindblad-simple-restate} and classical trajectory $\rhoc(t)$ solving the corresponding Fokker-Planck equation~\eqref{eq:fokker-planck-restate} with initial state $\rhoc(t\liq 0) = \WW[\rho(t\liq 0)]$ as in Definition~\ref{def:correspondingDynamics}.  Assume the initial state $\rho(t\liq 0)$ is a mixture of Gaussians states that are not too squeezed as in Assumption~\ref{assum:suitableStates}.
    Associated with the dynamics, let $\rds$ be the relative diffusion strength \eqref{eq:zz-def} from Definition~\ref{def:eff-inv-diff-strength} and let  $\harmErrConstCtt{\HC}{\LCk}$,
    $\harmErrConstQtt{\HC}{\LCk}$, and $\harmErrConstQfBare[\LCk,\hbar,\rds^{-1}\hbar]$ be the anharmonicity factors \eqref{eq:harmErrConstCtt-def}-\eqref{eq:harmErrConstQf-def}.
    Then there exists a quasiclassical quantum trajectory $\rhot(t)$ which is a 
    mixture of Gaussians which approximates $\rho$ and $\rhoc$ in the following sense:
    \begin{enumerate}[a.]
        \item \label{item:classical-error-main}
        $\rhot(t)$ approximates the corresponding classical trajectory $\rhoc(t)$ for all possible classical variables in the sense that 
        \begin{equation}\label{eq:main-classical}
            \Lonenorm{\WW[\rhot(t)]-\rhoc(t)} \le 
            14 d^{3/2}t\, 
            \rds^{-3/2}\hbar^{1/2}\,
            \harmErrConstCtt{\HC}{\LCk};
        \end{equation}
        and
        \item \label{item:quantum-error-main}
        $\rhot(t)$ approximates the true quantum trajectory $\rho(t)$ for \emph{all} possible quantum observables in the sense that
        \begin{align}\label{eq:main-quantum}
            \Trnorm{\rhot(t)-\rho(t)} \le  C_d
            t \,\rds^{-\frac32} \hbar^{\frac12}
            \left( \harmErrConstQtt{\HC}{\LCk}+
            \harmErrConstQfBare[\LCk,\hbar,\sqrt{\hbar/\rds}]
            \right).
        \end{align}
    \end{enumerate}
    Here, $C_d$ is a universal constant depending only on the dimension $d$.
\end{restatable}
Note that the classical error~\eqref{eq:main-classical} does \emph{not} include an unspecific constant $C_d$, and we can see  dimensional dependence is on the order $d^{3/2}$.  In contrast the dimensional constant $C_d$ appearing in~\eqref{eq:main-quantum} grows superexponentially in the dimension, and one can recover from our proof a bound\footnote{In the case treated in the companion paper \cite{hernandez2023decoherence1}, we find the analogs of both~\eqref{eq:main-classical} and~\eqref{eq:main-classical} come with dimensional-dependence of only $d^{3/2}$.  The dependence $(d!)^C$ comes from the bound on the Moyal product appearing in Proposition~\ref{prp:moyalBound}.} of the form 
$C_d \leq (d!)^C$.
Below we discuss how to recover the simplified Theorem~\ref{thm:informal} stated in the introduction, using Theorem~\ref{thm:mainResult}.  The argument is primarily a matter of notation.  
\begin{proof}[Proof of Theorem~\ref{thm:informal} assuming Theorem~\ref{thm:mainResult}]
Note the simplified Theorem~\ref{thm:informal} refers to Eq.~\eqref{eq:fp-simple}, which uses a coupling strength $\diffstrength$ multiplying the Lindblad terms.  In contrast, Theorem~\ref{thm:mainResult} refers to the Lindblad equation in Eq.~\eqref{eq:lindblad-simple-restate}, which does not include the $\diffstrength$ factor, or equivalently sets $\diffstrength=1$.  Of course, $\diffstrength$ can be absorbed into the definition of the $\{\LQk\}_k$ operators. (We find $\diffstrength$ is helpful for the introductory discussion but clutters the technical discussion.)

More concretely, to obtain Theorem~\ref{thm:informal} from Theorem~\ref{thm:mainResult}, first we restore the coupling strength $\diffstrength$ in the definition \eqref{eq:zz-def} of $\rds$ using the replacement $\LCk \mapsto \sqrt{\gamma} \LCk$ and likewise $\scD\mapsto\diffstrength\scD$, so that  $\rds^{-3/2}\sim \max\{\diffstrength^{-3/2},1\}$, up to constants depending on $\HC$ and $\LCk$ (but not on $\hbar$ or $\diffstrength$). Then the constants $\harmErrConstCtt{\HC}{\sqrt{\gamma}\LCk},$ $\harmErrConstQtt{\HC}{\sqrt{\gamma}\LCk}$ and $\harmErrConstQfBare[\sqrt{\gamma}\LCk,\hbar,\sqrt{\hbar/\rds}]$ are at most $\max\{1,\gamma\}$, again up to constants depending on $\HC$ and $\LCk$.  We then obtain
\begin{align}\label{eq:main-informal-rhot}
\begin{split}
    \Lonenorm{\WW[\rhot(t)]-\rhoc(t)} &\le  
    \erate t
    \\ 
    \Trnorm{\rhot(t)-\rho(t)} &\le    
    \erate  t
\end{split}
\end{align}
where 
\begin{align}
   \erate = C(H, \LCk)\, \hbar^{1/2} 
   \max\{ \diffstrength^{-3/2}, \diffstrength\}.
\end{align}

For any classical observable $A(x,p)$ and 
corresponding quantum observable $\hat{A}=\Op[A]$
we can also obtain a bound
that does not refer to $\rhot$: 
\begin{equation} \label{eq:smooth-observable-bound-repeat}
\left|\Tr[\rho(t) \hat{A}] - \int f(t) A \, \dd \alpha\right| \leq 
\erate t (\|A\|_{L^\infty} +
\|\hat{A}\|_{\text{op}}).
\end{equation}
This follows directly from Eq.~\eqref{eq:main-informal-rhot}, applying  Eqs.~\eqref{eq:Lonenorm-variational-defn} and \eqref{eq:tracenorm-variational-defn}.  Also, using the 
 Calder{\'o}n-Vaillancourt theorem \cite{zworski2022semiclassical},
the operator norm in Eq.~\eqref{eq:smooth-observable-bound-repeat} can be upper bounded 
as $\|\hat{A}\|_{\text{op}} < \|A\|_{L^\infty} + O(\hbar)$ for symbols $A$ that are smooth and independent of $\hbar$. Thus we arrive at Theorem \ref{thm:informal}.
\end{proof}

\section{Basic notation and definitions}\label{sec:notation}

This section recalls some basic notation and definitions that we will use. A glossary of our most important notation can be found in Table~\ref{tab:glossary}.
Some readers may wish to only skim this section before reading the proof in Section~\ref{sec:main_proof}, returning here as necessary for clarification.

\subsection{Indices and the symplectic form}\label{sec:symplectic-basics}

We consider the non-relativistic, first-quantized,  open-system quantum dynamics of a particle in $d$ spatial dimension with position operator $\XQ = (\XQ^1,\ldots,\XQ^d)$ and momentum operator $\PQ = (\PQ^1,\ldots,\PQ^d)$:
\begin{align}
    \label{eq:x-op-def}
    (\XQ^j\psi)(x) &= x^j \psi(x),\\
    \label{eq:p-op-def}
    (\PQ^j\psi)(x) &= -i\hbar (\partial_j \psi)(x),
\end{align}
for $\psi\in L^2(\bbR^d)$, $j=1,\ldots,d$.
We use $\RQ = (\RQ^1,\ldots,\RQ^{2d}) = (\XQ,\PQ) = (\XQ^1,\ldots,\XQ^d,\PQ^1,\ldots,\PQ^d)$ for the combined phase-space operator.   As shown, we use upper indices $a, b, \ldots =1,\ldots, 2d$ to access the elements of vectors like $\RQ$.  We parameterize the points in phase with $\alpha$, $\beta$, or $\gamma$ (as when integrating over it), where $\alpha = (\alpha^\x,\alpha^\p) = (\alpha^1,\ldots,\alpha^{2d})$.  The phase-space coordinate vector \emph{function} is denoted $r$, i.e., $\RC^a(\alpha) = \alpha^a$, so the mean of a distribution $f$ is $\langle \RC^a \rangle_f = \int\!\dd\alpha\, \alpha^a f(\alpha)$. (Note in particular that the index $a$ is not an exponent.) We use multi-indices $n:=(n_1,n_2,\cdots,n_{2d}) \in (\mathbb{Z}_{\ge 0})^{\times 2d}$ to write $\alpha^n := (x_1)^{n_1}\cdots (x_d)^{n_d} (p_1)^{n_{d+1}}\cdots (p_d)^{n_{2d}}$ and $\partial_\alpha^n E=\partial_{x_1}^{n_1}\cdots\partial_{x_d}^{n_d} \partial_{p_1}^{n_{d+1}}\cdots\partial_{p_d}^{n_{2d}}E$. We define the factorial $n!=\prod_{j=1}^{2d} (n_j)!$.

We use lower indices to access the elements of co-vectors (1-forms), like partial derivatives, with lower indices: $(\partial_a E)(\alpha)= \partial E(\alpha)/\partial \alpha^a$. 
Because of its special importance to dynamical systems, it will be useful to raise and lower indices with the \introduce{symplectic form},
\begin{align}\label{eq:symplectic-form}
    \sf =
    \left(\begin{matrix} 0 & \IdM \\ -\IdM & 0 \end{matrix} \right),
\end{align}
where $\IdM$ is the $d\times d$ identity matrix: $\RQ_a := \sf_{ab}\RQ^b$, $\partial^a := \sf^{ab}\partial_b$, where repeated indices are summed over. (Einstein notation is used throughout.) Our sign convention is (in $d=1$) $\sf^{\x\p} = +1 = \sf_{\p\x}$, $\sf^{\p\x} = -1 = \sf_{\x\p}$. All of the above applies similarly to (hat-less) phase-space vectors like $\alpha^a$ and $\beta^a$, as well as higher-order tensors like $\hh_{ab}(\alpha) = \partial_a \partial_b H(\alpha)$.  Note that, due to the antisymmetry of the symplectic form, $v^a w_a = - v_a w^b$ and hence that $v^a v_a = 0$ for any $v$, e.g., $\partial^a \partial_a = 0$.  
Phase-space vectors are thus contracted with the symplectic form as $\alpha_a\beta^a = \alpha^b \sf_{ab} \beta^a = \alpha^\x \cdot \beta^\p - \alpha^\p \cdot \beta^\x = \alpha^\tp \sf \beta $, where `$\,\cdot\,$' is the traditional inner product on $\bbR^{d}$.

We use left and right arrows on partial derivatives to indicate that they respectively act on everything to the left and right, extending beyond parentheses and brackets. Thus, $[\vecpartial{}_a A+B]C = \partial_a(AC)+BC$ and $A\cevpartial{}_a\vecpartial{}^a B = (\partial_a A)(\partial^a B)$, but $[\partial_a A + B]C = (\partial_a A)C+BC$. This allows us to write many expressions more clearly and compactly.

At times we will find it convenient to dispense with the index notation and rely on conventional matrix multiplication, in which case the elements of the un-indexed vectors and matrices are assumed to correspond to the indexed versions found in Table~\ref{tab:glossary}.
For phase-space vectors, we use the bare symbol and the transpose, e.g., $\alpha$ and $\beta^\tp$, producing scalars like $\beta^\tp \sf \alpha$. We reserve bra-ket notation for quantum states, e.g., $|\psi\rangle$, $\langle\phi|$, and $\langle\phi|\hat{E}|\psi\rangle$.

\begin{table}
     \centering
 \begin{tabular}{c|c|c}
    Notation & Meaning & Reference to definition \\
    \hline
    $d$ & Number of degrees of freedom  & -- \\
    $\sf^{ab}$ & Symplectic form & Definition~\ref{def:correspondingDynamics}  \\
    $X_a \text{ vs. } X^a$ & Index raising/lowering via $X^a = \omega^{ab} X_b$ & -- \\
    $\RQ^a$ & Phase space operator
    $\RQ = (\XQ,\PQ)$  & Eqs.~\eqref{eq:x-op-def}, \eqref{eq:p-op-def} \\
    $\alpha^a, \beta^b$ & Phase space coordinate $\alpha=(x,p) \in \mathbb{R}^{2d}$ & -- \\
    $\sigma^{ab}$ &  Covariance matrix  & Eq.~\eqref{eq:sigma-def} \\
    $\tauCas$ & Gaussian classical distribution   & Eq.~\eqref{eq:gaussian-wigner} \\
    $\tauQas$ & Gaussian quantum state  & Eq.~\eqref{eq:gaussian-state} \\
    $\hat{E} = \Op[E]$ & Weyl quantization of function $E$  & Eq.~\eqref{eq:weyl-op-def}
    \\
    $E = \Op^{-1}[\hat{E}]$ & Wigner transform of operator $\hat{E}$
    &  Eq.~\eqref{eq:wigner-weyl-def} \\
    $\LLQ$ & Lindbladian generator of quantum Markovian evolution &  
     Eq.~\eqref{eq:lindblad-simple-restate}\\
    $\LLC$ & Liovillian generator of classical Markovian evolution &  
    Eq.~\eqref{eq:fokker-planck-restate}\\
    $\HC$ & Hamiltonian function &  
    Definition~\ref{def:correspondingDynamics}\\
    $\LCk$ & Lindblad function & 
    Definition~\ref{def:correspondingDynamics}\\
    $\GC^a$ & Friction vector & Definition~\ref{def:correspondingDynamics},
    Eq.~\eqref{eq:friction-vector}\\
    $\D^{ab}$ & Diffusion matrix &  Definition~\ref{def:correspondingDynamics},
    Eq.~\eqref{eq:diffusion-matrix} \\
    $\rds$ & Relative diffusion strength &
    Definition~\ref{def:eff-inv-diff-strength}
    \\
    $\SNTS(\NTSm)$ & Set of NTS covariance matrices with squeezing ratio $0\le\NTSm\le 1$ & 
    Definition~\ref{def:NTS}\\
    $\UC^a$ & Deterministic drift 
    & Eq.~\eqref{eq:def-deterministic-drift} \\
    $\meandrift^a$ & Mean drift 
    & Eq.~\eqref{eq:def-mean-drift} \\
    $\hh^a_{\pha b}$ & 
    Hessian matrix of $\HC$ 
    &  Eq.~\eqref{eq:hh-def}\\
    $\GG^a_{\pha b}$ & Gradient matrix of $\GC$ 
    &  Eq.~\eqref{eq:gg-def}\\
    $\sdot^{ab}$ & Time derivative of $\sigma$ &  Eq.~\eqref{eq:sdot}\\
    $\scD^{ab}$ & Scaled diffusion matrix &  Eq.~\eqref{eq:delta-def} \\
    $\locD^{ab}$ & Localization matrix &  Eq.~\eqref{eq:localization-matrix-def} \\
    $\FDmax$ & Max strength ratio of $\hh$ to $\D$ &  
    Eq.~\eqref{eq:fdmax-def-recall} \\
    $\condmax$ & Max condition number of $\D$ &  
    Eq.~\eqref{eq:condmax-def-recall}\\
    $\constQ^{q,r}_{\hbar}$, $\constN^{q,r}_{\hbar;s,\nu}$
    & Anharmonicity seminorms &
    Eqs.~\eqref{eq:constQ-def}, \eqref{eq:constN-def}\\
    $\harmErrConstCttBare$, $\harmErrConstQttBare$, $\harmErrConstQfBare$ & Anharmonicity factors for $\HC$ and $\LCk$ &  Eqs.~\eqref{eq:harmErrConstCtt-def}, \eqref{eq:harmErrConstQtt-def}, \eqref{eq:harmErrConstQf-def} \\
    $|\cdot|_{C^k}$ & $C^k$ seminorm
    &  Eq.~\eqref{eq:c-k-seminorm-def}
    \\
    $\NTSZ$ & Symplectic coordinate-change matrix  & -- 
 \end{tabular}
     \caption{A glossary of notation used in this paper. All operators have hats except the quantum states $\rho$ and $\rhot$. All function above are real-valued except the complex-valued Lindblad function $\LCk$. Hats on a function $\hat{E}$ denote quantization with $\Op[\,\cdot\,]$, but the hat on the Gaussian state $\tauQas = \WW^{-1}[\tauCas] = (2\pi\hbar)^d \Op[\tauCas]$ differs by a factor of $(2\pi\hbar)^d$ because it is a Wigner function rather than a Wigner transform; see Section~\ref{sec:wigner-weyl}. When we use conventional matrix multiplication notation and consequently suppress indices on $\UC$, $\sdot$, $\hh$, and $\GG$ in Secs.~\ref{sec:main_proof} and \ref{sec:linear-algebra-facts}, they refer to $\UC^a$, $\sdot^{ab}$, $\hh^a_{\pha b} = \partial_b \partial^a \HC$, and $\GG^a_{\pha b} = \partial_b G^a$.}
     \label{tab:glossary}
 \end{table}

\subsection{Matrices}\label{sec:matrix-basics}

We use $\lambdamin$ and $\lambdamax$ respectively for the smallest and largest eigenvalue value of a matrix.  We also use the unsubscripted norm $\|\cdot\|$ for the operator norm of a matrix, and operator norms for operators on infinite-dimensional Hilbert space are written $\|\cdot\|_{\text{op}}$.

Associated with the symplectic form is the idea of a \introduce{symplectic} matrix $A$, characterized by preserving the symplectic form under conjugation: $A^\tp \sf A =\sf$. The set of all symplectic matrices is denoted $\mathrm{Sp}(2d,\bbR)$.  When a (non-singular) symplectic matrix is also symmetric, $A^\tp = A$, it satisfies $\sf^\tp A \sf = A^{-1}$.  

Additionally, we will consider \introduce{Hamiltonian} matrices (not to be confused with the Hamiltonian function $\HC$ of the dynamics), which instead satisfy $A^\tp = -\sf^\tp A \sf$, and \introduce{skew-Hamiltonian} matrices, which satisfy $A^\tp = \sf^\tp A \sf$.  Equivalently, $A$ is Hamiltonian (skew-Hamiltonian) when $A \sf$ is symmetric (antisymmetric), which means an arbitrary matrix $A$ can be uniquely decomposed as a sum of its Hamiltonian component $(A-\sf^\tp A \sf)/2$ and its skew-Hamiltonian component $(A+\sf^\tp A \sf)/2$. Symplectic matrices are closed under multiplication, while Hamiltonian and skew-Hamiltonian matrices are closed under both addition and the inverse.

As discussed further in Sec~\ref{sec:harmonic-gaussian}, symplectic positive definite matrices correspond to covariance matrices of pure Gaussian states.  The special role of Hamiltonian matrices for us is that they generate linear time evolution for such matrices.  More precisely, suppose ${\sigma}(t)$ is a time-dependent symmetric symplectic matrix.
In order that the symmetry condition $\sigma^\tp = \sigma$ is preserved, we must have $\dot{\sigma}^\tp = \dot{\sigma}$. Likewise, for the symplectic condition $\sigma \sf\sigma = \sf$ to be preserved, we must have
\begin{equation}
\begin{split}
0 &= \frac{\dd}{\dd t} (\sigma \sf\sigma) = \dot{\sigma} \sf\sigma +  \sigma\sf\dot{\sigma} = \dot{\sigma}\sigma^{-1} \sf + \sf\sigma^{-1}\dot{\sigma} = \dot{\sigma}\sigma^{-1} \sf - (\dot{\sigma}\sigma^{-1} \sf)^\tp
\end{split}
\end{equation}
i.e., $\dot{\sigma}\sigma^{-1}$ is Hamiltonian.  We can always express the dynamics as $\dot{\sigma} = A\sigma + \sigma A^\tp$ for Hamiltonian $A := \dot{\sigma}\sigma^{-1}/2$ (since $\sigma$ and $\dot{\sigma}$ are both symmetric).\footnote{More abstractly, the space $\mathrm{sp}(2d,\bbR)$ of Hamiltonian matrices is the Lie algebra that generates the Lie group $\mathrm{Sp}(2d,\bbR)$ of symplectic matrices.}

\subsection{Norms and seminorms}\label{sec:norms}

For a function $f(\alpha)$ of the phase space variable $\alpha = (x,p) \in \mathbb{R}^{2d}$, the \introduce{(Lebesgue) $L^q$ norm} is
\begin{align}
    \norm{f}_{L^q} := \left( \int \!\dd \alpha\, \left|f(\alpha)\right|^q\right)^{1/q}.
\end{align}
In this paper, we only need the case $q=2$ (used for 
wavefunctions) and $q=1$ (for Wigner functions and classical
probability distributions).  In the latter case we note that
\begin{equation} \label{eq:Lonenorm-variational-defn}
     \|f\|_{L^1} 
    = \sup_{\CkSN{\phi}{0}=1} 
    \int \phi(\alpha)f(\alpha)\dd\alpha.
\end{equation}
The supremum is over continuous functions $\phi$ bounded by $1$.
In particular, for probability distributions $f$ and $g$ 
the error 
$\|f-g\|_{L^1}$ represents the largest possible discrepancy
of a bounded classical observable $\phi$ with respect to the 
probability distributions $f$ and $g$. 

The analogous norm on the quantum side is the \introduce{trace norm} $\|\hat{A}\|_{\mathrm{Tr}} := \Tr[(\hat{A}^\dagger \hat{A})^{1/2}]$ of an operator $\hat{A}$, i.e., the sum of the singular values of $\hat{A}$. Just as in the classical case, there is an equivalent expression
\begin{equation}\label{eq:tracenorm-variational-defn}
\|\hat{A}\|_{\mathrm{Tr}} = \sup_{\|\hat{B}\|_{\mathrm{op}}=1} \Tr[\hat{A} \hat{B}].
\end{equation}
where $\|\hat{B}\|_{\mathrm{op}} := \sup_{\|\psi\| = \|\phi\| = 1} |\langle \psi|\hat{B}|\phi\rangle|$ is the traditional \introduce{operator norm}, i.e., the largest singular value of an operator $\hat{B}$.  In particular, $\Trnorm{\rho-\eta}$ gives a bound on the difference between two quantum states $\rho$ and $\eta$ as measured by any bounded observable. Thus, two classical states (quantum states) cannot be easily distinguished when they are close in $L^1$ norm (trace norm), no matter what measurement is performed.

We define the $C^k$ seminorm of a function $E$ to be\footnote{Note that, for the purpose of defining the $C^k$ seminorm, we have picked a particular norm $\OffDiag{Z} := \sup_{\|\beta_j\|=1}|(\beta_1\otimes\beta_2\otimes\cdots\otimes\beta_k)\cdot Z|$ on tensors $Z$ of order $k$. All norms on finite-dimensional tensors are equivalent up to an overall constant, but our bounds on the classical side in fact are sensitive to this constant. See Eqs.~(\ref{eq:constQ-def}--\ref{eq:main-classical}) and Sec.~\ref{sec:harmonic-error-classical}.}
\begin{align}
    \label{eq:c-k-seminorm-def}
    \CkSN{E}{k} := 
    \sup_\alpha \OffDiag{\nabla^k E(\alpha)}
    = \sup_\alpha \sup_{\|\beta_j\|=1}
    \left|\beta_1^{a_1} \cdots \beta_k^{a_k} \partial_{a_1}\cdots\partial_{a_k} E(\alpha)\right|
\end{align}
In particular, $\CkSN{E}{1}$ measures the largest gradient of $E$, and $\CkSN{E}{2}$ the maximum operator norm of its Hessian.

\subsection{Wigner-Weyl representation}\label{sec:wigner-weyl}
Here we recall the basic components of the Wigner-Weyl representation. We emphasize (linear) ``symplectic covariance'' and a careful handling of normalization factors since they play an important role in our main result. For more extensive review, see Refs.~\cite{dimassi1999spectral, zworski2022semiclassical, degosson2021quantum,folland1989harmonic}.

In order to compare\footnote{There are alternative mappings one can consider, each furnishing an alternative representation of quantum mechanics on phase space.  Most are associated with a particular convention for ordering mixed products of $\XQ$ and $\PQ$, with Wigner-Weyl corresponding to symmetric ordering \cite{agarwal1970calculus,cohen2013weyl}. The Wigner-Weyl representation has useful symmetry properties, and we have chosen it merely for convenience.  Our result does not depend on Wigner-Weyl being the ``correct'' phase-space representation of quantum mechanics.} 
quantum and classical systems, we use \introduce{Weyl quantization}\footnote{Alternatively, when $E$ is analytic, $\Op[E]$ can equivalently be defined by expanding $E$ as a power series and mapping $p^m x^n \mapsto 2^{-n}\sum_{r=0}^n  \binom{n}{r} \XQ^r \PQ^m \XQ^{n-r} = 2^{-m}\sum_{s=0}^m \binom{m}{s} \PQ^s \XQ^n \PQ^{m-s}$ \cite{mccoy1932function, curtright2014concise}. In particular, $\Op[\XC] = \XQ$ and $\Op[\PC] = \PQ$.}
of a \introduce{symbol} $E$ \cite{zworski2022semiclassical}:

\begin{align}
\label{eq:weyl-op-def}
\Op[E]:=\,& 
\frac{1}{(2\pi\hbar)^{2d}}
\int_{\mathbb{R}^{2d}}\! \dd \chi \int_{\mathbb{R}^{2d}}\! \dd \alpha\, E(\alpha) e^{i \chi_a (\RQ-\alpha)^a/\hbar}
\end{align}
This defines an invertible mapping between complex-valued functions on phase space $\bbR^{2d}$ and operators on the Hilbert space $L^2(\bbR^d)$ of complex-valued wavefunctions on configuration space $\bbR^d$.  
When it is unambiguous from context, we will for compactness use a hat\footnote{In the special case of Gaussian \emph{states} (see Section~\ref{sec:harmonic-gaussian}), we will in this paper also use hats slightly differently to distinguish the quantum Gaussian state $\tauQas$ from its Wigner function, the corresponding classical Gaussian state $\tauCas:=\WW[\tauQas] = \Op^{-1}[\tauQas]/(2\pi\hbar)^d$.} to denote the quantum operator corresponding to a classical function: $\hat{E} = \Op[E]$.
Weyl quantization obeys $\Op[E^*] = \Op[E]^\dagger$ \cite{dimassi1999spectral, zworski2022semiclassical, degosson2021quantum} and the trace identity\footnote{This does not extend to the trace of a product of three or more operators. For that, one must deploy the Moyal product described in the next subsection.}
\begin{align}\label{eq:trace-identity}
    \Tr[\hat{E}_1 \hat{E}_2] = \frac{1}{(2\pi\hbar)^d}  \int_{\mathbb{R}^{2d}}\! \dd \alpha\, E_1(\alpha)E_2(\alpha)
\end{align}
when $\hat{E}_1$, $\hat{E}_2$ are Hilbert-Schmidt operators\footnote{We expect that \eqref{eq:trace-identity} also holds when $\hat{E}_1$ is a polynomially bounded operator and $\hat{E}_2$ is a Schwartz operator as defined by Keyl et al \cite{keyl2016schwartz}.  (A operator, such as a  density matrix, is a Schwartz operator if and only if its Wigner function is rapidly decaying, in which case its Wigner function must be a Schwartz function \cite{hernandez2022rapidly}.) The special case where $\hat{E}_2$ is a mixture (convex combination) of Gaussian states is demonstrated in Lemma~\ref{lem:gaussian-trace}.} (Proposition 155 of Ref.~\cite{degosson2021quantum}.) In particular, $\Tr[\hat{E}] = (2\pi\hbar)^{-d}  \int\! \dd \alpha\, E(\alpha)$ when $\hat{E}$ is trace-class.
The action of a Weyl operator on a vector $\psi\in L^2(\bbR^d)$ is \cite{zworski2022semiclassical}
\begin{align}
    \label{eq:weyl-op-action-def}
    (\Op[E]\psi)(x) = (\hat{E}\psi)(x) =
    \frac{1}{(2\pi \hbar)^{d}}
    \int_{\mathbb{R}^{d}}\! \dd y \int_{\mathbb{R}^{d}}\! \dd p\, e^{i (x-y)\cdot p/\hbar}
    E\Big(\frac{x+y}{2},p\Big) \psi(y).
\end{align}
This form does not respect the symplectic covariance, but it is common and often useful for calculations.

The inverse of Weyl quantization is the \introduce{Wigner(-Weyl) transform} of 
$\hat{E}$, producing the symbol $E$, a scalar function on phase space: 
\begin{align}
    \label{eq:wigner-weyl-def}
    \Op^{-1}[\hat{E}](\alpha)
    =&\,  
    \frac{1}{(2\pi\hbar)^d}
    \int_{\mathbb{R}^{2d}}\! \dd \chi\, \Tr\left[e^{i \chi_a (\RQ-\alpha)^a/\hbar} \hat{E}\right].
\end{align}
A slightly different object is the \introduce{Wigner function} $\WW[\rho]$ of a density matrix (positive semidefinite trace-class operator) $\rho$.  In order to obtain relations like $\Tr[\XQ^n\rho] = \int x^n  \WW[\rho](x,p) \, \dd x \dd p$ and $\Tr[\PQ^m\rho] = \int p^m \WW[\rho](x,p)  \, \dd x \dd p$, we must define the Wigner function to differ\footnote{Although this seems a bit unusual, the normalization factor 
in \eqref{eq:weyl-op-def}
is fixed by the desideratum that 
$\Op[1] = \IdQ$ (the identity operator)
while the normalization factor in \eqref{eq:wigner-function-def} is fixed by the desideratum that $\Tr[\rho]=\int \WW[\rho](\alpha)\,\dd\alpha$. Indeed, $\Op$ preserves the physical units (e.g., meters for $\XQ = \Op[\XC]$), while for $\WW[\rho]$ to be a probability distribution over phase space it needs to have the same units as $\hbar^{-d}$ even though the operator $\rho$ has no units.} from the Wigner transform of $\rho$ by a factor of $(2\pi\hbar)^{-d}$:
\begin{align}
    \label{eq:wigner-function-def}
    \WW[\rho](\alpha) := \frac{\Op^{-1}[\rho](\alpha)}{(2\pi\hbar)^d} 
    = 
    \frac{1}{(2\pi\hbar)^{2d}}
    \int_{\mathbb{R}^{2d}}\! \dd \chi\, \Tr\left[e^{i \chi_a (\RQ-\alpha)^a/\hbar} \rho\right]
\end{align}
For compactness we will sometimes use the notation $W_\rho := \WW[\rho]$ when there is no chance of ambiguity. 

For a quantum state $\rho$ with Schwartz kernel $K_{\rho}(x,y) = \langle x |\rho|y\rangle$, 
\begin{align}
    (\rho\psi)(x) = \int\!\dd y\, K_{\rho}(x,y)\psi(y),
\end{align}
an alternative and maybe more recognizable expression for the Wigner function is\footnote{Note that $K_{\rho}$ need only be distribution valued in order
to make sense of $\WW[\rho]$ as a distribution, since the oscillatory integral can be considered as a distributional Fourier 
transform.}
\begin{align}
    \label{eq:wigner-weyl-kernel-def}
    \WW[\rho](x,p)
    = 
    \frac{1}{(2\pi \hbar)^{d}}\int_{\mathbb{R}^{d}} e^{i y\cdot p/\hbar} K_{\rho}(x+y/2,x-y/2) \dd y.
\end{align}
This expression is more amenable to direct computation than the equivalent expression \eqref{eq:wigner-function-def}, but breaks symplectic covariance by treating position and momentum differently.

\subsection{Moyal star product}

On the phase-space side, the 
\introduce{Moyal star product} 
$\star$ implements the equivalent of matrix multiplication,
i.e., $\Op[A\star B] = \Op[A]  \Op[B]$ for 
the symbols $A$ and $B$. 
The general definition is
\begin{align}\label{eq:moyal-integral-def}
	A \star B(\alpha) = \frac{1}{(2\pi \hbar)^{d}} \int e^{i\beta_a\xi^a/(2\hbar)} A(\alpha+\beta/2)B(\alpha+\xi/2) \diff\beta\diff\xi.
\end{align}
When $A$ and $B$ are analytic, it can alternatively be expressed as
\begin{align}\label{eq:moyal-series-def}
	A \star B=& A \exp\left[(i\hbar/2) \cev{\partial}{}_a \vec{\partial} {}^a\right] B \\
	\label{eq:moyal-star-expansion}
	=&\sum_{n=0}^\infty \frac{(i\hbar/2)^n}{n!}  (\partial_{a_1} \cdots \partial_{a_n} A)(\partial^{a_1} \cdots \partial^{a_n} B)\\
	=& A B + \frac{i\hbar}{2} \Poissonbracket{A,B} + O(\hbar^2)
\end{align}
where $\Poissonbracket{A,B} := (\partial_a A) (\partial^a B) = (\partial_x A) (\partial_p B)-(\partial_p A) (\partial_x B) $ is the Poisson bracket and $\sf^{ab}$ is the antisymmetric Levi-Civita symbol.
When one of the functions (say $A$) is a polynomial of degree $n$, then the summation in \eqref{eq:moyal-star-expansion} is naturally understood to terminate after the $n$-th term, and one can check that it agrees with the integral definition \eqref{eq:moyal-integral-def} so long as the other function ($B$) has derivatives defined through order $n$, even if $B$ is not analytic.

Likewise, the \introduce{Moyal bracket} is
\begin{align}
	\Moyalbracket{A, B} := & \frac{-i}{\hbar}(A \star B - B \star A) \\
	=& \Poissonbracket{A,B} + O(\hbar^2)
\end{align}
which reduces to the Poisson bracket for small $\hbar$ as expected.

\section{Technical preliminaries}\label{sec:tech-prelim}

This section collects previously known results in a common notation that will be used in our proof.  It also introduces the ``local harmonic approximation'' for quantum and classical Markovian dynamics which, in the quantum case, we were unable to find explicitly in the literature in full generality.  Some readers may wish to only skim this section before reading the proof in Section~\ref{sec:main_proof}, returning here as necessary for clarification.

\subsection{Classical limit of Lindblad dynamics: Fokker-Planck equation}\label{sec:corresponding-dynamics}

We assume our system follows Markovian dynamics so the density matrix $\rho$ of the system obeys a Lindblad equation $\partial_t \rho = \LLQ[\rho]$ with
\begin{align}
	\label{eq:lindblad}
	\LLQ[\rho] & = -\frac{i}{\hbar}[\HQ,\rho] + \frac{1}{\hbar} \sum_k \left( \LQk \rho \LQk^\dagger - \frac{1}{2} \{ \LQk^\dagger \LQk,\rho\} \right) \\
	\label{eq:lindblad-with-commutators}
	& =  -\frac{i}{\hbar}[\HQ,\rho] + \frac{1}{2\hbar} \sum_k \left( [ \LQk \rho, \LQk^\dagger] + [\LQk, \rho \LQk^\dagger] \right).
\end{align}
where $\HQ$ is the Hamiltonian and $\{\LQk\}$ some set of Lindblad operators.

In this section we recall how to heuristically identify the classical Liouville equation (for the dynamics of a probability distribution over phase space) that is associated with a Markovian quantum system in the limit $\hbar \to 0$.  We will do so by considering the quantum dynamics in the Wigner phase-space representation. Note that this is not a formal limit.  Indeed even when the quantum and classical Liouvillians are close according to an appropriate metric, the evolving states will often diverge exponentially fast in time, so that similar dynamics on an identical initial state can produce very different states at later times, including flagrantly non-classical states.

The Lindblad equation \eqref{eq:lindblad} is
transformed to the Wigner representation as $\partial_t W_\rho = \LLCq[W_\rho]$ by applying $\WW$ to both sides \cite{graefe2018lindblad, strunz1998classical, tzanakis1998generalized}:
\begin{align}
    \LLCq [W_\rho] :=&\, \WW[\LLQ[\rho]] \\ 
    =&\, -\frac{i}{\hbar} (\HC \star W_\rho - W_\rho \star \HC)  + \frac{1}{\hbar }\sum_k \left( \LCk  \star W_\rho \star \LCk^* - \frac{1}{2} \LCk^* \star \LCk \star W_\rho - \frac{1}{2} W_\rho\star \LCk^* \star \LCk   \right)\\
    =&\, \Moyalbracket{\HC,W_\rho} + \frac{i}{2}\sum_k \left( \Moyalbracket{\LCk  \star W_\rho, \LCk^*} + \Moyalbracket{\LCk, W_\rho \star \LCk^*} \right) 
\end{align}
where $W_\rho=\WW[\rho]$ is the Wigner function of $\rho$. 
We emphasize that $\LLCq  = \WW\circ\LLQ\circ\WW^{-1}$ is\footnote{As usual, ``$\circ$'' denotes function composition.} just a different representation of the exact quantum dynamics generated by $\LLQ$.
Using the series expression for the Moyal star product \eqref{eq:moyal-star-expansion} to expand in powers of $\hbar$, and making use of $\partial_a \partial^a = \sf^{ab} \partial_a \partial_b = 0$ (by symmetry), we have \cite{frigerio1984diffusion,tzanakis1998generalized,strunz1998classical,dubois2021semiclassical} (see also \cite{heller1976wigner, pule1979classical, brodier2004symplectic,bondar2016wigner})
\begin{align}
    \label{eq:quantum-fokker-planck-unsimplified}
    \LLCq [W] =&\, 
        (\partial_a \HC) (\partial^a W )+  \partial^a \left[W  \Imag \sum_k \LCk \partial_a \LCk^*\right] +\frac{\hbar}{2} \partial_a \left[(\partial_b W) \Real  \sum_k(\partial^a \LCk)(\partial^b \LCk^*)\right] + O(\hbar^2)\\
    \label{eq:quantum-fokker-planck}
     =& -\partial_a \left[\left(\partial^a \HC +   \GC^a\right) W\right]  +	\frac{\hbar}{2} \partial_a \left(\scD^{ab} \partial_b W\right) + O(\hbar^2)
\end{align}
where 
the friction vector $\GC^a:=\Imag \sum_k \LCk \partial^a \LCk^*$ and the \introduce{scaled diffusion matrix}\footnote{Instead of $\scD_{ab}$, many authors (e.g., Ref.~\cite{joos2013decoherence}) have traditionally used a ``localization matrix'' $\locD_{ab} = \hbar^{-1}\scD_{ab} = \hbar^{-2}\D_{ab} = \hbar^{-1} \Real \sum_k \ell_{k,a}^* \ell_{k,b}$ (or maybe with a factor of 2).  Some intuition for the physical meaning of these matrices can come from noting that a superposition of two wavepackets widely separated in phase space by the vector $\alpha$ decoheres at a rate $\alpha^a \locD_{ab} \alpha^b = \hbar^{-1} \alpha^a \scD_{ab} \alpha^b$, i.e., the off-diagonal components of the density matrix decay like $\sim \exp(-t \alpha^a \locD_{ab} \alpha^b)$. We choose to work with $\scD_{ab}$ rather than $\locD_{ab}$ because $\scD_{ab}$ has no $\hbar$ dependence (as we consider $\LCk$ and $\HC$ to be independent of $\hbar$) when, as we have done, Lindblad operators are defined so products of pairs of them have the same units as the Hamltonian. This makes it easier to read off the classical limit $\hbar\to 0$.}
\begin{align}\label{eq:delta-def-appendix}
    \scD^{ab} := \Real \sum_k(\partial^a \LCk)(\partial^b \LCk^*)
\end{align}
are functions on phase space.   
This shows that if we identify the \introduce{diffusion matrix}\footnote{There have long been competing \cite{risken1984fokkerplanck} conventions \cite{gardiner2009stochastic} on whether to include the factor of $1/2$ in front of the diffusion term in the Fokker-Planck equation, and there is no uniformity even within authors studying quantum Brownian motion specifically.  Our convention for the matrix $\D_{ab}$ agrees with, e.g., Di\'{o}si \& Kiefer \cite{diosi2000robustness, diosi2002exact} and Graefe et al.\ \cite{graefe2018lindblad}, but differs by a factor of 2 from, e.g., Isar et al.~\cite{isar1994open} and Dekker \& Valsakumar~\cite{dekker1984fundamental}.}  as $\D^{ab} :=  \hbar\scD^{ab}$ then the quantum dynamics in the Wigner representation take on the general form\footnote{Strictly speaking, one can consider the Kramer-Moyal expansion, a partial differential equation for $f$ with derivatives of arbitrary power.  However, by the Pawula theorem, if the expansion does not terminate by second order then it must contain an infinite number of terms in order that $f$ remain positive \cite{pawula1967approximation}. See Ref.~\cite{pawula1965generalizations} and Sections~1.2.7, 3.3.2, and 4.1 of Ref.~\cite{risken1984fokkerplanck} for further discussion.  In the case of a classical stochastic system that arises as the limit of a Lindblad equation, we see that the additional terms will correspond to higher powers of $\hbar$, which get small in the classical limit.  We have kept track of the $O(\hbar^1)$ terms because these are the necessary ones to produce the classical state $\rhoc$ that $\rhot$ will well approximate.  That is, adding higher-order terms would define different $\rhoc$, but they would all be close to $\rhot$, while dropping the $O(\hbar^1)$ terms would give non-diffusive (though generically still dissipative) dynamics that produce a $\rhoc$ that is \emph{not} well approximated by $\rhot$.} of the Fokker-Planck equation \cite{risken1984fokkerplanck, gardiner2009stochastic}
\begin{align}\label{eq:fokker-planck-appendix}
    \LLC [f] 
    \, &= -\partial_a [(\partial^a\HC + \GC^a) \cstate]  +	\frac{1}{2} \partial_a ( \D^{ab} \partial_b \cstate)\\
    \label{eq:def-deterministic-drift}
    &= -\partial_a (\UC^a \cstate) +	\frac{1}{2} \partial_a ( \D^{ab} \partial_b \cstate)
\end{align}
up to terms of order $O(\hbar^2)$, where we have introduced the \introduce{deterministic drift} $\UC^a := \partial^a \HC +  \GC^a$. This justifies our Definition~\ref{def:correspondingDynamics} for corresponding classical dynamics.

It's worth briefly noting that the Fokker-Planck equation is often written as
\begin{align}\label{eq:fokker-planck-alt}
    \LLC [f] 
    \, &= -\partial_a [(\partial^a\HC + \GC^a +\partial_b \D^{ab}/2) f]  + \frac{1}{2} \partial_a \partial_b( \D^{ab} f)\\
    \label{eq:def-mean-drift}
    \, &= -\partial_a (\meandrift^a \cstate)  +	\frac{1}{2} \partial_a \partial_b( \D^{ab} \cstate)
\end{align}
which has the advantage\footnote{On the other hand, the form \eqref{eq:fokker-planck-alt} has the advantage of taking the explicit divergence form $\partial_a (\D^{ab} \partial_b f)$ which is a self-adjoint operator with respect to the $L^2$ norm.  The form \eqref{eq:fokker-planck-alt} is associated with the It\^{o} stochastic calculus, while \eqref{eq:fokker-planck-appendix} is associated with the alternative formalism of Stratonovich.} of isolating the \introduce{mean drift} vector
$\meandrift^a := \partial^a \HC +  \GC^a + \partial_b \D^{ab}/2$ (usually called simply the drift).  The mean drift points in the direction of the \emph{mean} probability flow, i.e., the direction that a strongly localized distribution will move when averaging over the diffusion: 
$\frac{\dd}{\dd t} \langle r^a\rangle_\cstate = \int\!\dd \alpha\, \meandrift^a(\alpha) \cstate(\alpha)$. 
The mean drift and the deterministic drift differ by the \introduce{spurious drift} vector $\partial_b \D^{ab}/2 = \meandrift^a-\UC^a$ (also known as the noise-induced drift).  For the important case of harmonic dynamics, discussed in Section~\ref{sec:harmonic-dynamics}, the diffusion matrix $D$ is constant over phase space, so $\meandrift = \UC$, the spurious drift vanishes, and the two forms \eqref{eq:fokker-planck-appendix} and \eqref{eq:fokker-planck-alt} coincide.  For non-harmonic dynamics, we will be most interested in the deterministic drift $\UC^a$ because, as discussed in Section~\ref{sec:harmonic-gaussian}, it is the direction in which Gaussians states flow under the local harmonic approximation introduced in Section~\ref{sec:harmonic-approx}.

We emphasize that even though the diffusion term 
\begin{align}
    \frac{1}{2}\partial_a \left( \D^{ab} \partial_b\cstate\right) = \frac{\hbar}{2} \partial_a \left( \scD^{ab} \partial_b\cstate\right) = \frac{\hbar}{2} \partial_a \left(\Imag \sum_k (\partial^a \LCk)(\partial^b \LCk^*) \partial_b\cstate\right)
\end{align} 
in the Fokker-Planck equation \eqref{eq:fokker-planck-appendix} vanishes as $\hbar\to 0$ with the classical function $\LCk$ fixed, we do not generically recover closed-system dynamics in this limit: the friction vector $\GC^a = \Imag \sum_k \LCk \partial^a \LCk^*$ survives.
However, when the system is closed ($\LCk=0$), both the friction and the diffusion vanish and we recover the Liouville equation: $\partial_t f = (\partial_a \HC) (\partial^a f) = \Poissonbracket{\HC,f}$. 

\subsection{Harmonic Markovian dynamics: quadratic Lindblad equation}\label{sec:harmonic-dynamics}

It's widely known that when the Hamiltonian of a closed classical or quantum system is \emph{quadratic} in the phase space variables $\XC$ and $\PC$ (so $H = \hh_{ab}r^a r^b/2$ after an appropriate choice of the origin), the dynamics can be solved exactly for all time.  Such dynamics are often called ``linear'' because when the system is perturbed its response is proportional to the size of the perturbation.\footnote{More precisely, Hamilton's equation of motion $\partial_t r^a = \partial^a H = \hh^{a}_{\pha b} r^b$ is linear in the variable $r(t) = (x(t),p(t))$, so solutions (trajectories) are closed under linear combinations.  Equivalently, when eliminating $p$, the second order equation $[\partial_t^2 - \hh^{a}_{\pha a} \partial_t + \hh^{ab}\hh_{ab}/2]x(t)=0$ for $x$ is linear.}  To avoid confusion between the quadratic variables and the resulting linear response, we will call these ``harmonic'' dynamics.

It is less often appreciated that exact solutions also exist in the more general case of a Lindbladian open systems when, in addition to a \emph{quadratic} Hamiltonian, the Lindblad operators are \emph{linear} in $\XQ$ and $\PQ$ \cite{lindblad1976brownian, isar1994open, Riedel_MO_QBM}. (Introducing linear Lindblad operators, rather than quadratic ones,  is the natural way to generalize a quadratic Hamiltonian since the Lindblad operators appear together in pairs in the Lindblad equation.) We will call this \introduce{harmonic} (\introduce{Markovian}) \introduce{dynamics},\footnote{This is often called ``quantum Brownian motion'' (QBM), but that terminology is sometimes also applied to dynamics that feature non-quadratic Hamiltonians or that do not strictly obey the Markov property.} where the Hamiltonian and Lindblad operators take the form\footnote{Or, more explicitly, $\HQ = \hh_0 + \hh_\x \XQ + \hh_\p \PQ + \frac{1}{2} \hh_{\x\x} \XQ^2 + \hh_{\x\p} (\XQ \PQ+\PQ \XQ) +  \frac{1}{2} \hh_{\p\p} \PQ^2$ and $\LQk = \ell_{k,0} + \ell_{k,\x} \XQ + \ell_{k,\p} \PQ$.}
\begin{align}
	\label{eq:linear-operators}
	\HQ &= \hh_0 \IdQ + \hh_a \RQ^a + \frac{1}{2} \hh_{ab} \RQ^a\RQ^b,\\
 \label{eq:linear-lindblad}
	\LQk &= \ell_{k,0} \IdQ + \ell_{k,a} \RQ^a 
\end{align}
for real number $\hh_{0}$, $\hh_{a}$, and $\hh_{ab} = \hh_{ba}$ and complex numbers $\ell_{k,0}$ and $\ell_{k,a}$. 

The Lindblad equation \eqref{eq:lindblad-simple-restate} becomes\footnote{One way to simplify the manipulation is to make the Lindblad gauge transformation $\HQ \to \HQ + \Imag \sum_k \ell_{k,0} \LQk^\dagger$ and $\LQk \to \LQk - \ell_{k,0} \IdQ$ (which has effect $\hh_a \to \hh_a + \Imag \sum_k \ell_{k,0}  \ell_{k,a}^*$ and $\ell_{k,0} \to 0$).} \cite{dekker1981classical,isar1994open}
\begin{align}
    \LLQ^{\har} [\rho]
    & = -\frac{i}{\hbar}\bigg[\hh_a\RQ^a+ \frac{1}{2}\hh_{ab}\RQ^a\RQ^b+\Imag \sum_k \ell_{k,0} \ell_{k,a}^*\RQ^a,\rho \bigg] + \frac{1}{\hbar} \sum_k \ell_{k,a} \ell_{k,b}^*\left( \RQ^a \rho \RQ^b - \frac{1}{2} \left\{  \RQ^b \RQ^a,\rho\right\} \right) \\
    \label{eq:QBM-general}
    &= - \frac{i}{\hbar}\left[\RQ^a,\frac{1}{2}\left\{(\hh_a + \GG_a) + (\hh_{ab} +\GG_{ab})\RQ^b, \rho \right\} \right] - \frac{1}{\hbar}\frac{\scD_{ab}}{2} \left[ \RQ^a,\left[\RQ^b, \rho \right] \right],
\end{align}
where\footnote{Note that while both the real and imaginary parts of $\sum_k \ell_{k,a}^* \ell_{k,b}$ ($\scD_{ab}$ and $\GG_{ab}$) appear in the harmonic dynamics, only the imaginary part of $\sum_k \ell_{k,0} \ell_{k,a}^*$ appears. The real part does not contribute due to the form of the Lindblad equation, and the same is true for $\sum_k \ell_{k,0} \ell_{k,0}^*$ (which is real by construction).} 
\begin{align}\label{eq:harmonic-parameters}
	\GG_a = \Imag \sum_k \ell_{k,0} \ell_{k,a}^*, \qquad
        \scD_{ab} =  \Real \sum_k \ell_{k,a}^* \ell_{k,b} , \qquad 
	\GG_{ab} = \Imag \sum_k  \ell_{k,a}^* \ell_{k,b} 
\end{align}
are real-valued parameters.  Note that the scaled diffusion matrix $\scD_{ab}$ and the \introduce{friction gradient} $\GG_{ab}$ are the real and imaginary parts of the positive semidefinite matrix $\sum_k \ell_{k,a}^* \ell_{k,b}$, so they are respectively symmetric and antisymmetric, and $\scD_{ab}$ is furthermore positive semidefinite itself.
Eq.~\eqref{eq:QBM-general} is the most general possible open-system quantum dynamics of a single degree of freedom when the Hamiltonian is constrained to be no more than quadratic in $\XQ$ and $\PQ$ and the Lindblad operators are constrained to be no more than linear.

The dynamical equation for the Wigner function $W_\rho=\WW[\rho]=(2\pi\hbar)^{-d}\Op^{-1}[\rho]$ equivalent to Eq.~\eqref{eq:QBM-general} is\footnote{As discussed in Sec.~\ref{sec:symplectic-basics}, the arrow on the partial derivative $\vecpartial$ indicates that it acts on everything to the right, including $W(\alpha)$.}
\begin{align} \label{eq:QBM-wigner-evo}
    \WW[\LLQ^{\har}[\rho]](\alpha)
    =&\, \left[-\vecpartial{}_a\left(\hh^a+\GG^a + (\hh^a_{\pha b} + \GG^a_{\pha b})  \alpha^b\right) + \frac{\hbar}{2}\scD^{ab}\vecpartial{}_a \vecpartial{}_b \right]
    \WW[\rho]
    \\
    =&\, \left[-\vecpartial{}_a(\partial^a\HC +\GC^a) + \frac{1}{2}\D^{ab}\vecpartial{}_a \vecpartial{}_b \right]
    \WW[\rho]
    \\
    =&\,
    \LLC^{\har}[\WW[\rho]](\alpha)
\end{align}
where we have evaluated our correspondence definitions in this case of harmonic dynamics:\footnote{Since $\hh_{ab}$ and $\GG_{ab}$ are respectively symmetric and antisymmetric by construction, their index-raised forms $\hh^{a}_{\pha b} = \sf^{ac}\hh_{cb}$ and $\GG^{a}_{\pha b} = \sf^{ac}\GG_{cb}$ are the Hamiltonian and skew-Hamiltonian components of $\hh^{a}_{\pha b} + \GG^{a}_{\pha b}$.  As will be seen in Section~\ref{sec:harmonic-gaussian}, $\hh^{a}_{\pha b} + \GG^{a}_{\pha b}$ controls the non-diffusive component of the dynamics for the covariance matrix of harmonically evolving Gaussian states.}
\begin{align} 
    \partial^a H(\alpha) = \, & \hh^a+\hh^a_{\pha b}\alpha^b\\
    \GC^a(\alpha) = \, & \Imag \sum_k \LCk(\alpha) \partial^a \LCk^*(\alpha) = \GG^a + \GG^a_{\pha b}\alpha^b\\
    \D^{ab}(\alpha) = \, & \hbar\scD^{ab}(\alpha) 
\end{align}
Note in particular that the Hamiltonian drift $\partial^a H$ and the friction $\GC^a$ (and hence the deterministic drift $\UC^a$) are all linear on phase space. Furthermore there is no spurious drift $\partial_b \D^{ab}/2$ because the diffusion $\D^{ab}$ is constant, so the deterministic drift and mean drift coincide: $\UC^a = \meandrift^a$. 

From \eqref{eq:QBM-wigner-evo} we see that this dynamical equation for the Wigner function in quantum harmonic dynamics takes the \emph{exact} same form as a Fokker-Planck equation for classical harmonic dynamics, i.e., 
\begin{align}\label{eq:harmonic-equivalence}
    \LLC^{\har}\circ\WW =\WW\circ\LLQ^{\har}. 
\end{align}
This is because, unlike the general anharmonic case discussed in Section~\ref{sec:corresponding-dynamics}, there are no terms of order $O(\hbar^2)$ or higher.

\subsection{Gaussian states and their harmonic evolution}\label{sec:harmonic-gaussian}

We recall that the \introduce{covariance matrix} of a pure quantum state $\psi$ with zero mean position and momentum ($\langle \psi | \XQ |\psi \rangle = 0$, $\langle \psi | \PQ| \psi \rangle = 0$) is defined as
\begin{align}
        \label{eq:covariance-def}
	\sigma^{ab}\, 
	&= \begin{pmatrix}
		\sigma^{\x\x} & \sigma^{\x\p} \\
		\sigma^{\p\x} & \sigma^{\p\p}
	\end{pmatrix}
    =  \left\langle \psi\middle| \frac{\{\RQ^a,\RQ^b\}}{2} \middle|\psi \right\rangle
	= \left\langle \psi\middle|\begin{pmatrix}
		 \XQ^2   &  (\XQ\PQ+\PQ\XQ)/2  \\
		(\XQ\PQ+\PQ\XQ)/2  &  \PQ^2 
	\end{pmatrix}\middle|\psi \right\rangle \\
	&=  \left\langle \RC^a \RC^b \right\rangle_W
	= \begin{pmatrix}
		\langle  \XC^2 \rangle_W  & \langle  \XC\PC  \rangle_W \\
		\langle  \XC\PC  \rangle_W & \langle \PC^2  \rangle_W
	\end{pmatrix}
\end{align}
Here, $\RC^a(\alpha) = \alpha^a$ is the phase-space coordinate function, $W$ is the Wigner function of $\psi$, and expectation values are $\langle  f(\alpha) \rangle_W := \int \! \dd \alpha\, W(\alpha) f(\alpha)$. More generally, when the state $\rho$ is mixed and the means
$\bar{\RC}^a:= (\bar{\XC},\bar{\PC})^a := \Tr[\rho \RQ^a ] = \langle \RC^a \rangle_W$ are non-zero,
the covariance matrix is 
\begin{align}\label{eq:sigma-def}
	\sigma^{ab}
	:= \Tr[\rho \{(\RQ-\bar{\RC})^a,(\RQ-\bar{\RC})^b\}/2 ] =  \langle (\RC-\bar{\RC})^a(\RC-\bar{\RC})^b \rangle_W.
\end{align}

A \introduce{Gaussian distribution} over phase space takes the form
\begin{align}\label{eq:gaussian-wigner}
	\tauCas(\alpha+\beta) = \frac{\exp(-\beta^a\sigma^{-1}_{ab}\beta^b/2)}{(2\pi)^d\sqrt{\det\sigma}} 
\end{align}
for a positive semidefinite covariance matrix $\sigma^{ab}$ and mean $\alpha^a$. It is the Wigner function of the quantum state
\begin{align}\label{eq:gaussian-state}
	\tauQas = \WW^{-1}[\tauCas]= (2\pi\hbar)^{d}\Op[\tauCas] 
\end{align}
that generally takes the form $\tauQas \propto e^{-\RQ^a A^{ab} \RQ^b}$ where $A$ is a matrix determined by $\sigma$; when $\sigma$ corresponds to a pure state $\tauQas$ (see below), the corresponding $A$ diverges, and one would instead write $\tauQas=|\alpha,\sigma\rangle \langle \alpha,\sigma|$ where $|\alpha,\sigma\rangle $ has a Gaussian wavefunction.
The Gaussian distribution obeys the mixing relation $\tauC_{\alpha_1,\sigma_1}\ast\tauC_{\alpha_2,\sigma_2} = \tauC_{\alpha_1+\alpha_2,\sigma_1+\sigma_2}$, where ``$\ast$'' denotes the convolution, $f\ast g(\alpha) = \int\!\dd \alpha\, f(\alpha-\beta)g(\beta) = g\ast f(\alpha)$.  This can be extended to Gaussian states through linearity of the Wigner function:  $\tauQ_{\alpha_1,\sigma_1}\ast\tauC_{\alpha_2,\sigma_2} = \tauQ_{\alpha_1+\alpha_2,\sigma_1+\sigma_2} = \tauC_{\alpha_1,\sigma_1}\ast\tauQ_{\alpha_2,\sigma_2}$, which preserves the normalization and positive semidefinite conditions.

For our main result we will need an simple extension of the Weyl trace identity, \eqref{eq:trace-identity}, to Gaussian states:
\begin{restatable}[Trace formula for mixtures of Gaussians]{lem}{lemGaussianTrace}\label{lem:gaussian-trace} 
If $E(\alpha)$ is bounded by a polynomial and $\rho  = \int \hat{\tau}_{\alpha,\sigma}\diff \mu(\alpha,\sigma)$ is a mixture of Gaussians,
\begin{align}\label{eq:schwartz-trace-formula}
    \Tr[\hat{E}\rho] = \Tr[\rho \hat{E}] = 
    \int_{\mathbb{R}^{2d}}\! \dd \alpha\, E(\alpha)W_\rho(\alpha)
\end{align}
where $W_\rho:=\WW[\rho]$ is the Wigner function of $\rho$.
\end{restatable}
\begin{proof}
One can directly compute
\begin{equation}
\begin{split}
\Tr[\hat{E}\rho]
&= \int \langle\alpha,\sigma|\hat{E}|\alpha,\sigma\rangle
\diff\mu(\alpha,\sigma) \\
&= \int \left(\int E(\beta) \tauCas (\beta) \diff \beta \right) 
\diff\mu (\alpha,\sigma) \\
&= 
\int E(\beta) 
\left(\int \tauCas (\beta)\diff\mu(\alpha,\sigma)\right) \diff \beta \\
&= 
\int E(\beta) 
W_\rho(\beta) \diff \beta.
\end{split}
\end{equation}
The second line follows from an explicit calculation of 
the inner product against a Gaussian state.  The key 
point is that if $E$ is bounded by a polynomial,
then in particular it is a tempered distribution so
$\hat{E}$ is well-defined as a map from Schwartz class
functions to tempered distributions, and thus
$\langle \alpha,\sigma| \hat{E}|\alpha,\sigma\rangle$
is well-defined.
\end{proof}

The distribution $\tauCas$ and $\tauQas$ are always normalized, $\int\!\dd\alpha\,\tauCas(\alpha) = \Tr[\tauQas] = 1$, but $\tauQas$ is only a pure quantum state state ($\tauQas \ge 0$, $\Tr[\tauQas]=\Tr[\tauQas^2]=1$) when $\frac{\sigma}{\hbar/2}$ is additionally a symplectic matrix (i.e., $\frac{\sigma}{\hbar/2} \sf \frac{\sigma}{\hbar/2} = \sf$).  More generally, these equivalent conditions on a positive semidefinite matrix $\sigma$ ensure that $\tauQas$ is a (possibly mixed) quantum state \cite{simon1994quantumnoise,weedbrook2012gaussian}:
\begin{itemize}
\item $\tauQas \ge 0$.
\item $\sigma \ge \sigmat$ for some $\sigmat$ such that $\frac{\sigmat}{\hbar/2}$ is symplectic and positive semidefinite, i.e., $\tauQas$ can be expressed as a Gaussian mixture of pure Gaussian states $\tauQ_{\alpha, \sigmat}$.
\item $\frac{\sigma}{\hbar/2} + i\sf \ge 0$.
\item $\nu_i\ge 1$, where $\{\nu_i\}_{i=1}^{2d}$ are the Williamson symplectic eigenvalues \cite{williamson1936algebraic,ikramov2018symplectic} of $\frac{\sigma}{\hbar/2}$.
\end{itemize}
When $\text{rank}(\tauQas)=1$, the state $\tauQas$ is more specifically a \introduce{pure Gaussian state} (also called a ``squeezed coherent state''), in which case the inequalities above are saturated. 

The above demonstrates why some authors in Gaussian quantum information set $\hbar=2$, although we will not do so in this paper.
Instead, we will occasionally work with the rescaled matrix $\sigmas := \frac{\sigma}{\hbar/2}$ for convenience.

A powerful fact about harmonic dynamics is that, in both classical and quantum systems, Gaussians remain Gaussian for all time, with the centroids following the classical equations of motion and the covariance matrices obeying linear dynamics \cite{isar1994open,brodier2004symplectic,graefe2018lindblad} (see also \cite{strunz1998classical, halliwell1995quantum}):\footnote{The deterministic drift \eqref{eq:adot} describes the movement of the center of a Gaussian wavepacket; it includes the symplectic flow $\partial^a \HC$ from the Hamiltonian and the friction $\GC^a = \Imag \sum_k \LCk \partial^a \LCk^*$ from the Lindblad terms. Using traditional matrix multiplication, the change in the covariance matrix \eqref{eq:sdot} can be written more concisely as 
$\sdot = \K\sigma+\sigma\K^\tp+\D$ 
where $\K$ denotes the asymmetric matrix $\K^{a}_{\pha c}=\partial_c\UC^a = \partial_c\partial^a \HC +\partial_c \GC^a$. This non-diffusive ($\D\to 0$) component of $\dot{\sigma} = \sdot$ arises from the (uncertainty-area-preserving) local stretching and skewing, and can be derived intuitively by looking directly at the change in the covariance matrix $\sigma$ under the linear flow $\partial\UC$ of the probability mass. The diffusive component $\D$ arises of course from the (uncertainty-expanding) noise.}

\begin{restatable}[Gaussian harmonic evolution]{lem}{lemGaussHarmEvo}
    \label{lem:gaussian_evolution}
    Assume a Gaussian classical initial state $\cstate(0)=\tauC_{\alpha_0,\sigma_0}$, centered at $\alpha_0$ in phase space with covariance matrix $\sigma_0$, and let $\cstate(t)$ be a solution to classical harmonic dynamics, i.e., the Fokker-Planck equation \eqref{eq:fokker-planck-restate} with a deterministic drift $\UC^a = \partial^a\HC+\GC^a$ and diffusion matrix $\D^{ab}$ that are respectively linear and constant functions of the phase-space coordinates $\RC=(\XC,\PC)$. Then 
    $\cstate(t)=\tauC_{\alpha(t),\sigma(t)}$ where
    \begin{align}
        \label{eq:adot}
        \frac{\dd \alpha^a(t)}{\dd t} 
         =&\, \adot^a(\alpha)
         \\
        \label{eq:sdot}
        \frac{\dd \sigma^{ab}(t)}{\dd t} 
        =&\, \sdot^{ab}(\alpha,\sigma) 
        := \partial_c\UC^a(\alpha)\sigma^{cb}+\sigma^{ac}\partial_c\UC^b(\alpha)+ \D^{ab}(\alpha)
    \end{align}
    with $\alpha(0) = \alpha_0$, $\sigma(0) = \sigma_0$. Likewise, for Gaussian quantum initial state $\rho(0)=\tauQ_{\alpha_0,\sigma_0}=\WW^{-1}[\tauC_{\alpha_0,\sigma_0}]$ and $\rho(t)$ a solution to quantum harmonic dynamics (the Lindblad equation \eqref{eq:lindblad-simple-restate} with  $\HQ$ and $\LQk$ respectively quadratic and linear in $\RQ=(\XQ,\PQ)$), we have 
    $\rho(t)=\tauQ_{\alpha(t),\sigma(t)}=\WW^{-1}[\tauC_{\alpha(t),\sigma(t)}]$.
\end{restatable}
\begin{proof}
The classical case can be checked by direct computation.\footnote{Another approach is to first observe that the Fokker-Planck equation preserves the Gaussian property of distributions and then compute the time derivatives of the mean and covariance from their definition using integration by parts.}
First, we recall the Gaussian derivatives reviewed in Appendix~\ref{sec:gaussian-derivatives}, 
\begin{align}
\frac{\partial}{\partial \alpha^a}\tauCas(\beta) 
= \frac{\partial}{\partial \alpha^a}\tauC_{0,\sigma}(\beta-\alpha) 
= - \frac{\partial}{\partial \beta^a}\tauC_{0,\sigma}(\beta-\alpha)
= - \frac{\partial}{\partial \beta^a}\tauCas(\beta) 
= \sigma^{-1}_{ab}\beta^b\tauCas(\beta) ,\\
\frac{\partial}{\partial \sigma^{ab}} \tauCas(\beta) 
=\frac{1}{2}(\sigma^{-1}_{ac}(\beta-\alpha)^c\sigma^{-1}_{bd}(\beta-\alpha)^d-\sigma^{-1}_{ab})\tauCas(\beta) 
= \frac{1}{2}\frac{\partial}{\partial\beta^a}\frac{\partial}{\partial\beta^b} \tauCas(\beta).
\end{align}
Then we evaluate the time derivative with the chain rule:
\begin{align}
    \partial_t \tauC_{\alpha(t),\sigma(t)}(\beta)
    \,&= \left[\frac{\dd \alpha^a(t)}{\dd t}\frac{\vecpartial}{\partial \alpha^a} + \frac{\dd \sigma^{ab}(t)}{\dd t}\frac{\vecpartial}{\partial \sigma^{ab}}\right]\tauC_{\alpha(t),\sigma(t)}(\beta)\\
    \,&= \left[-\UC^a(\alpha)\frac{\vecpartial}{\partial \beta^a} + \Big(\partial_c\UC^a(\alpha)\sigma^{cb}+\sigma^{ac}\partial_c\UC^b(\alpha)+ \D^{ab}(\alpha)\Big)\frac{1}{2}\frac{\vecpartial}{\partial\beta^a}\frac{\vecpartial}{\partial\beta^b}\right]\tauC_{\alpha(t),\sigma(t)}(\beta)\\\,
    \begin{split}
    &= \bigg[-\frac{\vecpartial}{\partial \beta^a}\UC^a(\alpha) 
    -\frac{1}{2}\frac{\vecpartial}{\partial \beta^a}\partial_c\UC^a(\alpha)\sigma^{cb}\sigma^{-1}_{bd}(\beta-\alpha)^d
    \\
    &\hspace{7.5em}-\frac{1}{2}\frac{\vecpartial}{\partial \beta^b}\sigma^{ac}\partial_c\UC^b(\alpha)\sigma^{-1}_{ad}(\beta-\alpha)^d+\frac{1}{2}\D^{ab}(\alpha)\frac{\vecpartial}{\partial\beta^a}\frac{\vecpartial}{\partial\beta^b}\bigg]\tauC_{\alpha(t),\sigma(t)}(\beta)
    \end{split}\\
    &= \left[-\frac{\vecpartial}{\partial \beta^a}\UC^a(\alpha) 
    +\frac{\vecpartial}{\partial \beta^a}\partial_c\UC^a(\alpha)(\alpha-\beta)^c
    +\frac{1}{2}\D^{ab}(\alpha)\frac{\vecpartial}{\partial\beta^a}\frac{\vecpartial}{\partial\beta^b}\right]\tauC_{\alpha(t),\sigma(t)}(\beta)\\
    \label{eq:use-linearity-of-u}
    &= \left[-\frac{\vecpartial}{\partial\beta^a}\UC^a(\beta) + \frac{1}{2}\frac{\vecpartial}{\partial\beta^a} \D^{ab}(\beta) \frac{\vecpartial}{\partial\beta^b}\right]\tauC_{\alpha(t),\sigma(t)}(\beta)\\
    &= \LLC[\tauC_{\alpha(t),\sigma(t)}](\beta).
\end{align}
To to get \eqref{eq:use-linearity-of-u} we used the constancy of the diffusion, $\D^{ab}(\beta)=\D^{ab}(\alpha)$, and the linearity of the drift, $\UC^a(\beta) = \UC^a(\alpha) + \partial_c\UC^a(\alpha)(\beta-\alpha)^c$. 
The quantum case follows from \eqref{eq:harmonic-equivalence}, i.e., the equivalence $\WW^{-1} \circ \LLC=\LLQ\circ\WW^{-1}$ for harmonic dynamics:
\begin{align}
    \partial_t \tauQ_{\alpha(t),\sigma(t)} = \WW^{-1} [\partial_t \tauC_{\alpha(t),\sigma(t)}] =  \WW^{-1} \circ \LLC [\tauC_{\alpha(t),\sigma(t)}] = \LLQ\circ\WW^{-1}[\tauC_{\alpha(t),\sigma(t)}] = \LLQ[\tauQ_{\alpha(t),\sigma(t)}]
\end{align}

\end{proof}

\subsection{Local harmonic approximation}\label{sec:harmonic-approx}

We will now define a local harmonic approximation to both quantum and classical dynamics about an arbitrary point $\alpha$ in phase space. The quantum approximation is a natural extension of Heller's semiclassical approximation for closed quantum systems \cite{heller1975time, heller1976wigner}.  In particular, see  Ref.~\cite{heller1976wigner} for a discussion of the basic reason that expanding the Wigner function in powers of $\hbar$ and truncating is often not well-behaved, while the present technique is: expand the dynamics $\LLQ$ in powers of $\hbar$, truncate, and then evolve the Wigner function exactly with that.  Vladimirov \& Petersen considered a local harmonic approximation to Markovian open-system dynamics in (effectively) the special case of linear Lindblad operators \cite{vladimirov2012gaussian}, although we are unsure if it is equivalent to our definition in that case.

In multi-index notation,\footnote{Recall: $\partial_\alpha^n E=\partial_{x_1}^{n_1}\cdots\partial_{x_d}^{n_d} \partial_{p_1}^{n_{d+1}}\cdots\partial_{p_d}^{n_{2d}}E$, $\alpha^n := (x_1)^{n_1}\cdots (x_d)^{n_d} (p_1)^{n_{d+1}}\cdots (p_d)^{n_{2d}}$, and 
$n!=\prod_{j=1}^{2d} (n_j)!$ for $n:=(n_1,n_2,\cdots,n_{2d}) \in (\mathbb{Z}_{\ge 0})^{\times 2d}$.} the Taylor approximations about $\alpha$ of a function $E$ at an arbitrary order $m\in\mathbb{Z}_{\ge 0}$:
\begin{align}
	E^{[\alpha,m]}(\alpha+\beta) &:= \sum_{|n|\le m} \frac{
    (\partial_\alpha^n E)(\alpha)
    }{n!} \beta^n
\end{align}
with error
\begin{align}
	\delta E^{[\alpha,m]} &:= E(\alpha)-E^{[\alpha,m]}.
\end{align}
The Taylor remainder theorem gives the bound 
$\delta E^{[\alpha,m]}(\beta) \leq \frac{1}{m!}|\beta-\alpha|^{m+1}|E|_{C^{m+1}}$.

The Taylor approximation for the operator, and its error, are then naturally defined using Weyl quantization:
\begin{align}
	\hat{E}^{[\alpha,m]} := \Op[E^{[\alpha,m]}], \qquad \qquad \delta\hat{E}^{[\alpha,m]} := \hat{E}-\hat{E}^{[\alpha,m]}
\end{align}
We will in particular use the second-order approximation to the classical Hamiltonian,
\begin{align}
	\HQLat =& 
	\Op[\HCLat]\\
	=& \hh_0(\alpha) + \hh_a(\alpha) (\RQ-\alpha)^a + \frac{1}{2} \hh_{ab}(\alpha) (\RQ-\alpha)^a (\RQ-\alpha)^b
\end{align}
and the first- and second-order approximations to the classical Lindblad functions,
\begin{align}
	\VQLaok = \Op[\VCLaok] &= \LCLaok(\RQ) - \LCk(\alpha) = \ell_{k,a}(\alpha) (\RQ-\alpha)^a \\
	\VQLatk = \Op[\VCLatk] &= \LCLatk(\RQ) - \LCk(\alpha) = \ell_{k,a}(\alpha) (\RQ-\alpha)^a + \frac{1}{2} \ell_{k,ab}(\alpha) (\RQ-\alpha)^a (\RQ-\alpha)^b
\end{align}
where we have defined the shorthand $\VQk := \LQk-\LCk(\alpha)$, which is just the Lindblad operator with its classical value at $\alpha$ subtracted off.  We have introduced\footnote{Eqs.~\eqref{eq:hh-def}, \eqref{eq:ell-def}, and \eqref{eq:gg-def} reduce to \eqref{eq:linear-operators}, \eqref{eq:linear-lindblad}, and \eqref{eq:harmonic-parameters} in that special case where the dynamics are globally harmonic and the center of the approximation is set at the origin $\alpha=0$.}
\begin{align}
    \label{eq:hh-def}
    \hh_0(\alpha)  := &\, \HC(\alpha), & \quad \hh_a(\alpha)  := &\, \partial_a \HC(\alpha), & \quad \hh_{ab}(\alpha)  := &\, \partial_a\partial_b \HC(\alpha),\\
    \label{eq:ell-def}
    \ell_{k,0}(\alpha)  := &\,\LCk(\alpha), & \quad \ell_{k,a}(\alpha)  := &\, \partial_a \LCk(\alpha), & \quad \ell_{k,ab}(\alpha)  := &\, \partial_a \partial_b \LCk(\alpha)
\end{align}
where in particular $\hh_a$ and $\hh_{ab}$ are the local gradient and Hessian of the classical Hamiltonian $\HC$.
In a closed (i.e., Hamiltonian) system, $\hh^a = \sf^{ab}\hh_b$ is the classical flow and $\hh^a_{\pha b} = \sf^{ac}\hh_{bc}$ is the Jacobian.
For later use we also define the shorthand
\begin{align}
    \label{eq:gg-def}
    \GG^a_{\pha b}(\alpha) := \partial_b\GC^a(\alpha) =  \Imag \sum_k \left(\partial_b\LCk  \partial^a \LCk^* +\LCk  \partial_b\partial^a \LCk^*\right).
\end{align}

It is tempting to simply start with the Lindblad equation and replace $\HQ$ with its quadratic approximation $\HQLat := \Op[\HCLat]$ and $\LQk$ with its linear approximation $\LQLaok := \Op[\LCLaok]$, and indeed this would give harmonic dynamics, but it would not give the \emph{correct} harmonic dynamics.  The reason is that quadratic term in the Taylor approximation to $\LQk$ can still contribute at the same order as the quadratic part of $\HQ$, to wit, the quadratic term in $\LQk$ multiplied by the constant (zeroth order) term in $\LQk^\dagger$.

So instead, we re-write the \emph{exact} Lindblad equation \eqref{eq:lindblad} as
\begin{align}\label{eq:lindblad-modified}
	\begin{split}
		\LLQ[\rho]
		& = -\frac{i}{\hbar}\left[\HQ+ \Imag\sum_k \LCk(\alpha) (\LQk-\LCk(\alpha))^\dagger,\rho\right] \\
		&\qquad + \frac{1}{\hbar} \sum_k \left( (\LQk-\LCk(\alpha)) \rho( \LQk-\LCk(\alpha))^\dagger - \frac{1}{2} \left\{ (\LQk-\LCk(\alpha))^\dagger (\LQk-\LCk(\alpha)),\rho\right\} \right)
	\end{split}
	\\
	& = -\frac{i}{\hbar}\left[\HQ+ \Imag\sum_k \LCk(\alpha) \VQk^\dagger,\rho\right] + \frac{1}{\hbar} \sum_k \left( \VQk \rho \VQk^\dagger - \frac{1}{2} \left\{ \VQk^\dagger \VQk,\rho\right\} \right)
\end{align}
Here we are just observing the well-known fact that $\LLQ$ is invariant under the replacements $\HQ \to \HQ +\Imag\sum_k \LCk(\alpha) (\LQk-\LCk(\alpha))^\dagger$ and $\LQk \to \LQk-\LCk(\alpha)$, where  $\Imag\sum_k \LCk(\alpha) \VQk^\dagger$ is the contribution by the Lindblad operators to the Hamiltonian part of the dynamics. (Note that $\VQk$ will be different for different choices of $\alpha$, although we do not denote this dependence explicitly; Eq.~\eqref{eq:lindblad-modified} holds for any choice of $\alpha$. We emphasize that no approximation has yet been made.)

With this form we can now identify a harmonic approximation $\LLQLa$ to the Lindbladian $\LLQ$ near the point $\alpha$, where all terms are at most second-order in the phase-space operators:
\begin{align}
    \label{eq:q-lindblad-linearized}
    \LLQLa [\rho]
    \, :=& -\frac{i}{\hbar}\left[\HQLat+ \Imag\sum_k \LCLazk \VQLatk{}^\dagger,\rho\right] + \frac{1}{\hbar} \sum_k \left( \VQLaok \rho \VQLaok{}^\dagger - \frac{1}{2} \left\{ \VQLaok{}^\dagger \VQLaok,\rho\right\} \right)
    \\
    \label{eq:q-lindblad-linearized-classicalish}
    =&
    -\frac{i}{\hbar}\left[\RQ^a,\frac{1}{2}
    \left\{\widehat{\partial_a\HC^{[\alpha,2]}}+\GQLao_a, \rho\right\}\right] - 
    \frac{1}{2\hbar^2}  
    \left[ \RQ^a,\left[\DQLaz_{ab}\RQ^b, \rho\right] \right].
\end{align}

Our motivation to consider the second line comes from \eqref{eq:QBM-general}, and it should be compared to \eqref{eq:classical-fokker-planck-harmonic} below. It can be manipulated into this form directly\footnote{Recall that, per our notation, $\widehat{\partial_a\HC^{[\alpha,2]}}:=\Op[\partial_a\HC^{[\alpha,2]}] = \Op[(\partial_a\HC)^{[\alpha,1]}]=\Op[\partial_a\HC]^{[\alpha,1]} = \partial_a\HC(\alpha)+\partial_a\partial_b\HC(\alpha)(\RQ-\alpha)^b$. 
One can expand $\GCLao_a(\beta) = \GG_a + \GG_{ab}\beta^b$ for some real-valued $\GG_a$ and $\GG_{ab}$. One finds that $\GG_a = \Imag \sum_k \ell_{k,0} \ell_{k,a}^*$ and $\scD_{ab} =  \Real \sum_k \ell_{k,a}^* \ell_{k,b}$, in agreement with \eqref{eq:harmonic-parameters}, but that $\GG_{ab} = \Imag \sum_k  (\ell_{k,0} \ell_{k,ab}^* + \ell_{k,a}^* \ell_{k,b})$, which contains the extra term $\ell_{k,0} \ell_{k,ab}^*$ relative to \eqref{eq:harmonic-parameters} that vanishes in the (globally harmonic, so $\ell_{k,ab}=0$) case considered in Sec.~\ref{sec:harmonic-dynamics}. This term produces a non-zero symmetric component of $\GG_{ab}$.} (albeit laboriously).

By construction, these dynamics are harmonic.
Note the appearance of both $\VQLaok$ and $\VQLatk$, and also that $\LCLazk=\LCk(\alpha)$ is just a scalar.  Unlike simply replacing the Lindblad operators in the Lindblad equation with their linear approximations at $\alpha$, this definition correctly captures the complete harmonic dynamics near $\alpha$.

On the classical side, we would like a similar approximation to the Fokker-Planck equation \eqref{eq:fokker-planck-appendix}, 
$\LLC [\cstate] = -\partial_a [(\partial^a\HC + \GC^a) \cstate]  +	\frac{1}{2} \partial_a [ \D^{ab} \partial_b \cstate]$,
that best approximates $\LLC$ in the vicinity of a point $\alpha$ while preserving Guassianity in the distribution $f$.

The most general Gaussian-preserving Fokker-Planck equation is one where the drift vector $\partial^a\HC + \GC^a$ and diffusion matrix $\D^{ab}$ are, respectively, linear and constant functions on phase space.

With maybe less initial motivation\footnote{Per Lemma~\ref{lem:q-c-harm-equiv}, a Gaussian centered on $\alpha$ will, under $\LLCLa$, flow in the direction of the deterministic drift $\adot^a(\alpha)$. Alternatively, one might consider flowing them along the mean drift $\meandrift^a(\alpha)$ by including a linear approximation to the spurious drift $\meandrift^a(\alpha)-\adot^a(\alpha) = \partial_b\D^{ab}/2$ in \eqref{eq:fokker-planck-harmonic-2}. Although there are existing uses of the local harmonic approximation to the Fokker-Planck equation in the case of uniform diffusion (zero spurious drift), e.g., Ref.~\cite{ben-nun1992approximate}, we were unable to find a clear definition in the case of non-zero spurious drift. The present definition, without the spurious drift, is ultimately justified by Lemma~\ref{lem:q-c-harm-equiv}.} than the quantum case, we will consider the harmonic approximation 
$\LLCLa$ to the classical dynamics $\LLC$ near the point $\alpha$ to be given by taking the linear approximation to the transport terms $\partial^a\HC + \GC^a$ and the zero-th order approximation to the diffusion term $\D^{ab}$: 
\begin{align}\label{eq:classical-fokker-planck-harmonic}
	\LLCLa [\cstate]
        =& - \partial_a \left[\left(\partial^a\HC^{[\alpha,2]} +  \GC^{[\alpha,1]}{}^a \right) \cstate \right]  + \frac{1}{2} \partial_a[\D^{[\alpha,0]}{}^{ab} \partial_b \cstate]
        \\
    \begin{split}
        \label{eq:classical-fokker-planck-harmonic-detailed}
	:=\,& \left(\partial_a \HCLat+ \Imag \sum_k \LCLazk \partial_a  \VCLatk{}^*\right)(\partial^a \cstate)  + \partial^a\left(\cstate  \Imag \sum_k \VCLaok \partial_a  \VCLaok{}^* \right)   \\ 
	&\qquad + \frac{\hbar}{2}  \partial_a\left[ \Real \sum_k\big(\partial^a \VCLaok\big)\big(\partial^b  \VCLaok{}^*\big) \partial_b \cstate\right]
    \end{split}
\end{align}

We collect these approximations in the following definition:
\begin{restatable}[Harmonic Approximation]{defin}{definHarmApprox}
	\label{def:harmonic-approx}
	Given the classical Markovian dynamics
	\begin{align}\label{eq:fokker-planck-exact-2}
		\LLC [\cstate] =& -\partial_a [\cstate (\partial^a \HC + \GC^a) ]  +	\frac{1}{2} \partial_a (\D^{ab} \partial_b \cstate)
	\end{align}
	we define the \introduce{classical harmonic approximation} to the dynamics at $\alpha$ as
	\begin{align}\label{eq:fokker-planck-harmonic-2}
		\LLCLa [\cstate]
		:=& -\partial_a [\cstate (\partial^a \HCLat + \GCLao{}^{a}) ]
		+ \frac{1}{2}  \partial_a (\D^{[\alpha,0]ab} \partial_b \cstate)
	\end{align}
	where $E^{[\alpha,m]}$ denotes the $m$-th order Taylor approximation to the phase space function $E$ at $\alpha$.  Likewise, given quantum Markovian dynamics
	\begin{align}\label{eq:quantum-dynamics-exact-2}
		\LLQ[\rho] & = -\frac{i}{\hbar}[\HQ,\rho] + \frac{1}{\hbar} \sum_k \left( \LQk \rho \LQk^\dagger - \frac{1}{2} \{ \LQk^\dagger \LQk,\rho\} \right)
	\end{align}
	we define the \introduce{quantum harmonic approximation} to the dynamics at $\alpha$ as
	\begin{align}
		\LLQLa [\rho]
		& := -\frac{i}{\hbar}\left[\HQLat+ \Imag\sum_k \LQLazk \VQLatk{}^\dagger,\rho\right] + \frac{1}{\hbar} \sum_k \left( \VQLaok \rho \VQLaok{}^\dagger - \frac{1}{2} \left\{ \VQLaok{}^\dagger \VQLaok,\rho\right\} \right) \\
	\end{align}
	where $\VCk:=\LCk-\LCLazk$ and $\hat{E}^{[\alpha,m]} = \WW[E^{[\alpha,m]}]$. 
    The respective errors are denoted
    \begin{align}
	\LLCRa := \LLC - \LLCLa,\qquad \qquad
        \LLQRa := \LLQ - \LLQLa
    \end{align}
\end{restatable}

Importantly, evolving $\rhot$ with $\LLQLa$ is equivalent to evolving its Wigner function $W_\rho =\WW[\rho]$ with $\LLCLa$: 
\begin{restatable}[Quantum-classical harmonic equivalence]{lem}{lemQCHarmEquiv}\label{lem:q-c-harm-equiv}
    Consider the exact classical dynamics \eqref{eq:fokker-planck-exact-2} corresponding (in the sense of Definition~\ref{def:correspondingDynamics}) to the exact quantum dynamics \eqref{eq:quantum-dynamics-exact-2} with $\HC$ and $\LCk$ twice differentiable.  Then their respective harmonic approximations $\LLQLa$ and $\LLCLa$ at any point $\alpha$ are equivalent in the sense of being directly related by the Wigner transform:
	\begin{align}\label{eq:q-c-harm-equiv}
		\LLCLa \circ \WW = \WW \circ \LLQLa 
	\end{align}
\end{restatable}
\begin{proof}
    By Definition~\ref{def:correspondingDynamics}, $\LLC$ is the classical limiting dynamics corresponding to $\LLQ$ when $\GC^a =\Imag \sum_k \LCk \partial^a \LCk^*$ and $\D^{ab}= \hbar \Real \sum_k (\partial^a \LCk)(\partial^b \LCk^*)$, so
    \begin{align}
        \GCLao{}^a &= \Imag \sum_k (\LCLazk \partial^a  \VCLatk{}^* + \VCLaok \partial^a  \VCLaok{}^*), \\
        \D^{[\alpha,0]ab} &= \hbar \Real \sum_k (\partial^a \LCLaok)(\partial^b  \LCLaok{}^*)
    \end{align}    
    Then \eqref{eq:q-c-harm-equiv} can be checked through direct computation with the Moyal product \eqref{eq:moyal-series-def} 
    using, for example, 
    \begin{align}\begin{split}
        & \WW\left[\VQLaok \rho \VQLaok{}^\dagger - \frac{1}{2} \left\{ \VQLaok{}^\dagger \VQLaok,\rho\right\} \right] 
        \\
        & \qquad =  - \partial_a\left[\Imag \big( \VCLaok \partial^a \VCLaok {}^* \big) W_\rho \right] +  \frac{\hbar}{2}  \left[ \Real \big(\partial^a \VCLaok\big)\big(\partial^b  \VCLaok{}^*\big)\right](\partial_a \partial_b W_\rho).
    \end{split}\end{align}
\end{proof}

\section{Proof of Theorem~\ref{thm:mainResult}}\label{sec:main_proof}

In this section we give a detailed outline of the proof of Theorem~\ref{thm:mainResult}, up to lemmas that are deferred to later sections.

\subsection{Defining the Gaussian mixture \texorpdfstring{$\rhot(t)$}{}}\label{sec:def-rhot}

We will define a quantum trajectory
\begin{align} \label{eq:rho_tilde_mixture}
	\rhot(t) = \int_{\bbR^{2d}} \int_{\bbR^{d\times d}} \tauQas \dd \pmsrt(\alpha,\sigma)
\end{align}
for a probability measure $\pmsrt$ that we will construct to satisfy
\begin{equation}
\label{eq:rhot-constraint}
\frac{\dd}{\dd t} \rhot(t) = 
\iint \!  \LLQLa[\tauQas] \dd\pmsrt(\alpha,\sigma)
\end{equation}
where $\LLQLa$ is the harmonic approximation to the Lindbladian, as defined in Section~\ref{sec:harmonic-approx}. The double integral sign is used to emphasize that the integral is taken over both phase space, $\bbR^{2d}$, and the space of all covariance matrices for pure Gaussian states, i.e., positive semidefinite $\sigma$ where $\sigma/(\hbar/2)$ is symplectic. We suppress the explicit integration domains in \eqref{eq:rhot-constraint} and hereafter.

We now invoke our lemma from Section~\ref{sec:harmonic-gaussian}, restated here for convenience, about the evolution of Gaussian states under harmonic dynamics:
\lemGaussHarmEvo*
\noindent By Definition~\ref{def:harmonic-approx} and Lemma~\ref{lem:q-c-harm-equiv}, the dynamics $\LLQLa$ are harmonic and characterized by quadratic Hamiltonian $\HCLat$, linear friction $\GCLao$, and constant scaled diffusion $\scD^{[\alpha,0]}$.  So it follows from Lemma~\ref{lem:gaussian_evolution} and the chain rule that\footnote{The third equality in \eqref{eq:harmonic-evo-gaussian-extend} is simply because $E^{[\alpha,m]}(\alpha) = E(\alpha)$ for all $m\ge 0$.} 
\begin{align}\label{eq:harmonic-evo-gaussian-extend}
    \LLQLa [\tauQas] = \partial_t \tauQas
    = [
    \UCLao(\alpha) 
    \vecpartial{}_\alpha + \sdot^{[\alpha,0]}(\alpha,\sigma) \vecpartial{}_\sigma] \tauQas = [\UC(\alpha) \vecpartial{}_\alpha + \sdot(\alpha,\sigma) \vecpartial{}_\sigma] \tauQas
\end{align}
for all $\alpha$, 
where we have introduced the abbreviations $A  \partial_{\alpha}\partial_{\alpha} := A^{ab} \partial_{a}\partial_{b} = A^{ab} \partial^2/\partial \alpha^{a}\partial \alpha^{b}$ and $A  \partial_{\sigma} := A^{ab} \partial/\partial \sigma^{ab}$.
Indeed, for the rest of this section and in Sec.~\ref{sec:linear-algebra-facts} it will be simpler to work with implicit matrix multiplication rather than explicit indices, so we will use $\UC$, $\sdot$, $\hh$, and $\GG$ in place of $\UC^a$, $\sdot^{ab}$, $\hh^a_{\pha b} =
\partial_b \partial^a \HC$, and $\GG^a_{\pha b} = \partial_b G^a$.
Then our desired condition~\eqref{eq:rhot-constraint} becomes
\begin{equation}
\label{eq:pas-weak-eq}
\begin{split}
     \frac{\dd}{\dd t} \iint \tauQas \dd\pmsrt(\alpha,\sigma)
    = \iint \! \dpmsrtas 
    \left[\UC(\alpha) \vecpartial{}_\alpha + \sdot(\alpha,\sigma) \vecpartial{}_\sigma\right]
    \tauQas 
\end{split}
\end{equation}

We could integrate the right-hand side of~\eqref{eq:pas-weak-eq} by parts in $\sigma$ and $\alpha$ to obtain a transport equation for $\pmsrt$, 
but we would quickly lose control of the covariance matrix $\sigma$, which can be strongly stretched by the evolution.
Instead, we observe that a component of the flow in the $\sigma$ direction (toward increasing mixedness of the state) can also be interpreted
as diffusion in the $\alpha$ direction.\footnote{When $\pmsrt(\alpha,\sigma)$ has support only on a single value of the covariance matrix $\sigma$, it is the measure associated with the  Glauber-Sudarshan P function, and it has long been known that diffusive dynamics, which would increases the mixedness of a single Gaussian state, can often be re-cast as diffusion in the P function over pure states \cite{weidlich1967quantumI, weidlich1967quantumII, haken1967quantum, lax1967quantum, isar1991quasiprobability, diosi2000robustness}.  What makes the present approach distinct is that we are considering a more general distribution $\pmsrt$ supported on a large (but restricted) space of pure-state covariance matrices $\sigma$.  Increasing this allowed space to include $\sigma$ corresponding to mixed states may allow our main result to be generalized further, but we defer that to future work.}  In particular, for 
Gaussian states $\tauCas(\beta) = \exp[-(\beta-\alpha)^\tp \sigma^{-1}(\beta-\alpha)/2]/((2\pi)^d\sqrt{\det\sigma})$,
\begin{align}\label{eq:covar-diff-relation}
    \partial_{\sigma} \tauCas = \frac{1}{2} \partial_{\alpha}  \partial_{\alpha} \tauCas,
\end{align}
as reviewed in Appendix~\ref{sec:gaussian-derivatives}.
Therefore, for any decomposition $\sdot=\sdotD+\sdotZ$, we have
\begin{equation}\label{eq:gauss-deriv-iden}
\left[\UC\vecpartial_\alpha
+ \sdot \vecpartial_\sigma\right]
\tauQas = 
\left[\UC\vecpartial{}_\alpha
+ \sdotZ \vecpartial{}_\sigma 
+ \frac{1}{2} \sdotD \vecpartial{}_\alpha\vecpartial{}_\alpha \right]
\tauQas.
\end{equation}
Plugging this into~\eqref{eq:pas-weak-eq} and integrating by parts, we see that if $\pmsrt$ is a probability measure then~\eqref{eq:rhot-constraint} is satisfied so long as 
$\pmsrt$ solves\footnote{Recall that the arrow on the partial derivatives $\vecpartial$ indicates that they act on everything to the right, including $\pmsrt$.}
\begin{equation}
\label{eq:pas-evolution-eq}
\frac{\dd}{\dd t}\pmsrt = \left[-\vecpartial{}_\alpha \UC 
-\vecpartial{}_\sigma \sdotZ 
+ \frac{1}{2} \vecpartial{}_\alpha\vecpartial{}_\alpha \sdotD\right]\pmsrt
\end{equation}

The main question remaining is the definition of $\sdotZ$ and $\sdotD$ such that $\pmsrt$ remains a probability measure and supported on the pure NTS states.
This is deferred to Section~\ref{sec:linear-algebra-facts}, but we state here the primary condition.

\begin{restatable}[$\NTS$-preservation condition]{lem}{lemNTSpreservation}
    \label{lem:NTS-preserve}
    Suppose that an initial probability measure $\pmsr_0$ is supported on the set 
    $\bbR^{2d} \times\SNTS(\NTSm)$
    and that 
    \begin{align}\label{eq:nts-preserve-sdot-a-s}
        \sdot(\alpha,\sigma) = [\hh(\alpha)+\GG(\alpha)]\sigma + \sigma[\hh(\alpha)+\GG(\alpha)]^\tp + \D(\alpha)
    \end{align}
    for matrix-valued function $\hh$, $\GG$, and $\D$ that are Hamiltonian, skew-Hamiltonian, and positive semidefinite, respectively, and which satisfy $\D/\hbar + i \GG \sf \ge 0$ (as is guaranteed for all Lindbladian dynamics). Suppose also that $\NTSm$ satisfies
    \begin{align}\label{eq:nts-preserve-cond}
        2 \hbar \NTSm \FDmax +  \NTSm^2 \condmax < 1
    \end{align}
    where
    \begin{align}
        \FDmax \, & :=  
        \sup_\alpha \|\hh(\alpha)\| \|\D^{-1}(\alpha)\|,\\
        \condmax \, & := \sup_\alpha \|\D(\alpha)\| \|\D^{-1}(\alpha)\| ,
    \end{align}
    are the respective extremal ratios taken by 
    $\|\hh\|$ and $\|\D\|$ relative to the minimum eigenvalue $\lambdamin[\D] = \|\D^{-1}\|^{-1}$.  (The latter is just the maximum condition number taken by $\D$ over phase space.) It is also sufficient that $\GG = 0$ and $\hbar \lambdas \FDmax < 1$.
    Then there exists a decomposition $\sdot = \sdotZ + \sdotD$ such that when $\pmsrt$ is evolved according to $\partial_t \pmsrt = \LLP[\pmsrt]$ where
    \begin{align}
    \label{eq:llpas-def}
        \LLP[\pmsrt] := \left[-\vecpartial{}_\alpha \UC 
        -\vecpartial{}_\sigma \sdotZ 
        + \frac{1}{2} \vecpartial{}_\alpha\vecpartial{}_\alpha \sdotD\right] \pmsrt,
    \end{align}
    $\pmsrt$ remains a probability measure and supported on $\bbR^{2d} \times\SNTS(\NTSm)$ for all times $t\geq 0$.
\end{restatable}

The partial differential equation $\partial_t \mu_t = \LLP[\mu_t]$ preserves the positivity of $\mu_t$ so long as the diffusion matrix $S_D$ is non-negative.  This is a consequence of the parabolic maximum principle.  Although a priori one might only expect solutions $\mu_t$ to be valued in the space of distributions, the positivity of $\mu_t$ ensures that in fact $\mu_t$ remains a measure for all positive times.

Thus, under the assumptions of Theorem~\ref{thm:mainResult}, we have defined a trajectory $\rhot(t)$ that is at all times a mixture \eqref{eq:rho_tilde_mixture} of Gaussian states with covariances matrices from the set $\SNTS(\NTSm)$.  We now turn to proving the 
bounds \eqref{eq:main-quantum} and \eqref{eq:main-classical}

\subsection{Bounding \ToP{$\|\rho-\rhot\|_{\mathrm{Tr}}$}{the quantum error}}\label{sec:quantum-duhamel}

First observe what may be understood as a version of Duhamel's principle,
\begin{align}\label{eq:duhamel-quantum}
	\rhot(t)-\rho(t)	& = \int_0^t \dd s \,  e^{ (t-s) \LLQ} \left( \partial_s - \LLQ \right) [\rhot(s)]
\end{align}
because the anti-derivative of the integrand (with respect to $s$) is $e^{ (t-s) \LLQ}  [\rhot(s)]$ (and $\rhot(0)=\rho(0)$). Then 
\begin{align}
	\Trnorm{\rhot(t)-\rho(t)}
	& = \Trnorm{  \int_0^t \dd s \,  e^{ (t-s) \LLQ} \left( \partial_s - \LLQ \right) [\rhot(s)] } \\
	& \le  \int_0^t \dd s \,   \Trnorm{ e^{ (t-s) \LLQ} \left( \partial_s - \LLQ \right) [\rhot(s)] } \\
	\label{eq:dumb0}
	& \le  \int_0^t \dd s \,   \Trnorm{ \left( \partial_s - \LLQ \right) [\rhot(s)] } \\
	\label{eq:dumb1}
	& =  \int_0^t \dd s \,   \Trnorm{ \int \left(  \LLQLa - \LLQ \right)    [\tauQas] \dpmsr_s(\alpha,\sigma)} \\
	\label{eq:dumb2}
	& =  \int_0^t \dd s \,   \Trnorm{ \int  \LLQRa   [\tauQas] \dpmsr_s(\alpha,\sigma)} \\
	&\le \int_0^t \dd s \, \int \Trnorm{ \LLQRa [\tauQas] \dpmsr_s(\alpha,\sigma) }\\
	&
 \label{eq:dumb2andhalf}
 \le \max_\alpha \max_{\sigma\in\SNTS(\NTSm)} \Trnorm{ \LLQRa [\tauQas]} \int_0^t \dd s \, \int \dpmsr_s(\alpha,\sigma) \\
	\label{eq:dumb3}
	&\le t \max_\alpha \max_{\sigma\in\SNTS(\NTSm)} \Trnorm{ \LLQRa [\tauQas]}
\end{align}
where \eqref{eq:dumb0} follows from the fact that $e^{ (t-s) \LLQ}$ is a CP map and so cannot increase the trace norm,
\eqref{eq:dumb1} follows from dynamics \eqref{eq:rhot-constraint} for $\rhot$, in \eqref{eq:dumb2} we have used $\LLQ = \LLQLa+\LLQRa$, in~\eqref{eq:dumb2andhalf} we have used the fact that $\pmsrt$ is supported on $\SNTS(\NTSm)$ (which follows from Lemma~\ref{lem:NTS-preserve}), and in
\eqref{eq:dumb3} we have used that $\pmsrt$ is a probability measure so that $\int \dpmsrtas = 1$. We will now make use of the following lemma on the error introduced by the harmonic approximation, proved in Section~\ref{sec:quantum-bound-pf}:
\begin{restatable}[Error in harmonic approximation to quantum dynamics]{lem}{lemHarmErrorQ}
	\label{lem:HarmErrorQ}
	The error $\LLCRa$ in the local harmonic approximation to the quantum dynamics acting on coherent state $\tauCas$ satisfies
	\begin{align}
     \label{eq:main-trace-err}
		\Trnorm{ \LLQRa[\tauQas] }  \leq 
        &\, C_d\frac{\|\sigma\|^{3/2}}{\hbar} 
        \left(\harmErrConstQtt{\HC}{\LCk} + \harmErrConstQfBare[\LCk,\hbar,\|\sigma\|^{1/2}] \right)
	\end{align}
 where $\harmErrConstQtt{\HC}{\LCk}$ and 
 $\harmErrConstQfBare[\LCk,\hbar,\nu]$ are defined in~\eqref{eq:harmErrConstQtt-def} and~\eqref{eq:harmErrConstQf-def}.
\end{restatable}

Applying Lemma~\ref{lem:HarmErrorQ} gives
\begin{align}
    \Trnorm{ \LLQRa [\tauQas]} \,
    &\leq \frac{\|\sigma\|^{3/2}}{\hbar} 
        \left(\harmErrConstQtt{\HC}{\LCk} + \harmErrConstQfBare[\LCk,\hbar,\|\sigma\|^{1/2}] \right)\\
    \,&\leq \hbar^{1/2} \rds^{3/2}
        \left(\harmErrConstQtt{\HC}{\LCk} + \harmErrConstQfBare[\LCk,\hbar,\|\sigma\|^{1/2}] \right)
\end{align}
for all $\alpha$ and $\sigma\in\SNTS(\rds/2)$, recalling $\sigma\in\SNTS(\NTSm)$ satisfy $\norm{\sigma} \leq (2\NTSm)^{-1} \hbar$ and that $\harmErrConstQfBare[\LCk,\hbar,\vsc]$ is a monotonically increasing function of $\vsc$, so \eqref{eq:dumb3} implies
\begin{align}\label{eq:rhot-rho}
	\Trnorm{\rhot(t)-\rho(t)} \le 
    &\,  t \hbar^{1/2} (2\NTSm)^{-3/2} \left( \harmErrConstQtt{\HC}{\LCk} +  \harmErrConstQfBare[\LCk,\hbar, (2\NTSm)^{-1/2}\hbar^{1/2}] \right) 
\end{align}

\subsection{Bounding \ToP{$\|\WW[\rhot]-\rhoc\|_{\mathrm{L^1}}$}{the classical error}}\label{sec:classical-duhamel}
The calculation is similar to the that of the previous subsection. Again we observe a version of Duhamel's principle,
\begin{align}\label{eq:duhamel-classical}
	\WW[\rhot(t)]-\rhoc(t)	& = \int_0^t \dd s \,  e^{ (t-s) \LLC} \left( \partial_s - \LLC \right) [\WW[\rhot](s)]
\end{align}
which follows for a similar reason as \eqref{eq:duhamel-quantum}.
Then we have 
\begin{align}
	\Lonenorm{\WW[\rhot(t)]-\rhoc(t)}
	& = \Lonenorm{  \int_0^t \dd s \,  e^{ (t-s) \LLC} \left( \partial_s - \LLC \right) \WW[\rhot(s)] } \\
	& \le  \int_0^t \dd s \,   \Lonenorm{ e^{ (t-s) \LLC} \left( \partial_s - \LLC \right) \WW[\rhot(s)] } \\
	\label{eq:dumb0c}
	& \le  \int_0^t \dd s \,   \Lonenorm{ \left( \partial_s - \LLC \right) \WW[\rhot(s)] } \\
	\label{eq:dumb1c}
	& =  \int_0^t \dd s \,   \Lonenorm{ \int \left(  \LLCLa - \LLC \right) [\tauCas] \dpmsr_s(\alpha,\sigma)  } \\
	\label{eq:dumb2c}
	& =  \int_0^t \dd s \,   \Lonenorm{ \int  \LLCRa   [\tauCas] } \dpmsr_s(\alpha,\sigma) \\
	&\le \int_0^t \dd s \, \int    \Lonenorm{ \LLCRa [\tauCas] } \dpmsr_s(\alpha,\sigma)\\
	&\le 
 \label{eq:dumb2halfc}
 \max_\alpha \max_{\sigma\in\SNTS(\NTSm)} \Lonenorm{ \LLCRa [\tauCas]} \int_0^t \dd s \, \int \dpmsr_s(\alpha,\sigma) \\
	\label{eq:dumb3c}
	&\le t \max_\alpha \max_{\sigma\in\SNTS(\NTSm)} \Lonenorm{ \LLCRa [\tauCas]}
\end{align}
where \eqref{eq:dumb0c} follows from the fact that $e^{ (t-s) \LLC}$ (flow and diffusion) does not increase $L^1$ norm,
\eqref{eq:dumb1c} follows from $\WW\circ \LLQLa = \LLCLa\circ\WW$ (by Lemma~\ref{lem:q-c-harm-equiv}),  
\eqref{eq:dumb2c} follows from $\LLQ = \LLQLa+\LLQRa$, in~\eqref{eq:dumb2halfc} we used that $\pmsrt$ 
is supported on $\bbR^{2d}\times \SNTS(\NTSm)$ (by Lemma~\ref{lem:NTS-preserve}), and \eqref{eq:dumb3c} follows from $\int \dpmsr_s(\alpha,\sigma) = 1$. 
We will now make use of the following bound on the error in the classical dynamics introduced in the harmonic approximation, proved in Section~\ref{sec:harmonic-error}:
\begin{restatable}[Error in harmonic approximation to classical dynamics]{lem}{lemHarmErrorC}
	\label{lem:HarmErrorC}
	The error $\LLCRa := \LLC - \LLCLa$ in the local harmonic approximation to the classical dynamics acting on coherent state $\tauCas$ satisfies
	\begin{equation}
		\Lonenorm{\LLCRa [\tauCas]}
		\le 14 d^{\frac32}\frac{1}{\hbar}\norm{\sigma}^{\frac{3}{2}}  \harmErrConstCtt{\HC}{\LCk}
	\end{equation}
	with anharmonicity factor 
    \begin{align} \label{eq:harmErrConstCtt-def-lemma}
	\harmErrConstCtt{\HC}{\LCk} := 
    \left(\CkSN{H}{3} +  
     \CkSN{G}{2} + \CkSN{\scD}{1} 
     \right).
    \end{align}
     depending only on the classical Hamiltonian and Lindblad functions through $\GC^a=\Imag \sum_k \LCk \partial^a \LCk^*$ and $\scD^{ab} = \Real\sum_k (\partial^a \LCk)(\partial^b \LCk^*)$. 
\end{restatable}
By Lemma~\ref{lem:HarmErrorC} we have
\begin{align}
	\|\LLCRa [\tauCas]\|_{L^1} \le 14 d^{\frac32} \frac{1}{\hbar} \norm{\sigma}^{\frac{3}{2}}  \harmErrConstCtt{\HC}{\LCk} \leq  14
 d^{\frac32}
 \hbar^{1/2} (2\NTSm)^{-3/2}
 \harmErrConstCtt{\HC}{\LCk}
\end{align}
for all $\alpha$ and $\sigma\in\SNTS(\rds/2)$, recalling $\sigma\in\SNTS(\NTSm)$ satisfy $\norm{\sigma} \leq (2\NTSm)^{-1}\hbar$, so \eqref{eq:dumb3c} implies
\begin{align}\label{eq:Wrhot-rhoc}
	\Lonenorm{\WW[\rhot(t)]-\rhoc(t)} \le 14 d^{\frac32} 
 \hbar^{1/2} (2\NTSm)^{-3/2}t
 \harmErrConstCtt{\HC}{\LCk}
\end{align}

\subsection{Concluding the proof}

We construct $\rhot(t)$ as in Section~\ref{sec:def-rhot}.  We need to choose our class of NTS states such that the condition sufficiently strong diffusion in Lemma~\ref{lem:NTS-preserve} is satisfied: 
\begin{align}\label{eq:nts-condition-repeat}
    \begin{cases}
      2 \hbar \NTSm \FDmax +  \NTSm^2 \condmax < 1 & \text{if}\ \GC^a \neq 0 \,\,\,\, \text{(frictionful)} \\
      \hbar \NTSm \FDmax < 1 & \text{if}\ \GC^a = 0 \,\,\,\, \text{(frictionless)}
    \end{cases}
\end{align}
where we recall 
\begin{align}
    \label{eq:fdmax-def-recall}
    \FDmax \, & :=  
    \sup_\alpha \|\hh(\alpha)\| \|\D^{-1}(\alpha)\| =  
    \hbar^{-1}\left(\inf_\alpha 
    \frac
    {\lambdamin[\scD(\alpha)]}{\lambdamax[\nabla^2\HC(\alpha)]}
    \right)^{-1},\\
    \label{eq:condmax-def-recall}
    \condmax \, & := \sup_\alpha \|\D(\alpha)\| \|\D^{-1}(\alpha)\| =  
    \left(\inf_\alpha \frac{\lambdamin[\D(\alpha)]}{\lambdamax[\D(\alpha)]}\right)^{-1}.
\end{align}
Therefore we pick NTS states characterized $\sigma\in\SNTS(\NTSm = \rds/2)$, where we recall:
\defEffInvDiffStrength*
\noindent In other words, we choose 
\begin{align}
    \NTSm = \frac{\rds}{2} = \begin{cases}
      \min\left\{(4\hbar\FDmax)^{-1}, \condmax^{-1/2}\right\}, & \text{if}\ \GC^a \neq 0 \,\, \text{(frictionful)} \\
      \min\left\{(2\hbar\FDmax)^{-1}, 1 \right\}, & \text{if}\ \GC^a = 0 \,\, \text{(frictionless)}
    \end{cases}
\end{align}

This choice ensures \eqref{eq:nts-condition-repeat}, so we have by Lemma~\ref{lem:NTS-preserve} 
that the evolution of $\rhot$ given by~\eqref{eq:pas-evolution-eq} preserves the property that $\pmsrt$
is always supported on $\bbR^{2d}\times\SNTS(\rds/2)$.
Then by \eqref{eq:rhot-rho} we conclude
\begin{align}
\label{eq:rhot-rho-2}
\Trnorm{\rhot(t)-\rho(t)} 
    \le 
    &\, 
    t \,
    \hbar^{\frac12}\rds^{-\frac32}
    \left(
    \harmErrConstQtt{\HC}{\LCk}+\harmErrConstQfBare[\LCk,\hbar,\sqrt{\hbar/\rds}]
    \right)
\end{align}
Likewise by \eqref{eq:Wrhot-rhoc} we conclude
\begin{align}\label{eq:Wrhot-rhoc-2}
\Lonenorm{\WW[\rhot(t)]-\rhoc(t)} 
    \le 
    14 d^{\frac32}
    t \,
    \hbar^{\frac12}\rds^{-\frac32}
    \harmErrConstCtt{\HC}{\LCk}
\end{align}

This concludes the proof of Theorem~\ref{thm:mainResult}, our main result.  The proof depended on lemmas concerning the preservation of the NTS condition (Lemma~\ref{lem:NTS-preserve}, proven in Section~\ref{sec:linear-algebra-facts}) and the size of the error from the classical and quantum harmonic approximations (Lemma~\ref{lem:HarmErrorQ} and Lemma~\ref{lem:HarmErrorC}, proven in Section~\ref{sec:harmonic-error}).

\section{NTS Preservation}
\label{sec:linear-algebra-facts}

In this section we prove Lemma~\ref{lem:NTS-preserve}, which 
we now restate.
\lemNTSpreservation*

The proof of Lemma~\ref{lem:NTS-preserve} relies primarily on Lemma~\ref{lem:covar-diff-decomp} below, a 
statement about just linear algebra which we use to define the decomposition
$\sdot = \sdotZ+\sdotD$.  To state our decomposition we recall from Section~\ref{sec:matrix-basics} that a matrix $A$ is defined to be symplectic, Hamiltonian, or skew-Hamiltonian when it satisfies the respective conditions $A^\tp \sf A=\sf$, $A^\tp = -\sf^\tp A \sf$, $A^\tp = \sf^\tp A \sf$.  When $A$ is symmetric, symplectic, and invertible (as is true for the covariance matrix for all pure Gaussian states) it therefore satisfies $\sf^\tp A \sf = A^{-1}$.

As the proof of Lemma~\ref{lem:covar-diff-decomp} is cumbersome, the reader may prefer to first examine ``Step 1'' of the simpler analogous proof of Theorem 1 of our shorter companion paper \cite{hernandez2023decoherence1}. 

\begin{restatable}[Decomposition of covariance dynamics]{lem}{lemCovarDiffDecomp}
\label{lem:covar-diff-decomp}
Suppose $\hh$ is a Hamiltonian matrix ($\hh^\tp = -\sf^\tp \hh \sf$), $\scD + i\GG\sf \ge 0$ is a positive semidefinite matrix with real and imaginary parts $\scD$ and $\GG\sf$ satisfying $\scD \ge c_{\scD} \IdM$, and $\sdot$ is the function
\begin{equation}
	\label{eq:covar-diff-decomp-s}
    \sdot(\sigma) := (\hh+\GG)\sigma + \sigma(\hh+\GG)^\tp + \hbar\scD
\end{equation}
on positive definite matrices $\sigma$.
Suppose moreover that $\NTSm \in (0,1]$ obeys 
\begin{align}
    \label{eq:NTSl-constraint}
    c_{\scD} > 2\NTSm \|\hh\| + \NTSm^2 \|\scD\|. 
\end{align}
or, alternatively, that $\GG=0$ and $\NTSm$ satisfies the weaker condition $c_{\scD} > \NTSm \|\hh\|$. 
Then there exists a decomposition 
$\sdot=\sdotZ+\sdotD$ 
satisfying
\begin{itemize} 
\item ``Diffusion positivity'': $\sdotD(\sigma)$ is positive semidefinite whenever $\sigma \in \SNTS(\NTSm)$; 
\item ``Purity preservation'': $\sdotZ(\sigma)$ is symmetric and $\sigma^{-1}\sdotZ(\sigma)$ is Hamiltonian; and 
\item ``NTS preservation'': 
$v^\tp \sdotZ(\sigma)v > 0$ 
whenever $v$ is an eigenvector of $\sigma$ 
with eigenvalue $ \lambda \leq (\hbar/2)\NTSm$.
\end{itemize}
\end{restatable}

\begin{proof}

In the following proof, we will let an overline denote division by $\hbar/2$, so $\sigmas = \sigma/(\hbar/2)$. 
We work with these normalized quantities because $\sigmas = \sf^\tp\sigmas^{-1}\sf > 0$ 
is symplectic exactly when $\sigma$ is the covariance matrix of a pure Gaussian state. Likewise, $\sigma\in \SNTS(\NTSm)$ implies $\NTSm \IdM_{2d} \le \sigmas \le \NTSm^{-1} \IdM_{2d} $. 

The dynamics $\sdot(\sigma)$ generate a very general positivity-preserving linear dynamics for $\sigma$, and our goal is to break this up into a piece $\sdotZ$ that additionally preserves the symplectic property (``purity preservation'') and a remainder $\sdotD$ that is equivalent to diffusion of the state in phase space (``diffusion positivity''). The purity-preservation condition is that $\sigma^{-1}\sdotZ(\sigma) = \sigmas^{-1}\sdotZs(\sigma)$ is Hamiltonian, which is equivalent to the form $\sdotZs(\sigma) = \hht(\sigmas)\sigmas + \sigmas \hht(\sigmas)^\tp$ for some Hamiltonian matrix $\hht(\sigmas)$.  Intuitively,\footnote{Indeed, if the overall dynamics are pure Hamiltonian (i.e., if all Lindblad terms are zero), then we could just make the choice $\hht(\sigmas) = \hh$.} we want $\hht$ to include the Hamiltonian part $\hh$ of the overall dynamics $\sdot$ plus an extra piece $\Yh(\sigmas)$ that will fight against any squeezing that risks violating the NTS-preservation condition.  Therefore we look for $\hht(\sigmas) = \hh+\Yh(\sigmas)$,
\begin{align}
    \label{eq:diff-sigma-dot-zero}
    \sdotZs(\sigma) &= [\hh + \Yh(\sigmas)]\sigmas + \sigmas [\hh + \Yh(\sigmas)]^\tp, \\
    \label{eq:diff-sigma-dot-delta}
    \sdotDs (\sigma) &= 2\scD + [\GG -\Yh (\sigmas) ]\sigmas + \sigmas  [\GG - \Yh (\sigmas) ] ^\tp.
\end{align}

In the frictionful case ($\GG \neq 0$) we make the ansatz
\begin{align}\label{eq:y-ansatz-frictionful}
    \Yh(\sigmas) = \frac{1}{2}[{\scD}\sigmas^{-1} - \sigmas \sf^\tp { \scD} \sf],
\end{align}
which is the Hamiltonian part of ${\scD}\sigmas^{-1}$.  Because $\sigmas$ is symplectic, $\sf^\tp \sigmas \sf = \sigmas^{-1}$, so $\Yh(\sigmas)$ is Hamiltonian by construction and hence preserves purity. Furthermore,
\begin{align}
    \sdotDs (\sigma) \,
    &= {\scD} + \sigmas \sf^\tp {\scD} \sf \sigmas + \GG\sigmas+\sigmas\GG^\tp  \\
    & = \frac{1}{2} \left[(1+i\sf\sigmas)^\dagger({\scD} + i \GG \sf)(1+i\sf\sigmas) + \mathrm{tp.} \right]
\end{align}
is a positive semidefinite because ${\scD} + i \GG \sf$ is a positive semidefinite matrix. (Above, ``$\mathrm{tp.}$'' denotes the transpose of the preceding expression.)  This ensures diffusion positivity. Finally, for 
$v$ 
an eigenvalue of $\sigmas$ with eigenvalue $\bar{\lambda} \le \NTSm$, we consider
\begin{align}
    v^\tp\sdotZs(\sigma)v \, 
    &= v^\tp [\hh \sigmas + \sigmas \hh^\tp +  {\scD} -  \sigmas \sf^\tp {\scD} \sf \sigmas  ] v, \\
    &= 2\bar{\lambda} v^\tp \hh v +  v^\tp{\scD}v  -\bar{\lambda}^2 v^\tp \sft {\scD} \sf v , \\
    &\ge - 2\bar{\lambda} \|\hh\| +   c_{{\scD}} -  \bar{\lambda}^2 \|{\scD}\| 
\end{align}
This is guaranteed to be positive, and thus NTS preserving, when \eqref{eq:NTSl-constraint} holds because $0<\bar{\lambda} \le \NTSm \le 1$.

Alternatively, in the frictionless case ($\GG = 0$) we make the same ansatz \eqref{eq:y-ansatz-frictionful} except with
$\scD \to 2 c_{\scD}/(1-\NTSm^{2} )$, i.e., 
\begin{align}\label{eq:y-ansatz-frictionless}
    \Yh(\sigmas) = \left(\frac{c_{\scD}}{\NTSm}\right)\frac{\sigmas^{-1} - \sigmas}{\NTSm^{-1} - \NTSm} = \Yh(\sigmas)^\tp.
\end{align}
Again, $\Yh(\sigmas)$ is a Hamiltonian matrix and so preserves purity. Furthermore,
\begin{align}
    \sdotDs (\sigma) \,
    &= 2 \scD - 2c_{\scD}\left(\frac{\sigmas}{\NTSm}\right)\left(\frac{\sigmas^{-1} - \sigmas}{\NTSm^{-1} - \NTSm}\right)\\
    &\ge 2(\scD - c_{\scD} \IdM_{2d})\\
    &\ge 0,
\end{align}
ensuring diffusion positivity for $\sigma\in\SNTS(\NTSm)$ because $\NTSm \IdM \le \sigmas \le \NTSm^{-1} \IdM$. Lastly, with $v$ again an eigenvector of $\sigmas$ with eigenvalue $\bar{\lambda}\le \NTSm \le 1$, we have
\begin{align}
    v^\tp\sdotZs(\sigma)v \, 
    &= v^\tp [\hh \sigmas + \sigmas \hh^\tp +  2\Yh(\sigmas)\sigmas  ] v, \\
    &= 2\bar{\lambda} (v^\tp \hh v) +  2c_{\scD}\left(\frac{\bar{\lambda}}{\NTSm}\right)\frac{\bar{\lambda}^{-1} - \bar{\lambda}}{\NTSm^{-1} - \NTSm}\\\
    &\ge 2\bar{\lambda}\left(\frac{c_{\scD}}{\NTSm} -  \|\hh\|\right)
\end{align}
This is guaranteed to be positive, and thus NTS preserving, when $c_{\scD} > \NTSm \|\hh\|$.
\end{proof}

Intuitively, Lemma~\ref{lem:covar-diff-decomp} has established that the dynamics \eqref{eq:covar-diff-decomp-s} for the covariance matrix of our Gaussian states can always be reinterpreted as diffusion of the center of the Gaussian plus Hamiltonian (i.e., purity-preserving) dynamics that confine the covariance matrix to $\SNTS(\NTSm)$. We will now make this precise. 
\begin{proof}[Proof of Lemma~\ref{lem:NTS-preserve}]
First, observe that this Lemma's assumed form for $\sdot(\alpha,\sigma)$ and the constraint on $\NTSm$ (given by \eqref{eq:nts-preserve-sdot-a-s} and \eqref{eq:nts-preserve-cond}, respectively) ensures that, for any fixed $\alpha$, $\sdot(\alpha,\sigma)$ satisfies the form for $\sdot(\alpha)$ and the constraint for $\NTSm$ in Lemma~\ref{lem:covar-diff-decomp} (given by~\eqref{eq:covar-diff-decomp-s} and \eqref{eq:NTSl-constraint}, respectively) when taking $\scD = \hbar^{-1}\D$ as expected.  In other words, we can apply Lemma~\ref{lem:covar-diff-decomp} freely at all points $\alpha$ in phase space. 

We will show that for any probability measure $\pmsrt$ 
evolved by $\LLP$, \eqref{eq:llpas-def}, the total probability mass of $\NTS$ states,
\begin{equation}
    m_\NTS(t) := 
    \iint \indicNTS(\sigma) \dpmsrtas,
\end{equation}
is non-decreasing.  Here $\indicNTS(\sigma)$ is the indicator function enforcing the minimum-eigenvalue condition\footnote{Note that everywhere we work with covariance matrices of pure states (i.e., $\frac{\sigma}{\hbar/2}$ symplectic and positive definite), so it's not necessary to enforce those conditions with the indicator function.} for the NTS covariance matrices: 
\begin{equation}
\indicNTS(\sigma) := \heaviside(\lambdamin[\sigma] - (\hbar/2)\NTSm) = \begin{cases}
1, & \lambdamin[\sigma] \geq (\hbar/2)\NTSm \\
0, & \lambdamin[\sigma] < (\hbar/2)\NTSm
\end{cases},
\end{equation}
where $\heaviside$ is the Heaviside step function. 
Therefore 
\begin{align}\label{eq:deriv-indic-NTS}
\partial_\sigma \indicNTS(\sigma) = \delta(\lambdamin[\sigma]-(\hbar/2)\NTSm)
\partial_\sigma \lambdamin[\sigma].
\end{align}

We use this to 
compute the time derivative of
$m_{\NTS}(t)$:
\begin{align*}
\frac{\dd}{\dd t} m_{\NTS}(t) &=
\langle \indicNTS, \LLP[\pmsrt]\rangle \\
&= \langle \LLP^*[\indicNTS], \pmsrt\rangle.
\end{align*}
To show that $\frac{\dd}{\dd t}m_{\NTS}(t)$ is
nonnegative it therefore suffices
to show that $\LLP^*[\indicNTS]$ is positive.
Here the adjoint of $\LLP$ is
\begin{equation}
\LLP^*[f] := U \partial_\alpha f
+ S_0 \partial_\sigma f
+ \frac12 S_D \partial_\alpha\partial_\alpha f.
\end{equation}
Since $\indicNTS$ does not depend on $\alpha$, the only term that 
remains is the term with $\partial_\sigma$.  Thus by~\eqref{eq:deriv-indic-NTS} we have
\begin{equation}
\begin{split}
\LLP^*[\indicNTS](\alpha,\sigma) & = S_0(\alpha,\sigma)\partial_\sigma \indicNTS(\sigma) \\
&= S_0(\alpha,\sigma) \delta(\lambdamin[\sigma] - (\hbar/2)\NTSm)
\partial_\sigma\lambdamin[\sigma].
\end{split}
\end{equation}

To conclude we need to show that 
$S_0(\alpha,\sigma)\partial_\sigma \lambdamin[\sigma] \geq 0$.  Note that
\begin{equation}
    \sdotZ(\alpha,\sigma) \partial_\sigma \lambdamin[\sigma] =
    \left.\frac{\dd}{\dd t}
    \lambdamin[\sigma + t \sdotZ(\alpha,\sigma)]\right|_{t=0}
    =
    v^\tp \sdotZ(\alpha,\sigma) v
\end{equation}
when $\sigma v = \lambdamin[\sigma] v$ for 
unit eigenvector $v$.
Using the ``NTS preservation condition'' of Lemma~\ref{lem:covar-diff-decomp}, it follows
that 
$v^\tp \sdotZ(\alpha,\sigma) v \geq 0$ 
when $\lambdamin[\sigma]=(\hbar/2)\NTSm$.  Therefore
$\sdotZ(\alpha,\sigma)\partial_\sigma \lambdamin[\sigma] \geq 0$,
so it follows that 
\begin{equation}
    \frac{\dd}{\dd t} m_{\NTS}(t) \geq 0.
\end{equation}

We note that $\pmsrt\geq 0$ is guaranteed by the ``diffusion positivity'' condition in Lemma~\ref{lem:covar-diff-decomp} ($\sdotD(\alpha,\sigma)\ge 0$) and that $\int \! \dd \pmsrt = 1$ is conserved, so we have that $\pmsrt$ is a probability measure and $m_{\NTS}(t) = 1$ for $t\geq 0$. Combining this with the ``purity preservation condition'' in Lemma~\ref{lem:covar-diff-decomp}, we conclude that $\pmsrt$ is supported on the set $\bbR^{2d}\times\SNTS(\NTSm)$ for all times $t\geq 0$. 

\end{proof}

\section{Harmonic approximation error}\label{sec:harmonic-error}

In this section we prove Lemma~\ref{lem:HarmErrorC} about the error in the harmonic approximation to the classical dynamics and Lemma~\ref{lem:HarmErrorQ} about the error in the harmonic approximation to the quantum dynamics.  In both cases we find that the instantaneous error scales as $\frac{1}{\hbar}\norm{\sigma}^{\frac{3}{2}}$, where $\sigma$ is the covariance matrix of the pure state on which the dynamics act.

\subsection{Classical case}\label{sec:harmonic-error-classical}

Consider the classical state $\tauCas(\alpha+\beta) = \exp(-\beta^a\sigma^{-1}_{ab}\beta^b/2)/((2 \pi)^d\sqrt{\det\sigma})$, a Gaussian probability distribution over phase space centered on $\alpha$ that's equal to the Wigner function of the quantum state $\tauQas = |\alpha,\sigma\rangle\langle \alpha,\sigma|$.  We want to bound the error 
\begin{align}
	\Lonenorm{(\LLC-\LLCLa)[\tauCas]} =  \Lonenorm{\LLCRa[\tauCas]}
\end{align}
due to approximating the true classical dynamics, generated by
\begin{align}\label{eq:classical-fokker-planck-exact}
	\LLC[\cstate]=& - \left(\partial^a \HC \right)(\partial_a \cstate)  - \partial_a\left[\cstate  \Imag \sum_k \LCk \partial^a \LCk^* \right]   + \frac{\hbar}{2} \partial_a \left[ ( \partial_b \cstate) \Real \sum_k(\partial^a \LCk)(\partial^b \LCk^*)\right]\\
	=& - \left(\partial^a \HC \right)(\partial_a \cstate)  - \partial_a\left[\cstate  \GC{}^a \right]   + \frac{\hbar}{2} \partial_a \left[ ( \partial_b \cstate) \scD^{ab}\right]
\end{align}
acting on the state $\tauCas$, with the linearization $\LLCLa$ given by (see Section~\ref{sec:harmonic-approx})
\begin{align}\label{eq:classical-fokker-planck-harmonic-2}
	\LLCLa [\cstate]
    =&   -\partial_a\left[\cstate \left(\partial^a \HCLat+ \GCLao{}^{a} \right)\right] + \frac{\hbar}{2}  \scD^{[\alpha,0]ab}\partial_a \partial_b \cstate\\
    =&  -\partial_a\left[\cstate  \left(\partial^a \HCLat + \Imag \sum_k \left(\LCk (\alpha) \partial^a  \VCLatk{}^* + \VCLaok \partial^a  \VCLaok{}^* \right)\right)\right]   \\
	&\qquad + \frac{\hbar}{2}  \left[ \Real \sum_k(\partial^a \VCLaok)(\partial^b  \VCLaok{}^*)\right](\partial_a \partial_b \cstate)
\end{align}
where we recall $\VCk = \LCk-\LCk(\alpha)$, $\GC_a = \Imag\sum_k \LCk \partial_a \LCk^*$, and $\scD^{ab} = \Real\sum_k (\partial^a \VCk)^* (\partial^b\VCk)$. 
(As described earlier, $E^{[\alpha,m]}$ denotes the $m$-th order Taylor approximation to $E$ at $\alpha$.)
The main result of this section will be to prove Lemma~\ref{lem:HarmErrorC}, which we now re-state:

\lemHarmErrorC*

\begin{proof}
In the proof below, because we are after an explicit constant in~\eqref{eq:harmErrConstCtt-def-lemma}, we compute the Gaussian integrals explicitly.

We start by observing the following identity for the derivative of the Gaussian state $\tauCas$:
\begin{align}\begin{split}\label{eq:deriv-gaussian}
		(\partial_c \tauCas)(\alpha+\beta) &= \frac{\partial}{\partial \beta^c} \frac{\exp\left[-\beta^a\sigma^{-1}_{ab}\beta^b/2\right]}{(2 \pi)^d\sqrt{\det\sigma}}\\
		&= -\sigma^{-1}_{cd} \beta^d  \frac{\exp\left[-\beta^a\sigma^{-1}_{ab}\beta^b/2\right]}{(2 \pi)^d \sqrt{\det\sigma}}\\
		&= -\m_c \tauCas(\alpha+\beta),\\
\end{split}\end{align}
where 
$\m_a:= \sigma^{-1}_{ab} \beta^b$.  
Then expanding $\LLC[f]-\LLCLa[f]$ using
the expressions in \eqref{eq:classical-fokker-planck-exact} and \eqref{eq:classical-fokker-planck-harmonic-2} with $f=\tauCas$, and using the triangle inequality,
\begin{align}\begin{split}\label{eq:classical-bound-decomp}
		&\Lonenorm{\LLCRa [\tauCas]}
		\le
		\Lonenorm{ \tauCas \m^a(\partial_a \delta \HCLat)} + \Lonenorm{ \tauCas \m^a \GCRao_{a} } + \Lonenorm{ \tauCas \partial^a  \GCRao_{a} }\\
		&\qquad \qquad \qquad \qquad\qquad + \frac{\hbar}{2} \Lonenorm{ \tauCas \m_a \m_b \delta\scD^{[\alpha,0]ab}}
\end{split}\end{align}
where $\delta E^{[\alpha,m]} = E - E^{[\alpha,m]}$ denotes error on the $m$-th order Taylor approximation to $E$.

Starting with the first term on the right-hand side, we use Taylor's theorem on the derivative of $\HCRa$:
\begin{align}\begin{split}
		\label{eq:hamiltonian-deriv-taylor}
		(\partial_d \HCRat)(\alpha+\beta) &=(\partial_d \HC)(\alpha+\beta)- (\partial_d \HCLat)(\alpha+\beta) \\
		& = \frac{1}{2!} \beta^a \beta^b (\partial_a \partial_b \partial_d \HC)(\alpha+z\beta)
\end{split}\end{align}
for some ($\beta$-dependent) choice of $z\in[0,1]$. 
We can bound this derivative
\begin{align}\begin{split}
		\label{eq:third-deriv-ham-both-sigma-bound}
		\left|\m^a \partial_a \HCRa(\alpha)\right|^2
		= \frac{1}{4} \left|\m^a  \beta^b  \beta^c (\partial_a \partial_b \partial_c H)(\alpha) \right|^2
		\le \frac{1}{4} |\m|^2 |\beta|^4 \CkSN{H}{3}^2.
\end{split}\end{align}
using the $C^3$ seminorm defined in \eqref{eq:c-k-seminorm-def}, 
\begin{align}\begin{split}\label{eq:hbar3}
		\CkSN{H}{3} := 
        \sup_\alpha \OffDiag{\nabla^3 \HC(\alpha)}
        = \sup_\alpha \sup_{\|\beta_i\|=1} |\beta_1^a \beta_2^b\beta_3^c\partial_a \partial_b \partial_c \HC(\alpha)|
\end{split}\end{align}
which gives a global upper bound on the third derivatives of the classical Hamiltonian function.  Likewise norms like 
\begin{align}
	|m|^2 = m^a \IdM_{ab} m^b = \beta^a (\sigma^{-1})_{a}^{\pha c} \IdM_{cd} (\sigma^{-1})^{d}_{\pha b} \beta^b = \beta^a \sigma^{-2}_{ab} \beta^b = (\beta^\tp \sigma^{-2} \beta)
\end{align}
are computed not with the symplectic form but with the Euclidean inner product.\footnote{Note that, physically, this inner product depends on our choice of units.  See Appendix~\ref{sec:units-symplectic-covariance}.}

We can then perform the Gaussian integral
\begin{align}\label{eq:gaussian-int-m1-b2}
	\Lonenorm{\tauCas |\m| |\beta|^2 }^2
	&\le  \norm{\tauCas^{1/2}}_{L^2}^2 \norm{\tauCas^{1/2} |\m| |\beta|^2 }_{L^2}^2\\
	&= \int \!\dd \beta\, \tauCas(\alpha+\beta)  \left(\beta^\tp \sigma^{-2} \beta\right) |\beta|^4\\
	&=  \Tr[\sigma^{-1}](\Tr\sigma)^2 + 2\Tr[\sigma^{-1}] \Tr[\sigma^2]+4\Tr[\sigma]\Tr[\sigma^0]+8\Tr[\sigma] \\
	&\le  \hbar^{-2}[2^5(d+1)d^2]\|\sigma\|^3+[2^4(d+1)d]\|\sigma\|\\
	&\le  \hbar^{-2}[2^5(d+2)(d+1)d]\|\sigma\|^3\\
        &\le  \hbar^{-2}[2^6 3 d^3]\|\sigma\|^3
\end{align}
where in the first line we have used Cauchy-Schwartz inequality,\footnote{\label{fn:c-s-integral}More specifically, the Cauchy-Schwartz is $|(v,w)|^2\le (v,v)(w,w)$ and we choose  $v=\sqrt{\tauCas}$ and $w=\sqrt{\tauCas}|\beta|$ so $\|v w\|^2_{L^1} = [\int \!\dd \beta\, |v(\beta)w(\beta)|]^2 = |(v,w)|^2 \le (v,v)(w,w) = \int \!\dd \beta\, |w(\beta)|^2$ because $(v,v) = \int \!\dd \beta\, \tauCas(\beta) = 1$.}  
and in the last two lines 
we have used $d\ge 1$ and the fact that, by the uncertainty principle, $\|\sigma\| \ge \hbar/2$  because $\sigma$ is the covariance matrix of a pure Gaussian state.
(The Gaussian integrals we use in this section are recalled in Appendix~\ref{sec:gaussian-integrals}.)
Thus we can bound the first term on the right-hand side of \eqref{eq:classical-bound-decomp}:
\begin{align} \label{eq:harm-error-classical-Ham}
	\Lonenorm{ \tauCas \m^a(\partial_a \delta \HCLat)}
        &\le \hbar^{-1} \CkSN{H}{3} [2^4 3 d^3]^{1/2}\|\sigma\|^{3/2}
\end{align}

Now we bound the rest of the terms in \eqref{eq:classical-bound-decomp} in basically the same way. Using Taylor's theorem again, we have
\begin{align}\begin{split}
		\GCRao_{a}(\alpha+\beta) &= \frac{1}{2!}\beta^b\beta^c\partial_b\partial_c \GC_{a}(\alpha+z\beta)\\
		\partial^a \GCRao_{a}(\alpha+\beta) &= \delta(\partial^a \GC_{a})^{[\alpha,0]}(\alpha+\beta) = \beta^c  \partial_c \partial^a \GC_{a}(\alpha+z\beta)\\
		\delta\scD^{[\alpha,0]ab}(\alpha+\beta) &= \beta^c\partial_c \scD^{ab}(\alpha+z\beta)
\end{split}\end{align}
where on each line $z\in[0,1]$ depends separately on $\beta$ on that line. 
Then applying the $C^k$ seminorms as before
\begin{align}\begin{split}
		|\m^a \GCRao_{a} (\alpha+\beta)| &\le
		|\m| |\beta|^2 \CkSN{G}{2}/2!
		\\
		|\partial^a \GCRao_{a}(\alpha+\beta) | &\le
		|\beta| \CkSN{G}{2}
		\\
		|m_a m_b \delta\scD^{[\alpha,0]ab} (\alpha+\beta)| &\le
		|m|^2 |\beta| \CkSN{\scD}{1}
\end{split}\end{align}
where
analogously to $\CkSN{H}{3}$ in \eqref{eq:hbar3} we defined
\begin{align}\label{eq:gbar-deltabar}
	\CkSN{\GC}{2}
	:=& \sup_\alpha \OffDiag{\nabla^2 \GC(\alpha)} = \sup_\alpha \sup_{\|\beta_i\|=1}  |\beta_1^a \beta_2^b\beta_3^c \partial_a \partial_b \GC_c|\\
	\label{eq:wbar-expanded}
    \begin{split}
	=& 
    \sup_\alpha \sup_{\|\beta_i\|=1}
    \Bigg|\beta_1^a \beta_2^b\beta_3^c\Imag \sum_k \Big[(\partial_{a} \partial_{b} \LCk)^*(\alpha)(\partial_{c} \LCk)(\alpha)+2(\partial_{a} \LCk)^*(\partial_{b} \partial_{c} \LCk)(\alpha)
    \\
    &\hspace{22em}
    +(\LCk)^*(\alpha)(\partial_{a}\partial_{b} \partial_{c} \LCk)(\alpha)\Big]\Bigg|, 
    \end{split}
    \\
	\CkSN{\scD}{1}
    :=& \sup_\alpha \OffDiag{\nabla^1 \scD(\alpha)} = \sup_\alpha \sup_{\|\beta_i\|=1}  |\beta_1^a \beta_2^b\beta_3^c \partial_a \scD_{bc}|\\
    =&  \sup_\alpha \sup_{\|\beta_i\|=1} \left|\beta_1^a \beta_2^b\beta_3^c \Real \sum_k \Big[(\partial_{a} \partial_{b} \LCk)^*(\partial_{c} \LCk)+(\partial_{b} \LCk)^*(\alpha)(\partial_{a} \partial_{c} \LCk)(\alpha)\Big]\right|.
\end{align}
(Seminorms are reviewed in Section~\ref{sec:norms}.)

We again use Cauchy-Schwartz (see footnote~\ref{fn:c-s-integral}) to get
\begin{align}\begin{split}
		\label{eq:gaussian-int-m2-b1}
		\Lonenorm{\tauCas |\beta| }^2
		&\le \int \!\dd \beta\, \tauCas(\alpha+\beta)  |\beta|^2
		=  \Tr[\sigma ]
		\le  (2d) \|\sigma\| \le \hbar^{-2}(8d) \|\sigma\|^3
		\\
		\Lonenorm{\tauCas |\m|^2 |\beta| }^2
		&\le \int \!\dd \beta\, \tauCas(\alpha+\beta)  |\m|^4 |\beta|^2\\
		&=  (\Tr [\sigma^{-1}])^2\Tr[\sigma]+4\Tr[\sigma^{-1}]\Tr[\sigma^0]+ \Tr[\sigma] \Tr[\sigma^{-2}] + 8\Tr[\sigma^{-1}]\\
		&\le  \hbar^{-2} [2^4(2d^2+5d+4)d]\|\sigma\|^3\\
            &\le  \hbar^{-2} [2^4 11 d^3]\|\sigma\|^3
\end{split}\end{align}
where the relevant Gaussian integrals are recalled in Appendix~\ref{sec:gaussian-integrals}.

Pulling this all together we can now bound the error \eqref{eq:classical-bound-decomp} on the classical harmonic approximation for a Gaussian state:
\begin{align}\begin{split}\label{classical-error-bound}
		\Lonenorm{\LLCRa [\tauCas]}
            &\le \frac{\|\sigma\|^{3/2}}{\hbar}\Big[ (\CkSN{H}{3} + \CkSN{G}{2})  \left[2^4 3 d^3\right]^{1/2} + \CkSN{G}{2}  \left[8d\right]^{1/2} + (\hbar/2)\CkSN{\scD}{1}  \left[2^4 11 d^3\right]^{1/2} \Big] \\
		&\le \frac{\|\sigma\|^{3/2}}{\hbar} \harmErrConstCtt{\HC}{\LCk}
\end{split}\end{align}
where 
\begin{align} \label{eq:harmErrConstC-repeat}
	\harmErrConstCtt{\HC}{\LCk} = 14 d^{3/2}\left(\CkSN{H}{3} +  
     \CkSN{\GC}{2}  
     + 
     \CkSN{\scD}{1} 
     \right).
\end{align}
is a measure of the anharmonicity of the classical Hamiltonian and Lindblad functions $\HC$ and $\LCk$ (and in particular does not depent on $\hbar$ or $\sigma$).
It may seem unusual that $\harmErrConstC$ depends on $\CkSN{G}{2}$, which in turn can diverge if the Lindblad functions $\LCk$ becomes arbitrarily large without its third derivative vanishing (see the last term in \eqref{eq:wbar-expanded}). However, this may be expected due to the fact that the overall dynamics are \emph{not} invariant under $\LCk(\alpha) \to \LCk(\alpha) + L_0$ (in contrast to the case of the Hamiltonian shift $\HC(\alpha) \to \HC(\alpha) + H_0$, which does preserve the dynamics).

That concludes the proof of Lemma~\ref{lem:HarmErrorC}.
\end{proof}

\subsection{Quantum case}
\label{sec:quantum-bound-pf}

We recall from \eqref{eq:lindblad-modified} that, for any $\alpha$, we can express the exact quantum dynamics with $\VQk := \LQk - \LCk(\alpha)$ as
\begin{align}\label{eq:lindblad-modified-2}
	\LLQ[\rho]
	& = -\frac{i}{\hbar}\left[\HQ+\Imag \sum_k \LCk(\alpha) \VQk^\dagger ,\rho\right] + \frac{1}{\hbar} \sum_k \left( \VQk \rho \VQk^\dagger - \frac{1}{2} \left\{ \VQk^\dagger \VQk,\rho\right\} \right)
\end{align}
We emphasize that although we express $\LLQ$ above in terms of $\alpha$, the object is independent of $\alpha$. As discussed in Section~\ref{sec:harmonic-approx}, the linearized dynamics $\LLQLa$ at $\alpha$, which of course \emph{do} depend on $\alpha$, are
\begin{align}
	\LLQLa[\rho]
	& = -\frac{i}{\hbar}\left[\HQLat+\Imag \sum_k \LCk(\alpha) \VQLatk{}^\dagger ,\rho\right] + \frac{1}{\hbar} \sum_k \left( \VQLaok \rho \VQLaok{}^\dagger - \frac{1}{2} \left\{ \VQLaok{}^\dagger \VQLaok,\rho\right\} \right)
\end{align}
We want a global (independent of $\alpha$) bound on the error
\begin{align}
	\label{eq:quant-err-tracenorm}
	\Trnorm{(\LLQ-\LLQLa)[\tauQas]} =  \Trnorm{\LLQRa[	\tauQas]}.
\end{align}
for the Gaussian state $\tauQas = |\alpha,\sigma\rangle\langle \alpha,\sigma|$ with covariance matrix $\sigma$ located at $\alpha$. We can now re-state the lemma concerning the harmonic approximation which we prove in this subsection:

\lemHarmErrorQ*

Our proof of Lemma~\ref{lem:HarmErrorQ} is more
more involved than the classical case (Lemma~\ref{lem:HarmErrorC}) in the previous subsection.  We use many variations of the same basic trick, and we expect (for reasons related to the above discussion) that a more abstract  understanding of how Lindbladians and Liouvillians are Taylor approximated would make tighter, simpler bounds possible. 

In the proof of Lemma~\ref{lem:HarmErrorQ} we will not keep track of the constant $C_d$, choosing instead to use the notation $A\lsim B$ to mean that $A\leq CB$ for some constant $C$ depending only on dimension.  The implicit constant can change from line to line. 

\begin{proof}[Proof of Lemma~\ref{lem:HarmErrorQ}]
	We have
	\begin{align}\begin{split}
		\LLQRa[\tauQas] &= \LLQ[\tauQas] - \LLQLa[\tauQas]\\
		&= \frac{-i}{\hbar} \left[\HQRa + \Imag \sum_k \LCk(\alpha) \VQRatk ,|\alpha,\sigma\rangle\langle\alpha,\sigma |\right]  \\
		&\qquad + \frac{1}{\hbar} \sum_k \Bigg[ \VQRaok \ket{\alpha,\sigma }\bra{\alpha,\sigma } \VQLaok{}^\dagger + \VQLaok \ket{\alpha,\sigma }\bra{\alpha,\sigma } \VQRaok{}^\dagger  \\
		& \qquad\qquad\qquad\qquad + \VQRaok \ket{\alpha,\sigma }\bra{\alpha,\sigma } \VQRaok{}^\dagger -\frac{1}{2} \left\{\VQRaok{}^\dagger \VQLaok, \ket{\alpha,\sigma }\bra{\alpha,\sigma } \right\}  \\
		& \qquad\qquad\qquad\qquad -\frac{1}{2} \left\{\VQLaok{}^\dagger \VQRaok, \ket{\alpha,\sigma }\bra{\alpha,\sigma } \right\} -\frac{1}{2} \left\{\VQRaok{}^\dagger \VQRaok, \ket{\alpha,\sigma }\bra{\alpha,\sigma } \right\} \Bigg]
	\end{split}\end{align}
    For any state vectors $\ket{\psi}$ and $\ket{\phi}$, we have $\Trnorm{\ket{\psi}\bra{\phi}}^2 = \Tr[(\ket{\psi}\!\langle \phi|\phi\rangle\!\bra{\psi})^{1/2}]^2 = \langle \psi|\psi\rangle\langle \phi|\phi\rangle = \|\psi\|^2 \|\phi\|^2$, where the unlabeled norms denote Hilbert space norm.
	Therefore,\footnote{For a tighter bound, but only by a constant, we can compute more exactly.  
    Note $\norm{|v\rangle\langle w| - |w\rangle\langle v|} = \norm{v}\norm{w} (1- |\langle v|w\rangle|^2/(\norm{v}\norm{w}))^{1/2}$, so $\|[\HQ,\tauQas]\| = \|\HQ|\alpha\rangle \langle \alpha| - |\alpha\rangle \langle \alpha | \HQ \| = \|\HQ|\alpha\rangle\| (1-\langle \alpha | \HQ |\alpha \rangle^2/\langle \alpha | \HQ^2 | \alpha\rangle )^{1/2} = (\langle \alpha | \HQ^2 | \alpha \rangle - \langle \alpha | \HQ | \alpha \rangle^2)^{1/2} = (\textrm{Var}(\HQ)_{|\alpha\rangle})^{1/2}$.  And also $\||v\rangle\langle w| - |w\rangle\langle v|\|_1 = 2 \||v\rangle\langle w| - |w\rangle\langle v|\|$. } 
	\begin{align}
        \begin{split}\label{eq:quantum-error-initial-expansion}
			\Trnorm{ \LLQRa[\tauQas] }  \leq & \frac{2}{\hbar} \norm{ \HQRat|\alpha,\sigma \rangle } + \frac{2}{\hbar}\sum_k \LCk(\alpha) \norm{ \VQRatk|\alpha,\sigma \rangle }   \\
			&\quad + \frac{1}{\hbar} \sum_k \Bigg[  2 \norm{ \VQRaok  |\alpha,\sigma \rangle} \norm{ \VQLaok |\alpha,\sigma \rangle}   +\norm{ \VQRaok  |\alpha,\sigma \rangle}^2 +\norm{ \VQRaok{}^\dagger \VQLaok |\alpha,\sigma \rangle} \\
			&\qquad\qquad\qquad   +\norm{ \VQLaok{}^\dagger \VQRaok |\alpha,\sigma \rangle}  +\norm{ \VQRaok{}^\dagger \VQRaok |\alpha,\sigma \rangle}   \Bigg]
	\end{split}
        \end{align}
        
	These terms all become small in the classical limit for essentially the same reason: a Gaussian state of scale $\hbar$ cannot easily ``see'' the third order corrections to the harmonic approximation. However, since our current techniques are not strong enough show this rigorously in one fell swoop, we will bound each of these terms individually.  
    To simplify the bounds we will use the quantities 
    $\constQ^{q,r}_{\hbar}[E]$ and 
    $\constN^{q,r}_{\hbar;s,\nu}[E]$ defined in~\eqref{eq:constQ-def} and~\eqref{eq:constN-def} and recalled below for convenience:
\begin{align*}
\constQ^{q,r}_{\hbar}[E]
&:= \sum_{j=q}^r \hbar^{(j-q)/2}\CkSN{E}{j} \\
\constN^{q,r}_{\hbar;s,\nu}[E](\alpha)
&:= 
\sum_{j=q}^r \hbar^{(j-q)/2} 
\sup_\beta 
\frac{\TsrNrm{\nabla^jE(\alpha+\beta)}}{(1+\nu^{-1}|\beta|)^s}.
\end{align*}    
With this notation we will show that we can bound the terms of~\eqref{eq:quantum-error-initial-expansion} by
	\begin{align}
		\label{eq:HQR-term}
		\|\HQRat |\alpha,\sigma\rangle\|
		&\lsim
		\|\sigma\|^{3/2} 
        \constQ^{3,2d+4}_\hbar[\HC]
        \\
		\label{eq:VQR-term}
		|\LCk(\alpha)|
		\norm{ \VQRatk|\alpha,\sigma \rangle }
		&\lsim
        \|\sigma\|^{3/2}
        \sup_{\beta}
		 |\LCk(\beta)|
   \constN^{3,2d+6}_{\hbar;1,\|\sigma\|^{1/2}}[\LCk](\beta)
		\\
		\label{eq:VQlinear-term}
		\norm{ \VQRaok  |\alpha,\sigma \rangle}
		&\lsim
		\|\sigma\|
		\constQ^{2,2d+3}_{\hbar}[\LCk]
		\\
		\label{eq:Vmixed-term}
		\norm{ \VQLaok |\alpha,\sigma \rangle}
		&\lsim
        \|\sigma\|^{1/2}
        \constQ^{1,1}[\LCk]
		\\
		\label{eq:linear-outside-term}
		\begin{split}
			\norm{ \VQRaok{}^\dagger \VQLaok |\alpha,\sigma \rangle}
			&\lsim
			\|\sigma\|^{3/2}
			\constQ^{1,1}[\LCk] 
   \constQ^{2,2d+3}_{\hbar}[\LCk]
		\end{split}\\
		\label{eq:linear-inside-term}
		\begin{split}
			\norm{\VQLaok{}^\dagger \VQRaok |\alpha,\sigma \rangle}
			&\lsim
			\|\sigma\| ^{3/2}
			\constQ^{1,1}[\LCk] 
   \constQ^{2d+3}_{\hbar}[\LCk]
		\end{split}\\
		\label{eq:triple-moyal-term}
		\begin{split}
			\norm{ \VQRaok{}^\dagger \VQRaok |\alpha,\sigma \rangle}
			&\lsim
			\|\sigma\|^2
            \big(\constQ^{2,4d+6}_{\hbar}[\LCk]\big)^2
		\end{split}
	\end{align}
	Assuming these bounds, 
	the error \eqref{eq:quantum-error-initial-expansion} on the quantum harmonic approximation for a Gaussian state becomes
	\begin{align}\begin{split}\label{eq:quantum-error-bound}
			\Trnorm{ \LLQRa[\tauQas] }  
            & \lsim
			\frac{\|\sigma\|^{3/2}}{\hbar}
            \Bigg[
			\constQ^{3,2d+4}_{0;\hbar}[\HC]
			+ \sum_k
			\bigg(
            \constQ^{1,1}[\LCk]
			\constQ^{2,2d+3}_{\hbar}[\LCk]
            +
			\sup_{\beta} |\LCk(\beta)| 
   \constN^{3,2d+6}_{\hbar;1,\|\sigma\|^{1/2}}[\LCk](\beta)
            \\
            &\hspace{26em}
            +
			(\constQ^{2,4d+6}_{\hbar}[\LCk])^2 
            \|\sigma\|^{1/2}
			\bigg) 
			\Bigg]
            \\
            & \lsim
            \frac{\|\sigma\|^{3/2}}{\hbar}
            \left(\harmErrConstQtt{\HC}{\LCk} + \harmErrConstQfBare[\LCk,\hbar,\|\sigma\|^{1/2}]\right)
	\end{split}\end{align}
    for the anharmonicity factors $\harmErrConstQtt{\HC}{\LCk}$ and $\harmErrConstQfBare[\LCk,\hbar,\vsc]$ defined by \eqref{eq:harmErrConstQtt-def} and \eqref{eq:harmErrConstQf-def}.
    
Thus, to complete the proof of Lemma~\ref{lem:HarmErrorQ}, all we have to do is demonstrate the bounds (\ref{eq:HQR-term}--\ref{eq:triple-moyal-term}).
To do this we will use the trace formula\footnote{Recall that $\hat{E}:=\Op[E]$ denotes the Weyl quantization and $\star $ denotes the Moyal product.}
\begin{equation}
    \label{eq:trace-to-moyal}
    \|\hat{E} \ket{\alpha,\sigma}\|^2
    = \Tr[ \hat{E}^2 \tauQas]
    = \int (E\star E)(\beta) \tauCas(\beta) \dd \beta,
\end{equation}
which holds for any polynomially bounded operator $\hat{E}$ by Lemma~\ref{lem:gaussian-trace}. To bound this Moyal product in turn we use the following proposition, 
which is technically involved and is proved separately in Section~\ref{sec:moyal}.   The point of this proposition
is to show that if $A$ vanishes to order $m$ at $\alpha$ (meaning in 
particular that $|A(\alpha+\beta)| \lsim |\beta|^{m+1}$) then
$|A^*\star A(\alpha+\beta)|\lsim |\beta|^{2m+2} + \hbar^{m+1}$.  The reason we cannot simply use the standard
series approximation to the Moyal product with $O(\hbar^{m+1})$
remainder is that we are dealing with symbols that may grow 
at infinity, so we need a way of suppressing the dependence 
on the symbol far from $\alpha$.  This is taken care
of by the quantity $\constN_{\hbar;s,\vsc}^{q,r}[A](\alpha)$
in the statement of the lemma below.  The parameter $s$ controls how much growth in the symbol $A$ we want to allow.
	\begin{restatable}[Moyal Product Bound]{prp}{prpMoyalBound}
		\label{prp:moyalBound}
		Let $\delta E^{[\alpha,m]} = E - E^{[\alpha,m]}$ be the remainder to the $m$-th Taylor approximation 
		$E^{[\alpha,m]}(\alpha+\beta) = \sum_{k=0}^{m} \beta^{a_1}\cdots \beta^{a_k} (\partial_{a_1} \cdots \partial_{a_k} E)(\alpha)/k!$ 
		at $\alpha$ of some function $E$ over $2d$-dimensional phase space.
		Then, for any non-negative integer $s$ and phase-space length scale $\vsc>0$,
		\begin{equation}
  \begin{split}
			\label{eq:remainder-moyal-bd}
			|\delta E^{[\alpha,m]} {}^*\star \delta E^{[\alpha,m]}(\alpha+\beta)| 
			&\lsim_m
			(1+\vsc^{-2s}|\beta|^{2s})\\&\qquad (|\beta|^{2m+2} + \hbar^{m+1})
			[\constN_{\hbar;s,\vsc}^{m+1,2d+2+m+s}[E](\alpha)]^2
   \end{split}
		\end{equation}
		where $\constN_{\hbar;s,\vsc}^{q,r}[E](\alpha)$, defined in \eqref{eq:constN-def}, 
		is an upper bound on the $q$-th through $r$-th derivatives 
		of $E$ near $\alpha$, weighted by an $s$-th order polynomial decay in the distance $\beta$ from $\alpha$
  (and is particular is bounded for symbols
  in $S((1+|\alpha|)^s)$ as defined in~\eqref{eq:stupid-S-def})
		Likewise for non-negative integer $s$ we have
		\begin{align}
			\label{eq:FFFF-poly-bd-new}
			\begin{split}
				|E^{[\alpha,m]}{}^*\star E^{[\alpha,m]}\star E^{[\alpha,m]}{}^*\star E^{[\alpha,m]}(\alpha+\beta)|
				&\lsim_m
				(1+\nu^{-4s}|\beta|^{4s}) \\
    &\qquad (|\beta|^{4m+4} + \hbar^{2(m+1)})
				[\constN_{\hbar;2s,\vsc}^{m+1,2(2d+2+m+s)}[E](\alpha)]^4
			\end{split}
		\end{align}
		for the thrice iterated Moyal product.
	\end{restatable}
	
	We will predominantly make use of Proposition~\ref{prp:moyalBound} in the special case\footnote{Which, anyway, is the easy case to prove directly from the asymptotic series for the Moyal product.} of $s=0$, in which case the right hand side simplifies to 
	\begin{align}
		\label{eq:remainder-moyal-bd-szero}
		|\delta E^{[\alpha,m]} {}^*\star \delta E^{[\alpha,m]}(\alpha+\beta)| 
		\lsim
		(|\beta|^{2(m+1)} + \hbar^{m+1})
		(\constQ^{m+1,2d+m+2}_{\hbar}[E])^2
	\end{align}
	and 
	\begin{align}
		\label{eq:FFFF-poly-bd-new-szero}
		\begin{split}
			&|E^{[\alpha,m]}{}^*\star E^{[\alpha,m]}\star E^{[\alpha,m]}{}^*\star E^{[\alpha,m]}(\alpha+\beta)| 
			\lsim
			(|\beta|^{4(m+1)} + \hbar^{2(m+1)})
			(\constQ^{m+1,2(2d+m+2)}_{\hbar}[E])^4.
	\end{split}
\end{align}

We break up the rest of the proof into parts corresponding to Eqs.~(\ref{eq:HQR-term}--\ref{eq:triple-moyal-term}), and we apply Proposition~\ref{prp:moyalBound} in all but one.
For the purposes of this proof, we introduce the shorthand $\ell_a = \partial_a \LCk(\alpha)$ since we consider just one Lindblad operator at a time and $\alpha$ is just a fixed point we are expanding around.

Also, because we are not carefully computing the constants involved as we did in the classical case, we will mostly simply use the 
following bound for the Gaussian integral:
\begin{equation}
\label{eq:crude-gaussian-bd}
    \int \diff\beta \tauCas(\alpha+\beta) |\beta|^{2k} \lsim \|\sigma\|^k.
\end{equation}
\noindent\textbf{\textit{Proof of~\eqref{eq:HQR-term}}, \eqref{eq:VQlinear-term}, and~\eqref{eq:triple-moyal-term}:}

Using~\eqref{eq:trace-to-moyal},
\begin{align}\begin{split}\label{eq:delta-h-vec-norm-squared}
		\norm{\HQRat|\alpha,\sigma \rangle}^2 &= \langle\alpha,\sigma |(\HQRa)^2|\alpha,\sigma \rangle = \Tr\left[\tauQas(\HQRat)^2\right] \\
		&= \int \!\dd \beta\, \tauCas(\alpha+\beta)\left(\HCRat \star \HCRat\right)(\alpha+\beta)
	\end{split}
\end{align}
where $\tauCas(\alpha+\beta) = \exp(-\beta^a\sigma^{-1}_{ab}\beta^b/2)/(2 \pi\sqrt{\det\sigma})$, a positive-valued function on phase space,
is the Wigner function of the pure Gaussian state $|\alpha,\sigma \rangle$.  (Note that $\HQRa$ on the left-hand side is an operator while $\HCRa$ on the right-hand side is just a classical scalar function of the phase-space location $\beta$.)
Applying Proposition~\ref{prp:moyalBound} with $m=2$ gives
\begin{align}
	\begin{split}
		\norm{\HQRat|\alpha,\sigma \rangle}^2
		&\lsim 
		(\constQ^{3,2d+4}_{\hbar}[E])^2 \int \!\dd \beta\, \tauCas(\alpha+\beta) (|\beta|^{6} + \hbar^{3}) \\
  &\lsim (\constQ^{3,2d+4}_{\hbar}[\HC])^2 \|\sigma\|^3.
\end{split}\end{align}

	The proof of~\eqref{eq:VQlinear-term} follows the same strategy, first applying~\eqref{eq:trace-to-moyal}, then Proposition~\ref{prp:moyalBound} with $m=1$, then performing the Gaussian integral, and finally applying elementary inequalities:
	\begin{align}
		\norm{\VQRaok|\alpha,\sigma \rangle }^2
		&= \int \!\dd \beta\, \tauCas(\alpha+\beta)\left(\VCRaok \star \VCRaok\right)(\alpha+\beta)\\
		&\lsim 
		(\constQ^{2,2d+3}_{\hbar}[\LCk])^2 \int \!\dd \beta\, \tauCas(\alpha+\beta) (|\beta|^{4} + \hbar^{2})\\	
		&\lsim 
  (\constQ^{2,2d+3}_{\hbar}[\LCk])^2 \|\sigma\|^2 
	\end{align}
	For \eqref{eq:triple-moyal-term}, we deploy the triple-Moyal-product part of Proposition~\ref{prp:moyalBound} with $m=1$, giving
	\begin{align}
		\norm{ \VQRaok{}^\dagger  \VQRaok |\alpha,\sigma \rangle}^2 
		&= \langle \alpha,\sigma | \VQRaok{}^\dagger \VQRaok \VQRaok{}^\dagger \VQRaok | \alpha,\sigma \rangle \\
		&= \int \!\dd \beta\, \tauCas(\alpha+\beta) \left(\VCRaok{}^* \star\VCRaok\star \VCRaok{}^* \star\VCRaok \right)(\alpha+\beta)\\
		&\lsim
		(\constQ^{2,4d+6}_{\hbar}[\LCk])^4
		\int \!\dd \beta\,  \tauCas(\alpha+\beta)
		\left(\hbar^{4}+|\beta|^{8}\right)\\
		&\lsim  
  (\constQ^{2,4d+6}_{\hbar}[\LCk])^4 \|\sigma\|^4
	\end{align}
	
	\noindent\textit{\textbf{Proof of~\eqref{eq:VQR-term}}}:
	
	The presence of leading term $\LCk(\alpha)$ on the left-hand side of \eqref{eq:VQR-term} introduces a complication for our goal of bounding that side with a constant independent of $\alpha$.  In particular, we want our bound to hold in the special case of linear Lindblad operators, $\LCk(\alpha) = \ell_{k; a} \alpha^a$, so we cannot bound $|\LCk(\alpha)|$ and $\| \VQRatk|\alpha,\sigma \rangle \|$ separately.  It is for this term, and this term only, that we will use the form of Proposition~\ref{prp:moyalBound} with $s=1$ rather than $s=0$,
	\begin{align}
		\norm{\VQRatk|\alpha,\sigma \rangle }^2
		&= \int \!\dd \beta\, \tauCas(\alpha+\beta)\left(\VCRatk{}^* \star \VCRatk\right)(\alpha+\beta)\\
		& \lsim     
		(\constN^{3,2d+6}_{\hbar;1,\vsc}[\LCk] (\alpha))^2 \int \!\dd \beta\, \tauCas(\alpha+\beta)  (|\beta|^{6} + \hbar^{3} + \vsc^{-2}|\beta|^{8} + \vsc^{-2}\hbar^3|\beta|^{2})
		\\
		& \lsim      
		(\constN^{3,2d+6}_{\hbar;1,\vsc}[\LCk] (\alpha))^2 \left[
  \|\sigma\|^3 + \vsc^{-2} \|\sigma\|^4
		\right]
	\end{align}
    Then
    \begin{align}
        |\LCk(\alpha)|\norm{\VQRatk|\alpha,\sigma \rangle }
        &\lsim |\LCk(\alpha)| \constN^{3,2d+6}_{\hbar;1,\vsc}[\LCk] (\alpha)
        \left[\|\sigma\|^3 + \vsc^{-2} \|\sigma\|^4
		\right]^{1/2}
        \\
        &\lsim |\LCk(\alpha)| \constN^{3,2d+6}_{\hbar;1,\vsc}[\LCk] (\alpha) 
        \left[ \|\sigma\|^{3/2} + \vsc^{-1}\|\sigma\|^2\right]
    \end{align}
    Choosing $\nu=\|\sigma\|^{1/2}$ we
    obtain~\eqref{eq:VQR-term}.
	
	\noindent\textit{\textbf{Proof of~\eqref{eq:Vmixed-term}}}:

	For this we do not actually need Proposition~\ref{prp:moyalBound}
	because $\VCLaok(\beta) = \ell_a(\beta^a-\alpha^a)$ is just linear so, by the explicit Moyal product \eqref{eq:moyal-star-expansion},
	\begin{align}
		\VCLaok{}^*\star\VCLaok(\alpha+\beta) &= \left|\VCLaok(\alpha+\beta)\right|^2 + \frac{i\hbar}{2}(\partial_a\VCLaok{}^*)(\alpha+\beta)(\partial^a\VCLaok)(\alpha+\beta) \\
		&=  |\ell_a\beta^a|^2 + \frac{i\hbar}{2} \ell_a^* \ell^a      \\
		&\le |\ell|^2(|\beta|^2+\hbar/2)\\
		&\le (\constQ^{1,1}[\LCk])^2(|\beta|^2+\hbar/2)
	\end{align}
	where we have used the Cauchy-Schwartz inequality and $|\ell| = |\partial \LCk(\alpha)| \le C^{1,1}_{\LCk;0} = \sup_\beta [|\partial_\x \LCk(\beta)|+|\partial_\p \LCk(\beta)|]$.   Therefore
	\begin{align}
		\norm{\VQLaok|\alpha,\sigma \rangle }^2
		&= \int \!\dd \beta\, \tauCas(\beta)\left(\VCLaok{}^* \star \VCLaok\right)(\alpha+\beta)\\
		&\le (\constQ^{1,1}[\LCk])^2 \int \!\dd \beta\, \tauCas(\alpha+\beta)(|\beta|^2+\hbar/2)\\
		&= (\constQ^{1,1}[\LCk])^2 (\Tr[\sigma] + \hbar/2 )\\
		&\lsim (\constQ^{1,1}[\LCk])^2 \|\sigma\|
	\end{align}
	as desired.
	
	\noindent\textit{\textbf{Proof of~\eqref{eq:linear-outside-term}
			and~\eqref{eq:linear-inside-term}:}}
	
	We note that the left hand sides of~\eqref{eq:linear-inside-term}
	and~\eqref{eq:linear-outside-term} are related by
	\begin{align}\label{eq:linear-inside-outside-term-relation}
		\|\VQLaok{}^\dagger \VQRaok \ket{\alpha,\sigma}\!\|
		\leq
		\|\VQRaok \VQLaok{}^\dagger \ket{\alpha,\sigma}\!\|
		+
		\|\big[\VQLaok{}^\dagger,  \VQRaok\big] \ket{\alpha,\sigma}\!\|.
	\end{align}
	Lets start by expanding the first part of the right-hand side:
	\begin{equation}\label{eq:linear-outside-integral}
		\|\VQRaok \VQLaok{}^\dagger \ket{\alpha,\sigma}\!\|^2
		=
		\int
		\left( \VCRaok{}^* \star \VCRaok\right) (\alpha+\beta)
		\left( \VCLaok{}^* \star \tauCas \star \VCLaok\right)(\alpha+\beta) \diff \beta,
	\end{equation}
	The first term inside the integrand of \eqref{eq:linear-outside-integral} can be bounded with Proposition~\ref{prp:moyalBound}.
	\begin{equation}\label{eq:linear-outside-integral-part-1}
		\left( \VCRaok{}^* \star \VCRaok\right) (\alpha+\beta)
		\lsim
		(\constQ^{2,2d+3}_{\hbar}[\LCk])^2 (|\beta|^4 + \hbar^2).
	\end{equation}
	For the second term in the integrand of \eqref{eq:linear-outside-integral}, we note that since
	$\VCLaok$ is linear,  $ \VCLaok{}^* \star \tauCas$ can be computed explicitly with the Moyal product \eqref{eq:moyal-star-expansion} as
	\begin{equation}
		\VCLaok{}^*\star \tauCas (\alpha+\beta) =
		\ell_a^* \beta^a \tauCas(\alpha+\beta) + \frac{i\hbar}{2} \ell_a^* \partial^a \tauCas(\alpha+\beta).
	\end{equation}
	Now we recall \eqref{eq:deriv-gaussian},
	\begin{align}
		(\partial_c \tauCas)(\alpha+\beta) &=  -\m_c \tauCas(\alpha+\beta)
	\end{align}
	where $\m^a:= (\sigma^{-1})^a_{\pha b} \beta^b = \sf^{ac}\sigma^{-1}_{cb} \beta^b$, so
	\begin{align}
		\VCLaok{}^*\star \tauCas(\alpha+\beta)
		&= \ell_a^* \left[\beta^a - \frac{i\hbar}{2} m^a\right] \tauCas(\alpha+\beta)\\
		&= \ell_a^* \left[\delta^a_{\pha b} - \frac{i\hbar}{2} (\sigma^{-1})^a_{\pha b}\right]\beta^b \tauCas(\alpha+\beta).
	\end{align}
	Then we can apply the Moyal product with $\VCLaok$ on the right to get
	\begin{align}\label{eq:v-tau-v}\begin{split}
			&\VCLaok{}^* \star \tauCas \star \VCLaok (\alpha+\beta)
			\\
			&\qquad =
			\tauCas(\alpha+\beta)\left[
			|\ell_a\beta^a|^2
			- \frac{i\hbar}{2}(\ell_a^*(\sigma^{-1})^a_{\pha b} \beta^b)( \ell_c\beta^c) \right]
			-
			\frac{i\hbar}{2} \ell_c \left[ \partial^c ( \VCLaok{}^* \star \tauCas)\right](\alpha+\beta) .
	\end{split}\end{align}
	We can compute
	\begin{align}
		\left[ \partial_c ( \VCLaok{}^*\star \tauCas)\right](\alpha+\beta)
		= \left[\ell_c^* - \frac{i\hbar}{2} \ell_a^* (\sigma^{-1})^a_{\pha c} -
		\left(\ell_a^*\beta^a - \frac{i\hbar}{2}\ell_a^* (\sigma^{-1})^a_{\pha b}\beta^b\right)
		\sigma^{-1}_{cd}\beta^d
		\right]\tauCas(\alpha + \beta)
	\end{align}
	so now inserting this into \eqref{eq:v-tau-v} we have 
	\begin{equation}\label{eq:linear-outside-integral-part-2}
		\begin{split}
			\VCLaok{}^* \star \tauCas \star \VCLaok (\alpha+\beta)
			&= \tauCas(\alpha+\beta)
			\Big[
			|\ell_a\beta^a|^2 +
			i\frac{\hbar}{2}
			\big(\ell^c \ell^*_c - \ell^*_a\beta^a \ell^c\sigma^{-1}_{cb}\beta^b
			+  \ell^a{}^* \sigma^{-1}_{a b} \beta^b \ell_c\beta^c \big) \\
			&\qquad \qquad
			+ \frac{\hbar^2}{4} ( \ell^a{}^* \sigma^{-1}_{a b}\beta^b \ell^c\sigma^{-1}_{cd}\beta^d
			-  \ell^c{}^* (\sigma^{-1})_{ca} \ell^a)\Big]\\
			&= \tauCas(\alpha+\beta)
			\Big[
			|\ell_a\beta^a|^2 +
			\hbar
			\Imag \left(\ell^*_a\beta^a \ell^c\sigma^{-1}_{cb}\beta^b -\ell^c \ell^*_c/2 \right) \\
			&\qquad \qquad
			+ \frac{\hbar^2}{4} \left( |\ell^a \sigma^{-1}_{a b}\beta^b|^2
			-  \ell^c{}^* \sigma^{-1}_{ca} \ell^a\right)\Big]\\
			&= \tauCas(\alpha+\beta)
			\left[
			{\ell^a} \left(\frac{i\hbar}{2} \sf_{ab} + \frac{\hbar^2}{4} \sigma^{-1}_{ab}\right) \ell^b{}^* + \left|\ell^a \left(\sf_{ab}-\frac{i\hbar}{2}\sigma^{-1}_{ab}\right)\beta^a\right|^2 \right]\\
			&= \tauCas(\alpha+\beta)
			\left[
			v_0 
			+ w_a^* w_b \beta^a \beta^b \right]
		\end{split}
	\end{equation}
	where
	\begin{align}
		w_b :=& \ell^a \left(\sf_{ab}-\frac{i\hbar}{2}\sigma^{-1}_{ab}\right) \in \mathbb{C}^{2d},\\
		v_0 :=& 
		\frac{i\hbar}{2} {\ell^a} \left(\sf_{ab} -\frac{i\hbar}{2}  \sigma^{-1}_{ab}\right) \ell^b{}^* 
		=\frac{i\hbar}{2} w_b \ell^b{}^*\in \bbR 
	\end{align}
	which obey 
	\begin{align}
		|v_0| \le & \, (\hbar/2)|\ell |^2   \left(1 + (\hbar/2)\| \sigma^{-1}\| \right) 
		\le (\constQ^{1,1}[\LCk])^2  \left(\hbar/2  + \|\sigma\|\right) 
		\le 2 (\constQ^{1,1}[\LCk])^2  \|\sigma\|
		\\
		(w^\dagger \sigma^n w) =& w_a^* (\sigma^n)^{ab} w_b = (\ell^\dagger \sigma^n \ell)+\frac{\hbar^2}{4} (\ell^\dagger \sigma^{n-2} \ell) 
		\le (\constQ^{1,1}[\LCk])^2 \left( \|\sigma^n\|+\hbar^2 \|\sigma^{n-2}\|/4 \right) 
		\le 2 (\constQ^{1,1}[\LCk])^2 \|\sigma^n\|
	\end{align}
	If we insert \eqref{eq:linear-outside-integral-part-2} and \eqref{eq:linear-outside-integral-part-1} into \eqref{eq:linear-outside-integral} we get
	\begin{align}\begin{split}
			\|\VQRaok{}^\dagger \VQLaok\ket{\alpha,\sigma}\|^2
			& \leq 
			[C^{2,2d+3}_{\LCk;0;\hbar}]^2
			\int \tauCas(\alpha+\beta)\left[\hbar^2 v_0  + \hbar^2 w_a^* w_b \beta^a \beta^b + v_0 |\beta|^4 + w_a^* w_b \beta^a \beta^b |\beta|^4  \right]\diff \beta
	\end{split}\end{align}
	and then perform the Gaussian integral
	\begin{align}\begin{split}\label{eq:inside-bound}
			\frac{\|\VQRaok{}^\dagger \VQLaok\ket{\alpha,\sigma}\|^2}
			{(\constQ^{2,2d+3}_{\hbar}[\LCk])^2}
			&\leq 
			\int \tauCas(\alpha+\beta)\left[\hbar^2 v_0  + \hbar^2 w_a^* w_b \beta^a \beta^b + v_0 |\beta|^4 + w_a^* w_b \beta^a \beta^b |\beta|^4  \right]\diff \beta\\
			&= 
			v_0 \hbar^2 +  \hbar^2 (w^\dagger \sigma w) + v_0 \left((\Tr\sigma)^2 + 2 \Tr[\sigma^2]\right)  \\
			&\qquad \qquad    +  \left((w^\dagger \sigma w)(\Tr\sigma)^2+ 2 (w^\dagger \sigma w) \Tr[\sigma^2] + 4 \Tr[\sigma] (w^\dagger \sigma^2 w) + 8 (w^\dagger \sigma^3 w)\right) \\
			&= 
			(\constQ^{1,1}[\LCk])^2
			4\left[\hbar^2 \|\sigma\| + 8(d+1)^2\|\sigma\|^3\right]\\
			&\lsim (\constQ^{1,1}[\LCk])^2 
			\|\sigma\|^3.
	\end{split}
 \end{align}
	which proves \eqref{eq:linear-outside-term}.
	
	To prove \eqref{eq:linear-inside-term}, 
	we need to handle the commutator in \eqref{eq:linear-inside-outside-term-relation}.
	Note that $\VQLaok = \ell_a (\RQ^a-\alpha^a)$ is linear in the phase space variables $\RQ = (\XQ,\PQ)$, so the Wigner transform of (i.e., symbol for) the commutator $[\VQLaok{}^\dagger,  \VQRaok]$ can be computed directly with the Moyal product \eqref{eq:moyal-star-expansion} to be $ i \hbar \ell_a^* \partial^a \VCRaok$.  This is just the remainder from the zeroth order Taylor approximation to the function $i\hbar \ell_a^* \partial^a \LCk =i\hbar \ell_a^* \partial^a \VCk$:
	\begin{align}
		\ell_a^* \partial^a \delta \VCLaok 
		= \ell_a^* \partial^a \delta \LCLaok 
		= \delta(\ell_a^* \partial^a  \LCk)^{[\alpha,0]}, 
	\end{align}
	so that it satisfies the prerequisites of Proposition~\ref{prp:moyalBound} with $m=0$, which we can apply to get
	\begin{align}
		\norm{[\VQLaok{}^\dagger,  \VQRaok]|\alpha,\sigma \rangle }^2
		& = \hbar^2 \int \!\dd \beta\, \tauCas(\alpha+\beta) 
		\left( \delta(\ell_a^* \partial^a  \LCk)^{[\alpha,0]}{}^* \star  \delta(\ell_a^* \partial^a  \LCk)^{[\alpha,0]}\right)(\alpha+\beta)\\
		&\lsim  \hbar^2 
		\left(\constQ^{1,2d+2}_{\hbar}[\ell_a^* \partial^a \LCk]\right)^2 
		\int \!\dd \beta\, \tauCas(\alpha+\beta) (|\beta|^{2} + \hbar^{1})\\
		&\lsim  \hbar^2 (\constQ^{1,1}[\LCk] \constQ^{2,2d+3}_{\hbar}[\LCk])^2 
		\left(\Tr[\sigma] +\hbar^{1}\right) \\
		&\lsim
  \hbar^2 (\constQ^{1,1}[\LCk] \constQ^{2,2d+3}_{\hbar}[\LCk])^2 
		\|\sigma\| \\
		\label{eq:commutator-bound}
		&\lsim (\constQ^{1,1}[\LCk] \constQ^{2,2d+3}_{\hbar}[\LCk])^2  \|\sigma\|^3
	\end{align}
	where we have used 
	\begin{align}
		\constQ^{q,r}_{\hbar}[\ell_a^*\partial_a \LCk] \leq |\ell| \constQ^{q+1,r+1}_{\hbar}[\LCk] \le \constQ^{1,1}[\LCk] \constQ^{q+1,r+1}_{\hbar}[\LCk]
	\end{align}
	Taking the square roots of \eqref{eq:inside-bound} and \eqref{eq:commutator-bound} and inserting into  \eqref{eq:linear-inside-outside-term-relation} gives \eqref{eq:linear-inside-term}.

    Having now demonstrated all the bounds \eqref{eq:HQR-term}--\eqref{eq:linear-inside-term} with the help of Proposition~\ref{prp:moyalBound}, Lemma~\ref{lem:HarmErrorQ} is proved.
\end{proof}

The only remaining task to complete the demonstration of our main result is to justify Proposition~\ref{prp:moyalBound}, which is addressed in the next section.

\section{Moyal product bound}\label{sec:moyal}

In this section\footnote{Note that, due to the regrettably finite size of alphabets, in this section we have re-used variables previously defined for other purposes elsewhere in the paper. This section should be considered self-contained.} we prove Proposition~\ref{prp:moyalBound}, whose statement we now recall:

\prpMoyalBound*

We use the following integral formulation of the Moyal product:
\begin{equation}
	\label{eq:moyal-def}
	E\star G(\alpha) =
	(2\pi \hbar)^{-d}
	\int e^{i\beta_a\gamma^a/(2\hbar)} E(\alpha+\beta/2)G(\alpha+\gamma/2) \diff\beta\diff\gamma.
\end{equation}

The proof of Proposition~\ref{prp:moyalBound} splits into two main parts.  First, in Section~\ref{sec:main-lemma} we
state Lemma~\ref{lem:main-moyal} giving a bound for $F\star F(\alpha)$ in terms of a convolution of derivatives of $F$.
Then in Section~\ref{sec:applications} we show how Lemma~\ref{lem:main-moyal} implies Proposition~\ref{prp:moyalBound}.
The proof of Lemma~\ref{lem:main-moyal} is deferred to Section~\ref{sec:proof}.

\subsection{Main lemma}
\label{sec:main-lemma}

In the following lemma $\rho_K$ is the convolution kernel
\[
\rho_K(\alpha) := \hbar^{-d} (\hbar^{-1/2}|\alpha| + 1)^{-K}.
\]
Note that when $K>2d$
\begin{align*}
	\int \rho_K(\alpha) |\alpha|^j \diff \alpha
	&= \hbar^{-d} \int (\hbar^{-1/2}|\alpha|+1)^{-K}|\alpha|^j \diff \alpha \\
	&= \hbar^{j/2} \int (|\alpha| + 1)^{-K} |\alpha|^j \diff \alpha\\
	&= \hbar^{j/2} C_{d,K,j}
\end{align*}

Thus when $K>2d$ the convolution $\rho_K\ast E$ is well-defined:
\begin{equation}
	\rho_K\ast E(\alpha) :=
	\int \rho_K(\beta) E(\alpha-\beta)\diff\beta.
\end{equation}

The main bound we need to prove Proposition~\ref{prp:moyalBound} is a ``localized'' pointwise
bound for the Moyal product of two symbols.  In particular,
we need to bound $F\star G(\alpha)$ in a way that ideally
only depends on the values of $F$ and $G$ near $\alpha$.   The parameters $K_F$ and $K_G$ determine how much our bound depends
on the values of $F$ and $G$ far from $\alpha$.   In order 
to make the bound more local in $F$ we require higher order
derivative bounds on $G$, and vice-versa.
\begin{restatable}{lem}{lemMainMoyal}
	\label{lem:main-moyal}
	Let $F,G\in C^\infty(\bbR^{2d})$ be smooth functions,
	and let $K_F,K_G > 2d$ be nonnegative integers.  Then
	\begin{equation}
		\label{eq:main-moyal-bd}
		\begin{split}
			|F\star G(\alpha)|
			&\lsim			\left(\sum_{k=0}^{K_F}\big(\rho_{K_G}\ast \TsrNrm{\hbar^{k/2}\nabla^k F}\big)(\alpha)\right)
			\left(\sum_{k=0}^{K_G} \big(\rho_{K_F} \ast \TsrNrm{\hbar^{k/2}\nabla^k G}\big)(\alpha)\right).
		\end{split}
	\end{equation}
 where the implicit constant hidden by $\lsim$ depends on $K_F$ and $K_G$.
\end{restatable}

\begin{restatable}[Iterated Moyal bound]{cor}{corIteratedMoyal}
	\label{cor:iterated-moyal}
	Let $F\in C^\infty(\bbR^{2d})$ be a smooth function and let $K$ be a nonnegative integer.
	Then
	\begin{equation}
	|F\star F \star F \star F (\alpha)| \lsim\left[\rho_K \ast
	\left(\sum_{m=0}^{2K} \rho_K \ast \TsrNrm{\hbar^{m/2}\nabla^{m} F}\right)^2\right]^2(\alpha)
	\end{equation}
\end{restatable}
\begin{proof}[Proof using Lemma~\ref{lem:main-moyal}]
 Using Lemma~\ref{lem:main-moyal} we have
	\[
	|F\star F \star F\star F(\alpha)|
	\lsim\left(\sum_{k=0}^K (\rho_K \ast \TsrNrm{\hbar^{k/2}\nabla^k (F\star F)})(\alpha) \right)^2.
	\]
	To bound $\nabla^k (F\star F)$ we use the following product rule for the
	partial derivative $\partial^{\vec{n}}$, defined as 
$\partial_1^{n_1}\partial_2^{n_2}\cdots\partial_{2d}^{n_{2d}}$
	\[
	\partial^{\vec{n}} (F\star F) = 
	\sum_{\vec{m}\leq\vec{n}} \prod_{j=1}^{2d} \binom{n_j}{m_j} \partial^{\vec{m}} F\star 
	(\partial^{\vec{n}-\vec{m}} F).
	\]
	Applying Lemma~\ref{lem:main-moyal} to the terms in the right hand side we obtain
	\begin{equation}
		\begin{split}
			&\hbar^{k/2}
			\TsrNrm{(\nabla^j F)\star (\nabla^{k-j} F) (\alpha)}\\
			&\qquad\qquad \lsim
			\left(\sum_{m=0}^K \rho_K \ast \TsrNrm{\hbar^{(m+j)/2} \nabla^{m+j} F} (\alpha)\right)
			\left(\sum_{m=0}^K \rho_K \ast \TsrNrm{\hbar^{(m+k-j)/2} \nabla^{m+k-j} F} (\alpha)\right) \\
			&\qquad\qquad \lsim \left(\sum_{m=0}^{2K} \rho_K \ast \TsrNrm{\hbar^{m/2} \nabla^{m/2} F}(\alpha)\right)^2.
		\end{split}
	\end{equation}
\end{proof}

\subsection{Proof of Proposition~\ref{prp:moyalBound} from Lemma~\ref{lem:main-moyal}}
\label{sec:applications}
First we prove a simple estimate for convolutions of functions against the kernel $\rho_K$
in terms of the following weighted supremum
\begin{equation}
	M_\hbar^q[F](\alpha) := \sup_{\beta} (1 + \hbar^{-1/2}|\beta-\alpha|)^{-q} |F(\beta)|.
\end{equation}
\begin{restatable}{lem}{lemPolyConvEst}
	\label{lem:poly-conv-est}
	If $K \geq 2d+1$ then
	\begin{equation}
		\label{eq:rho-convolve-bd}
		(\rho_K \ast F)
		(\alpha) \leq C M_\hbar^{K-2d-1}[F](\alpha)
	\end{equation}
	for some absolute constant $C$.
\end{restatable}
\begin{proof}
	Set $q=K-2d-1$.  Then
	\begin{equation}
		\begin{split}
			|\rho_K\ast F (\alpha)|
			&\leq \hbar^{-d} \int (1 + \hbar^{-1/2}|\beta-\alpha|)^{-K} |F(\beta)|\diff\beta \\
			&\leq \hbar^{-d} M_\hbar^q[F](\alpha) \int (1+\hbar^{-1/2}|\beta-\alpha|)^{2d-1} \diff\beta \\
			&\leq CM_\hbar^q[F](\alpha).
		\end{split}
	\end{equation}
\end{proof}

We also note two quick facts about the weighted supremum.  The first is the weighted supremum of a monomial
$m_k(\alpha) = |\alpha|^k$, using that $(a+b)^k \leq 2^k (a^k + b^k)$.  For $q\geq k$ we have
\begin{equation}
	\label{eq:maximal-monomial}
	\begin{split}
		M_\hbar^q[m_k](\alpha)
		&= \sup_\beta (1+\hbar^{-1/2}|\beta-\alpha|)^{-q} |\beta|^k \\
		&\leq 2^k (|\alpha|^k +
		\sup_\beta (1+\hbar^{-1/2}|\beta-\alpha|)^{-q} |\alpha-\beta|^k) \\
		&\leq 4^k (|\alpha|^k + \hbar^{k/2}),
	\end{split}
\end{equation}
where in the last step we used that 
$\sup_t (1+\hbar^{-1/2}t)^{-q}t^k \leq 2^k$
when $q\geq k$.

The second quick fact that we need is the following product rule for the weighted supremum:
\begin{equation}
	\label{eq:maximal-prod-rule}
	M_\hbar^{q_1+q_2}[FG] \leq M_\hbar^{q_1}[F] M_\hbar^{q_2}[G].
\end{equation}

We will also use $M_\hbar^q[A]$ to refer
to $M_\hbar^q[\TsrNrm{A}]$ when $A$ is a tensor-valued
or vector-valued quantity. 

Finally, we introduce the weighted supremum at $\vsc$-scale,
\begin{equation}
    M_\vsc^q[F](\alpha)
    := \sup_{\beta}(1 + \nu^{-1}|\beta-\alpha|)^{-q}|F(\beta)|.
\end{equation}

Now we are ready to prove Proposition~\ref{prp:moyalBound}.
\begin{proof}[Proof of Proposition~\ref{prp:moyalBound} from Lemma~\ref{lem:main-moyal}]
	
	The first step is to prove~\eqref{eq:remainder-moyal-bd}.
	First we need a bound for $\delta E^{[\alpha,k]}$ that follows from the Taylor remainder formula,
	\begin{equation}
		\label{eq:third-order-remainder}
		f(t) =
		\sum_{j=0}^{k_0} \frac{t^j}{j!} f^{(j)}(0)
		+ \frac{1}{k_0!} \int_0^t (t-s)^{k_0} f^{(k_0+1)}(s)\diff s,
	\end{equation}
	which holds for functions in $C^{k_0+1}(\bbR)$.
	Applying this with $f(t) = E(t\alpha)$ and evaluating at $t=1$ we obtain 
	\begin{equation}
		\label{eq:third-remainder-alpha}
		E(\alpha) = E^{[0,k]}(\alpha)
		+ \frac{1}{k!} \int_0^1 (1-s)^{k} \alpha^{\otimes ({k+1})}\cdot \nabla^{k+1} E(s\alpha)\diff s.
	\end{equation}
	Thus we obtain the formula
	\[
	\delta E^{[0,k]}(\alpha) =
	\frac{1}{k!} \int_0^1 (1-s)^{k} \alpha^{\otimes (k+1)} \cdot \nabla^{k+1} E(s\alpha)\diff s.
	\]
	Above $\alpha^{\otimes(k+1)} \nabla^{k+1}E$ is shorthand for 
	$\alpha^{a_1}\cdots\alpha^{a_k}
	\partial_{a_1}\cdots \partial_{a_k}E$ with an implicit
	sum over all indices.
 
	Simply using the triangle inequality and estimating naively
, this produces the bound
	\[
	|\delta E^{[0,k]}(\alpha)| \leq
	\sup_{|\beta|\leq |\alpha|} |\nabla^{k+1}E(\beta)|
	|\alpha|^{k+1}
	\leq  M_\vsc^s[\nabla^{k+1}E](0) (1+\vsc^{-1}|\alpha|)^{s} |\alpha|^{k+1}.
	\]
	We also need bounds for the derivatives with respect to $\alpha$.  When $0\leq j\leq k$ we have 
	\begin{equation}
		\label{eq:local-low-bds}
		\begin{split}
			\TsrNrm{\nabla^j \delta E^{[0,k]}(\alpha)} &\leq
			\sum_{j'=0}^j
			\sup_{|\beta|\leq|\alpha|} \TsrNrm{\nabla^{k+1+j'} E(\beta)}
			\|\alpha\|^{k+1+j'-j} \\
			&\leq 
			\sum_{j'=0}^j
			M_\vsc^s[\nabla^{k+1+j'}E](0) (1+\vsc^{-1}|\alpha|)^s |\alpha|^{k+1+j'-j}.
		\end{split}
	\end{equation}
	For higher order derivatives we note that $\nabla^{k+1} E^{[0,k]} = 0$ and therefore
	$\nabla^{k+1} \delta E^{[0,k]} = \nabla^{k+1} E$, so that for $j\geq k+1$ we have
	\begin{equation}
		\label{eq:higher-grad-bds}
		\TsrNrm{\nabla^j \delta E^{[0,k]}(\alpha)} \leq \max_{|\beta|\leq |\alpha|} \TsrNrm{\nabla^jE(\beta)}
		\leq M_\vsc^s[\nabla^jE](0) (1+\vsc^{-1}|\alpha|)^s.
	\end{equation}
	
	We will estimate $\rho_K \ast |\hbar^{j/2}\nabla^j \delta E^{[0,k]}|(\alpha)$ in terms of the
	quantity
	\[
	Q_{k_0,s}[E] := \sum_{j'=k_0+1}^{2d+k_0+s+2} \hbar^{(j'-k-1)/2}M_\vsc^s[\nabla^{j'}E](0).
	\]
	
	We first work with $j\leq k$.  We combine
	Combining~\eqref{eq:local-low-bds}
	and Lemma~\ref{lem:poly-conv-est} along with~\eqref{eq:maximal-monomial} and
	the product rule~\eqref{eq:maximal-prod-rule}
	to see that for $K\geq 2d+k+s+1$ and $j\leq k_0$ we have
	\begin{equation}
		\begin{split}
			\rho_K \ast \TsrNrm{\hbar^{j/2}\nabla^j\delta E^{[0,k]}}(\alpha)
			&\leq C \hbar^{j/2}M_\hbar^{K-2d-1}[\nabla^j\delta E^{[0,k]}](\alpha) \\
			&\leq C \hbar^{j/2} \sum_{j'=0}^j M_\vsc^s[\nabla^{k+1+j'}E](0)
			M_\hbar^{k+s+1} [(1+\vsc^{-1}|\alpha|)^s|\alpha|^{k+1+j'-j}] \\ 
			&\leq C \hbar^{j/2} \sum_{j'=0}^j M_\vsc^s[\nabla^{k+1+j'}E](0)
			M_\hbar^{s} [(1+\vsc^{-1}|\alpha|)^s]
			M_\hbar^{k+1}[|\alpha|^{k+1+j'-j}] \\ 
			&\leq C (1+\vsc^{-1}|\alpha|)^s
			\sum_{j'=0}^j \hbar^{j/2}
			M_\vsc^s[\nabla^{k+1+j'} E](0)
			\left(|\alpha|^{k+1+j'-j} + \hbar^{(k+1+j'-j)/2}\right) \\
			&\leq C (1+\vsc^{-1}|\alpha|)^s Q_{k_0,s}[E]
			\sum_{j'=0}^J \left(\hbar^{(j-j')/2}|\alpha|^{j-j'}  |\alpha|^{k+1} + \hbar^{(k+1)/2}\right)
		\end{split}
	\end{equation}
    for some constant $C$.
	In the final line, note that either $\hbar^{1/2} > |\alpha|$, in which case the term $\hbar^{(k_0+1)/2}$ dominates,
	or else $\hbar^{1/2} < |\alpha|$, in which case $(\hbar^{1/2}|\alpha|^{-1})^{(j-j')} < 1$ so that the first
	term in the sum is bounded by $|\alpha|^{k+1}$.  Therefore we can simplify the above bound to
	\begin{equation}
		\rho_K \ast |\hbar^{j/2}\nabla^j \delta E^{[0,k]}|(\alpha)
		\leq C Q_{k_0,s}[E] (1+\vsc^{-1}|\alpha|)^s (|\alpha|^{k+1}+\hbar^{(k+1)/2})
	\end{equation}
	For larger derivatives $j>k$, we simply use~\eqref{eq:higher-grad-bds} to see that
	\[
	\rho_K \ast |\hbar^{j/2}\nabla^j \delta E^{[0,k]}|(\alpha) \leq
	\hbar^{j/2} M_\vsc^s[\nabla^j E] (1+\vsc^{-1}|\alpha|)^s
	\leq \hbar^{(k+1)/2} Q_{k_0,s}[E] (1+\vsc^{-1}|\alpha|)^s.
	\]
	Taking $K=2d+k_0+s+2$ we have
	\begin{equation}
		\label{eq:direct-bd}
		\sum_{j=0}^{K} (\rho_K \ast |\hbar^{j/2} \nabla^j  \delta E^{[0,k]}|)(\alpha)
		\leq C Q_{k_0,s}[E] (1+\vsc^{-1}|\alpha|)^s)(|\alpha|^{k_0+1} + \hbar^{(k_0+1)/2}).
	\end{equation}
	The proof of~\eqref{eq:remainder-moyal-bd} now follows from an application of Lemma~\ref{lem:main-moyal}.
	
	To prove the bound on the triple Moyal product~\eqref{eq:FFFF-poly-bd-new},
	we use the stronger quantity
	\[
	Q'_{k_0,s}[E] := \sum_{j=k_0+1}^{4d+2k_0+2s+4} \hbar^{(j-k_0-1)/2}
	M_\vsc^s[\nabla^{j'}E](0).
	\]
	Then taking \eqref{eq:direct-bd} with  $K'=4d+2k_0+2s+4$, squaring both sides, and convolving with $\rho_{K'}$,  we have 
	\begin{equation}
		\rho_{K'}\ast \Bigg(\sum_{j=0}^{K'} \rho_{K'}\ast \TsrNrm{\hbar^{j/2}\nabla^j\delta E^{[\alpha,k_0]}}\Bigg)^2
		\leq C (Q'_{k_0,s}[E])^2 (1+\vsc^{-2s}|\alpha|^{2s})(|\alpha|^{2k_0+2} + \hbar^{k_0+1}).
	\end{equation}
	Then~\eqref{eq:FFFF-poly-bd-new} follows from Corollary~\ref{cor:iterated-moyal}.
\end{proof}

\subsection{Proof of the Moyal product bound}
\label{sec:proof}

\begin{proof}[Proof of Lemma~\ref{lem:main-moyal}]
	The main idea is to use the following identity to integrate by parts in the
	$\gamma$ variables in order to obtain decay in the $\beta$ variables:
	\begin{equation}
		\label{eq:int-by-pts}
		e^{i\beta_a\gamma^a/(2\hbar)} =
		-2i\hbar (\beta_a)^{-1} \partial_{\gamma^a} e^{i\beta_a\gamma^a/(2\hbar)}.
	\end{equation}
	Symmetrically, we can integrate by parts in the $\beta$ variables to obtain decay in
	the $\gamma$ variables.
	
	To do this we introduce a partition of unity
	\[
	1 = \chi_0(t) + \sum_{j=1}^\infty \chi_j(t),
	\]
	where $\chi_0\in C_c^\infty(\bbR)$ is a smooth function supported in $[-\hbar^{1/2},\hbar^{1/2}]$
	and $\chi_j\in C_c^\infty(\bbR)$ are supported in $\{t\in\bbR \mid
	2^{j-1}\hbar^{1/2} \leq |t| \leq 2^{j+2} \hbar^{1/2}\}$.  We choose this partition of unity so that it satisfies the bounds
	\begin{equation}
		\label{eq:chi-grad-bds}
		\sup_t \left|\frac{\dd^k}{\dd t^k} \chi_j(t) \right| \leq C_k 2^{-jk} \hbar^{-k/2}.
	\end{equation}
	
	For such $\chi_j$, we also have
	\begin{equation}
		\label{eq:chi-t-bds}
		\sup_t \left|\frac{\dd^k}{\dd t^k} (t^{-a} \chi_j(t)) \right| \leq C_{a,k} 2^{-j(k+a)} \hbar^{-(k+a)/2}.
	\end{equation}
	
	Applying this partition of unity to each variable we we obtain the identity 
	\begin{equation}
		1 = \prod_{a=1}^{2d} \left(\sum_{j=0}^\infty \chi_j(\beta_a)\right)
		\prod_{b=1}^{2d} \left(\sum_{j=0}^\infty \chi_j(\gamma^a)\right)
		= \sum_{\mbf{j}^\beta,\mbf{j}^\gamma:[2d]\to\bbN} \prod_{a=1}^{2d} \chi_{j^\beta_a}(\beta_a),
		\prod_{a=1}^{2d} \chi_{j^\gamma_a}(\gamma^a).
	\end{equation}
	The latter sum is over pairs of tuples
	$\mbf{j}^\beta = (j^\beta_1,j^\beta_2,\cdots,j^\beta_{2d})$,
	$\mbf{j}^\gamma = (j^\gamma_1,j^\gamma_2,\cdots,j^\gamma_{2d})$.
	
	We use this to split the moyal product $F\star G$ into terms indexed by $\mbf{j}^\beta$ and $\mbf{j}^\gamma$,
	\begin{equation}
		\label{eq:j-decomp}
		F\star G = \sum_{\mbf{j}^\beta,\mbf{j}^\gamma} (F\star G)_{\mbf{j}^\beta,\mbf{j}^\gamma},
	\end{equation}
	where
	\begin{equation}
		(F\star G)_{\mbf{j}^\beta,\mbf{j}^\alpha}(\alpha) := (2\pi \hbar)^{-2d}
		\int e^{i \beta_a\gamma^a/(2\hbar)}
		F(\alpha+\beta/2)G(\alpha+\gamma/2)
		\prod_{a=1}^{2d} \chi_{j^\beta_a}(\beta_a)
		\prod_{a=1}^{2d} \chi_{j^\gamma_a}(\gamma_a)
		\diff \beta_a\diff\gamma^a.
	\end{equation}
	
	We estimate the quantity $(F\star G)_{\mbf{j}^\beta,\mbf{j}^\gamma}$ differently depending
	on whether $\mbf{j}^\beta=0$ and/or $\mbf{j}^\gamma = 0$.  We thus split into four terms:
	\[
	F\star G = (F\star G)_{\mbf{0},\mbf{0}}
	+ \sum_{\mbf{j}^\beta \not= 0}(F\star G)_{\mbf{j}^\beta,\mbf{0}}
	+ \sum_{\mbf{j}^\gamma \not= 0}(F\star G)_{\mbf{0},\mbf{j}^\gamma}
	+ \sum_{\mbf{j}^\beta,\mbf{j}^\gamma} (F\star G)_{\mbf{j}^\beta,\mbf{j}^\gamma}.
	\]
	
	The first term, with $\mbf{j}^\beta=\mbf{j}^\gamma = \mbf{0}$ being all zeros, can be bounded simply using the
	triangle inequality:
	\begin{equation}
		\begin{split}
			|(F\star G)_{\mbf{0},\mbf{0}}(\alpha)|
			&\leq  \hbar^{-2d} \int |F(\alpha + \beta/2)| |G(\alpha+\gamma/2)| \prod \chi_0(B_a) \diff B \\
			&\leq
			\big(\hbar^{-d} \int_{|\alpha'-\alpha|\leq \hbar^{1/2}} |F(\alpha')|\diff \alpha'\big)
			\big(\hbar^{-d} \int_{|\alpha'-\alpha|\leq \hbar^{1/2}} |G(\alpha')|\diff \alpha'\big) \\
			&\leq  2^{K_G+K_F}(\rho_{K_G}\ast |F|(\alpha)) (\rho_{K_F}\ast |G|(\alpha)).
		\end{split}
	\end{equation}
	In the last line the factor
	$2^{K_G+K_F}$ appears from the use of the
	fact that
	$2^K\rho_K(\alpha)\geq 1$ when $|\alpha|\leq\hbar^{1/2}$.
	
	For the remaining terms we integrate by parts using~\eqref{eq:int-by-pts}.  We will assume for this part that
	$\mbf{j}^\beta\not=\mbf{0}$ and $\mbf{j}^\gamma\not= \mbf{0}$ (this being the most technical case to handle).
	Let $a_0 = \argmax\mbf{j}^\beta$ and $b_0=\argmax\mbf{j}^\gamma$ be the indices for which $J^\beta := j^\beta_{a_0}$
	and $J^\gamma := j^\gamma_{b_0}$ are maximized.
	
	We integrate by parts first $K_G$ times in the $\gamma^{a_0}$ variable to obtain decay in $\beta_{a_0}$, obtaining
	\begin{equation}
		\begin{split}
			(F\star G)_{\mbf{j}^\beta,\mbf{j}^\alpha}(\alpha) =
			(2\pi \hbar)^{-2d}
			(2i\hbar)^{K_G}
			&\int e^{i \beta_a\gamma^a/(2\hbar)}
			\beta_{a_0}^{-K_G}
			F(\alpha+\beta/2)
			\partial_{\gamma^{a_0}}^{K_G}\big(\chi_{j^\gamma_{a_0}}(\gamma^{a_0})
			G(\alpha+\gamma/2)\big) \\
			&\qquad\qquad\times\prod_{a=1}^{2d} \chi_{j^\beta_a}(\beta_a)
			\prod_{b\not=a_0} \chi_{j^\gamma_b}(\gamma^b)
			\diff \beta_a\diff\gamma^a.
		\end{split}
	\end{equation}
	
	Next we integrate by parts $K_F$ times in $\beta_{b_0}$ to obtain decay in $\gamma^{b_0}$, then
	use the product rule to split up the derivatives
	\begin{equation}
		\begin{split}
			(F\star G)_{\mbf{j}^\beta,\mbf{j}^\alpha}(\alpha) &=
			(2\pi \hbar)^{-2d}
			(2i\hbar)^{K_G}
			(2i\hbar)^{K_F}
			\int e^{i \beta_a\gamma^a/(2\hbar)}
			\partial_{\beta_{b_0}}^{K_F}\big(
			\beta_{a_0}^{-K_G}
			\chi_{j^\beta_{b_0}}(\beta_{b_0})
			F(\alpha+\beta/2)\big) \\
			& \qquad\qquad\qquad
			\qquad\qquad\qquad
			\qquad\times
			(\gamma^{b_0})^{-K_F}
			\partial_{\gamma^{a_0}}^{K_G}\big(\chi_{j^\gamma_{a_0}}(\gamma^{a_0}) G(\alpha+\gamma/2)\big) \\
			&
			\qquad\qquad\qquad
			\qquad\qquad\qquad
			\qquad\times
			\prod_{a\not=b_0} \chi_{j^\beta_a}(\beta_a)
			\prod_{b\not=a_0} \chi_{j^\gamma_b}(\gamma^b)
			\diff \beta_a\diff\gamma^a \\
			&=
			(2\pi \hbar)^{-2d}
			(2i\hbar)^{K_G}
			(2i\hbar)^{K_F}
			\int e^{i \beta_a\gamma^a/(2\hbar)}
			\sum_{k=0}^{K_F}
			\binom{K_F}{k}
			\partial_{\beta_{b_0}}^{K_F-k}\big(
			\beta_{a_0}^{-K_G}
			\chi_{j^\beta_{b_0}}(\beta_{b_0})\big)
			\partial_{\beta_{b_0}}^{k} F(\alpha+\beta/2) \\
			& \qquad\qquad\qquad
			\qquad\qquad\qquad
			\qquad\times
			(\gamma^{b_0})^{-K_F}
			\sum_{k'=0}^{K_G}
			\binom{K_G}{k'}
			\partial_{\gamma^{a_0}}^{K_G-k'}\chi_{j^\gamma_{a_0}}(\gamma^{a_0})
			\partial_{\gamma^{a_0}}^{k'}G(\alpha+\gamma/2)\big) \\
			&
			\qquad\qquad\qquad
			\qquad\qquad\qquad
			\qquad\times
			\prod_{a\not=b_0} \chi_{j^\beta_a}(\beta_a)
			\prod_{b\not=a_0} \chi_{j^\gamma_b}(\gamma^b)
			\diff \beta_a\diff\gamma^a.
		\end{split}
	\end{equation}
	In the quantity
	$\partial_{\beta_{b_0}}^{K_F-k}
	(\beta_{a_0}^{-K_G} \chi_{j^\beta_{b_0}}
	(\beta_{b_0}))$
	there are two cases to consider.  If
	$a_0=b_0$ then we use~\eqref{eq:chi-t-bds},
	and if $a_0\not=b_0$ then the derivative
	only falls on $\chi$ and we use~\eqref{eq:chi-grad-bds}.  Regardless we have the bound
	(using $\beta_{a_0} \geq 2^{J^\beta}\hbar^{1/2}$)
	\begin{equation}
		|\partial_{\beta_{b_0}}^{K_F-k}
		(\beta_{a_0}^{-K_G} \chi_{j^\beta_{b_0}}
		(\beta_{b_0}))|
		\leq C_K
		2^{-J^\beta K_G} \hbar^{-(K_F+K_G-k)/2}.
	\end{equation}
	Now we use the estimates~\eqref{eq:chi-t-bds} and~\eqref{eq:chi-grad-bds}
	(using that $\gamma^{b_0} \sim 2^{J^\gamma}\hbar^{1/2}$ and $\beta_{a_0}\sim 2^{J^\beta}\hbar^{1/2}$) to
	bound the derivatives hitting $\chi$, and then apply the triangle inequality
	\begin{equation}
		\begin{split}
			|(F\star G)_{\mbf{j}^\beta,\mbf{j}^\alpha}(\alpha)| &\leq C_K
			\hbar^{K_G+K_F-2d}
			\int \sum_{k=0}^{K_F}
			2^{-J^\beta K_G}
			\hbar^{-(K_F-K_G-k)/2}
			|\partial_{\beta_{b_0}}^{k} F(\alpha+\beta/2)| \\
			& \qquad\qquad\qquad\qquad\qquad\times
			2^{-J^\gamma K_F}\hbar^{-K_F/2}
			\\
			& \qquad\qquad\qquad\qquad\qquad\times
			\sum_{k'=0}^{K_G}
			\hbar^{-(K_G-k')/2}
			|\partial_{\gamma^{a_0}}^{k'}G(\alpha+\gamma/2)\big| \\
			&\qquad\qquad\qquad\qquad\qquad\times
			\prod_{a} \wtild{\chi}_{j^\beta_a}(\beta_a)
			\prod_{b} \wtild{\chi}_{j^\gamma_b}(\gamma^b)
			\diff \beta_a\diff\gamma^a.
		\end{split}
	\end{equation}
	Above $\wtild{\chi}_a$ is the indicator
	function for the support of $\chi_a$.
	Collecting the constants, using
	$\binom{K}{k}\leq 2^K$ and combining factors
	of $2$ and $\hbar$ we arrive at
	\begin{equation}
		\begin{split}
			|(F\star G)_{\mbf{j}^\beta,\mbf{j}^\alpha}(\alpha)| &\leq
			\sum_{k=0}^{K_F} \sum_{k'=0}^{K_G}
			\hbar^{-2d}
			\int
			2^{-J^\beta K_G}
			\hbar^{k/2}
			|\partial_{\beta_{b_0}}^{k} F(\alpha+\beta/2)| \\
			& \qquad\qquad\qquad\qquad\qquad\qquad\times
			2^{-J^\gamma K_F}
			\hbar^{k'/2}
			|\partial_{\gamma^{a_0}}^{k'}G(\alpha+\gamma/2)\big| \\
			&\qquad\qquad\qquad\qquad\qquad\qquad\times
			\prod_{a} \wtild{\chi}_{j^\beta_a}(\beta_a)
			\prod_{b} \wtild{\chi}_{j^\gamma_b}(\gamma^b)
			\diff \beta_a\diff\gamma^a.
		\end{split}
	\end{equation}
	Since $J^\beta > j^\beta_a$ for any other index $a$ and on the support of the integrand above $\beta_a \sim \hbar^{1/2} 2^{j^\beta_a}$
	(and similarly for $\gamma$, it holds that
	\[
	2^{-KJ^\beta}
	\leq C^K (\hbar^{-1/2}|\beta|+1)^{-K}
	\leq \hbar^d \rho_K(\beta)
	\]
	and similarly
	\[
	2^{-KJ^\gamma}
	\leq C^K (\hbar^{-1/2}|\gamma|+1)^{-K}
	\leq \hbar^d \rho_K(\gamma).
	\]
	Recalling that $\rho_K(\beta) = \hbar^{-d} (\hbar^{1/2}|\beta| + 1)^{-K}$,
	and then noting that we can sum over all multi-indices $\mbf{j}^\beta$
	and $\mbf{j}^\gamma$ using
	\[
	\sum_{\mbf{j}^\beta}
	\prod_a \wtild{\chi}_{j^\beta_a}(\beta_a) \leq 4^d
	\]
	we can simplify the above bound to
	
		\begin{equation}
		\begin{split}
			|(F\star G)(\alpha)| 
			&\leq \hbar^{-2d}
			\sum_{\mbf{j}^\beta,\mbf{j}^\alpha}
			\int 2^{-J^\beta K_G}\hbar^{k/2}
			|\partial_{\beta_{b_0}}^k F(\alpha+\beta/2)|\\
			&\qquad\qquad\qquad \qquad 
			\times 2^{-J^\gamma K_F} \hbar^{k'/2}
			|\partial^{k'}_{\gamma^{a_0}}
			G(\alpha+\gamma/2)|
			\prod_a \wtild{\chi}_{j^\beta_a}(\beta_a)
			\prod_a \wtild{\chi}_{j^\gamma_b}(\gamma^b)
			\dd \beta_a \dd\gamma^a\\
			&\lsim
			\int \rho_{K_G}(\beta)\hbar^{k/2}
			(\sum_{k=0}^{K_F}\OffDiag{\nabla^k F(\alpha+\beta/2)})\\
			&\qquad\qquad\qquad \quad 
			\times \rho_{K_F}(\gamma) \hbar^{k'/2}
			(\sum_{k'=0}^{K_G}\OffDiag{\nabla^{k'} G(\alpha+\gamma/2)})
			\sum_{\mbf{j}^\beta,\mbf{j}^\alpha}
			(\prod_a \wtild{\chi}_{j^\beta_a}(\beta_a)
			\prod_b \wtild{\chi}_{j^\gamma_b}(\gamma^b))
			\dd \beta_a \dd\gamma^a\\
			&\lsim
			\int \rho_{K_G}(\beta)\hbar^{k/2}
			(\sum_{k=0}^{K_F}\OffDiag{\nabla^k F(\alpha+\beta/2)})\\
			&\qquad\qquad\qquad \quad
			\times
			\rho_{K_F}(\gamma) \hbar^{k'/2}
			(\sum_{k'=0}^{K_G}\OffDiag{\nabla^{k'} G(\alpha+\gamma/2)})
			\dd \beta_a \dd\gamma^a\\
			&\lsim
			\sum_{k=0}^{K_F} \sum_{k'=0}^{K_G}
			\big(\rho_{K_G} \ast \OffDiag{\hbar^{k/2} \nabla^k F}(\alpha)\big)
			\big(\rho_{K_F} \ast \OffDiag{\hbar^{k'/2} \nabla^{k'} G}(\alpha)\big),
		\end{split}
	\end{equation}
	as desired. The remaining terms
	(with $\mbf{j}^\beta\not=0$ and
	$\mbf{j}^\gamma=0$ or vice versa)
	are handled similarly.
\end{proof}

\appendix

\section{Physical units, symplectic covariance, and a corollary}\label{sec:units-symplectic-covariance}

In this section we offer some informal discussion of symplectic symmetry and the relationship to units.  To illustrate this, we then define some preferred choices of units and use them to state and prove Corollary~\ref{cor:MinDiff} of Theorem~\ref{thm:mainResult}. This corollary generalizes the main result from our companion paper \cite{hernandez2023decoherence1} to Hamiltonians not restricted to the form $\HQ = \PQ^2/2m+V(\XQ)$ at the expense of introducing the uncomputed constant $C_d$.

\subsection{Symplectic transformations of the main result}

For a symplectic matrix $\NTSZ$ ($\NTSZ^\tp\sf\NTSZ=\sf$) representing a linear symplectic transformation, we will use $\NTSZ$ as a superscript on scalar functions over phase space to denote the composition equivalent to the change of coordinates associated with the matrix: $E^\NTSZ(\alpha) := (E\circ\NTSZ)(\alpha) = E(\NTSZ\alpha)$. The same notation is used for tensor functions, except we must additionally transform the indices, e.g., $(E^\NTSZ)^{a}_{\pha b}(\alpha) := (\NTSZ^{-1})^{a}_{\pha c} \NTSZ^d_{\pha b} E^{c}_{\pha d}(\NTSZ\alpha)$.  

The general Fokker-Planck equation for a classical open system,
\begin{equation}
    \label{eq:fp-sympl}
    \partial_t \cstate = -\partial_a [\cstate (\partial^a \HC + \GC^a)]  +	 \frac12 \partial_a (D^{ab} \partial_b \cstate),
\end{equation}
and the Wigner representation of the Lindblad equation
\begin{align}
    \label{eq:lindblad-sympl}
	\partial_t W_\rho 
	=&\, -\frac{i}{\hbar} (\HC \star W_\rho - W_\rho \star \HC)  + \frac{1}{\hbar }\sum_k \left( \LCk  \star W_\rho \star \LCk^* - \frac{1}{2} \LCk^* \star \LCk \star W_\rho - \frac{1}{2} W_\rho\star \LCk^* \star \LCk   \right)
\end{align}
are both covariant under linear symplectic transformations. 
This means \eqref{eq:fp-sympl} is unchanged under $\HC \to \HC^\NTSZ$, $\GC \to \GC^\NTSZ$, $\D \to \D^\NTSZ$, and $\cstate\to \cstate^\NTSZ$ because $\partial_a \cstate^\NTSZ = \NTSZ^b_{\pha a} \partial_a \cstate$. 
Likewise \eqref{eq:lindblad-sympl} is unchanged under under $\HC \to \HC^\NTSZ$, $\LCk \to \LCk^\NTSZ$, and $W_\rho \to W_\rho^\NTSZ$ because $A^\NTSZ \star B^\NTSZ = (A \star B)^\NTSZ$, a basic property of the Moyal product.

If the data $(\HC, \{\LCk\}_{k=1}^K)$ are admissible under Assumption~\ref{assum:suitableLindblad}, then the transformed data $(\HC^\NTSZ, \{\LCk^\NTSZ\}_{k=1}^K)$ are also admissible. Now suppose  the initial state can expressed as a mixture of NTS states ($\sigma \ge \frac{\rds}{2} \sigmaco$) \emph{and also} as a mixture of $\NTSZ$-transformed NTS states ($\sigma \ge \frac{\rds_\NTSZ}{2} \NTSZ^{-1}\sigmaco\NTSZ^{-\tp}$), where $\rds$ and $\rds_\NTSZ$ are the relative diffusion strengths \eqref{eq:zz-def} computed with the original and transformed data. Then we can apply Theorem~\ref{thm:mainResult} to the original and transformed data and \emph{both} sets of bounds will hold. These will generally be distinct bounds because the anharmonicity measures $\harmErrConst_\cl$, $\harmErrConst_\qu$, and $\harmErrConst_{\qu'}$ are not invariant under linear symplectic transformations of the data.\footnote{Note that $\HC^\NTSZ$ is quadratic if and only if $\HC$ is quadratic, and $\LCk^\NTSZ$ is linear if and only if $\LCk$ is linear, so whether the anharmonicity measures vanish will generally be invariant under linear symplectic transformations.}

\subsection{Units}

Given a phase-space vector with mixed units like $\tilde{\alpha} = (x_0 \,\mathrm{m}, p_0\,\mathrm{kg}\!\cdot\!\mathrm{m}/\mathrm{s})$, where $x_0$ and $p_0$ are dimensionless numbers, we can transform this to a vector with uniform units using a symplectic matrix like $\NTSZ^{-1} = \diag(\eta,\eta^{-1})$ with $\eta = \sqrt{(\mathrm{kg}\!\cdot\!\mathrm{m}/\mathrm{s})/(\mathrm{m})} = \sqrt{\mathrm{kg}/\mathrm{s}}$. Specifically, $\NTSZ^{-1}\tilde\alpha = (x_0,p_0)[\mathrm{m}\sqrt{\mathrm{kg}/\mathrm{s}}]$.  Given such a choice of $\NTSZ$ and a real, physical Hamiltonian $H(x,p)$ taking
as input dimensionful positions $x$ and momenta $p$, we can then apply Theorem~\ref{thm:mainResult} to the transformed functions
$H^\NTSZ(\alpha):=H(\NTSZ\alpha)$ and $\LCk^\NTSZ(\alpha):=\LCk(\NTSZ\alpha)$, which will accept vectors with uniform units.
Now, there is generally no symplectic matrix that can make an arbitrary unitful vector unitless.  Still, a choice like $\NTSZ$ above is sufficient to ensure that all of the manipulations in this paper (such as taking the Euclidean\footnote{An alternative way to think about this is that the choice $\NTSZ$ defines an inner product: $\langle\tilde\alpha,\tilde\beta\rangle_\NTSZ := {\tilde\alpha}^\tp(\NTSZ\NTSZ^\tp)^{-1}\tilde\beta = \alpha^\tp \cdot \beta$ for uniform-unit vectors $\alpha = \NTSZ^{-1}\tilde\alpha$, $\beta = \NTSZ^{-1}\tilde\beta$.} 
norm of $\NTSZ\alpha$ for mixed-unit vector $\alpha$) are physically meaningful once such a choice of units has been made.  

The symplectic matrices are closed under multiplication (being a group) and so the choice of units does not exhaust the freedom to choose $\NTSZ$.  For instance, the skewing matrix $\NTSZ=\left(\begin{smallmatrix}1 & 1\\0&1\end{smallmatrix}\right)$ is symplectic and mixes position and momentum in a way that does not correspond to a choice of units.\footnote{In order that an arbitrary symplectic matrix $\NTSZ^{-1}$ correctly makes all units uniform, the entries $\NTSZ_{\x_i,\x_j}$ and $\NTSZ_{\p_i,\x_j}$ must have units of [length/momentum]${}^{1/2}$ and the entries $\NTSZ_{\x_i,\p_j}$ and $\NTSZ_{\p_i,\p_j}$ must have units of [momentum/length]${}^{1/2}$.  This ensures that all elements of the vector $\NTSZ\alpha$ have units of [length\,$\cdot$\,momentum]${}^{1/2}$, where $\alpha$ is a physical phase-space vector with units of [length] in the first $d$ elements and [momentum] in the other $d$ elements.}
Thus, the fact that multiple bounds can be derived using Theorem~\ref{thm:mainResult} by applying difference choices of $\NTSZ$ is not removed by a choice of units.
Although the trajectory $\rhot(t)$ constructed in different cases will generally be different, note that the existence of such a trajectory is a units-independent statement; the bounds \eqref{eq:main-classical}, \eqref{eq:main-quantum} are on unitless norms, which in turn constrains the maximum (unitless) difference in outcome probability for any measurement.

Ideally the choice of $\NTSZ$ could be optimized for the best bound
(since coherent states $\tauQa$ would correspond to covariance matrices satisfying 
$\sigma = \frac{\hbar}{2} \NTSZ^\tp \NTSZ$).  Alternatively, to name a concrete choice, one might for example choose
\begin{equation} \label{eq:Z0}
    \begin{split}
        \NTSZ{}_0 = \left(\begin{matrix} r\IdM_d & 0 \\0 & r^{-1}\IdM_d\end{matrix}\right)
    \end{split}
\end{equation}
where $r=\sqrt{r_x/r_p}$ for characteristic scales $r_x, r_p$ defined by $r_x^{-2} = \sup_\alpha \sup_{w \in \mathbb{R}^d, \norm{w}=1} (w_a \partial_{x_a})^2 H(\alpha)$ and $r_p^{-2} = \sup_\alpha \sup_{w \in \mathbb{R}^d, \norm{w}=1} (w_a \partial_{p_a})^2 H(\alpha)$, which quantify the maximum second derivatives of $H$ with respect to position and momentum, respectively.

\subsection{Corollary}

To illustrate the above informal discussion, we will apply Theorem~\ref{thm:mainResult} to a special case with linear Lindblad operators (diffusion matrix homogeneous over phase space) and no friction.  We will ask: given a physical system with (unitful) Hamiltonian $\HC$, how much environmental noise must we add to ensure that the quantum and classical dynamics cannot be distinguished up to some tolerable error?  We will first identify the relevant characteristic timescales and action scales of the Hamiltonian. They will be constructed from the Hamiltonian's derivatives, and in particular we recall from Eq.~\eqref{eq:constQ-def} the seminorm
\begin{align}
\constQ^{q,r}_{s}[E]
&:= 
\sum_{j=q}^r s^{(j-q)/2}\CkSN{E}{j}
= \sum_{j=q}^r s^{(j-q)/2} \sup_\alpha \sup_{\|\beta_\ell\|=1}
    \left|\beta_1^{a_1} \cdots \beta_j^{a_j} \partial_{a_1}\cdots\partial_{a_j} E(\alpha)\right|.
\end{align}
which bounds the $q$-th through $r$-th derivatives. For our main result, we used $s=\hbar$ to get the tightest bound, but here it will be instructive to use a (macroscopic) action scale $s \ge\hbar$.

\begin{restatable}[Characteristic classical scales]{defin}{defCharScales}
    \label{def:characteristicScales}
    For any classical Hamiltonian $\HC$ with bounded partial derivatives of degree $k=2,\ldots,2d+4$
    and any symplectic matrix $\NTSZ$, we define the 
    \introduce{harmonic time} $\tharm$ as the inverse of the maximum operator norm taken by the Hessian of $\HC^\NTSZ$ over phase space,\footnote{For linear Hamiltonians, $\tharm=\infty$ by convention. This indicates that there is no natural time scale because the (real or imaginary) frequency of the local harmonic dynamics is zero everywhere in phase space.} 
    \begin{align}
        \frac{1}{\tharm} := |\HC^\NTSZ|_{C^2} = \constQ^{2,2}[\HC^\NTSZ] 
        = \sup_\alpha \sup_{\|\beta\|=\|\xi\|=1} \left|\beta^{a}\xi^{b} \partial_{a}\partial_{b} \HC^\NTSZ(\alpha)\right|.
    \end{align} 
    The \introduce{anharmonic action} $\ho$ of $H$ is given by the ratio of the largest second and third directional derivative:
    \begin{align}
     \ho := 
        \left(\tharm |H^\NTSZ|_{C^3}\right)^{-2} = 
        \left(\frac{|H^\NTSZ|_{C^2}}{|H^\NTSZ|_{C^3}}\right)^2  = 
        \left(\frac{\constQ^{2,2}[\HC^\NTSZ]}{\constQ^{3,3}[\HC^\NTSZ]}\right)^2,
    \end{align}
    We use the anharmonic action
    $\ho$ to define the \introduce{modified anharmonic action}\footnote{For quadratic Hamiltonians, $\ho = \hH = \infty$ by convention. This is the case of harmonic dynamics where quantum and classical evolution is identical on phase space.} of $\HC$
    \begin{align}
    \label{eq:char-action-def}
        \hH := 
        \left(\tharm \constQ^{3,2d+4}_{\ho}[\HC^\NTSZ]\right)^{-2} = 
        \left(\frac{\constQ^{2,2}[\HC^\NTSZ]}{\constQ^{3,2d+4}_{\ho}[\HC^\NTSZ]}\right)^2.
    \end{align}
    Given $\tharm$, $\NTSZ$, and $\hH$, we furthermore have a natural choice of \introduce{characteristic diffusion matrix} given by $\Dc = (\hH/\tharm)\NTSZ\NTSZ^\tp$.
\end{restatable}

Intuitively, the harmonic time $\tharm$ is the \emph{shortest} timescale associated with the local harmonic approximation at any point in phase space.
When the harmonic time is long, the classical dynamics are slow compared to the quantum scale set by $\hbar$, and we expect they well approximate the quantum dynamics they correspond to.

The anharmonic action $\ho$ and modified anharmonic action $\hH$ are \emph{not} measures of the accessible phase space. Rather, they measure the phase-space scale on which the anharmonicity of the potential is important over the harmonic time $\tharm$.  For cubic potentials, $\hH = \ho$ because the higher-order seminorms in~\eqref{eq:char-action-def} vanish.  Introducing the higher order terms increases the denominator in $\hH$, so $\hH \le \ho$ always holds.

We now prove a corollary of Theorem~\ref{thm:mainResult} making use of the scales defined in Definition~\ref{def:characteristicScales}.  The seminorm $\constQ^{3,2d+4}_\hbar[\HC^\NTSZ]$ naturally arises in Theorem~\ref{thm:mainResult}, but through $\hH$ we will upper bound it with the factor of $\constQ^{3,2d+4}_{\ho}[\HC^\NTSZ] \ge \constQ^{3,2d+4}_\hbar[\HC^\NTSZ$ (assuming $\ho\ge\hbar$). This loosens the bound, essentially throwing out the detailed information about how the anharmonic factors depend on higher powers of the action scale, but has the benefit of isolating the leading $\hbar^{4/3}$ dependence, with everything else expressed in terms of ($\hbar$-independent) macroscopic properties of the classical Hamiltonian.

\begin{restatable}[Minimum diffusion for correspondence]{cor}{corMinDiff}
    \label{cor:MinDiff}
    For $d$ degrees of freedom, let $\HQ$ be a quantum Hamiltonian function with bounded partial derivatives of degree $j=2,\ldots,2d+4$ and corresponding classical Hamiltonian $\HC=\Op^{-1}[\HQ]$. Let $\NTSZ$ a symplectic matrix  
    and let 
    $\tharm=1/\constQ^{2,2}[H^\NTSZ]$, 
    $\ho=(\constQ^{2,2}[H^\NTSZ]/\constQ^{3,3}[H^\NTSZ])^{2}$, 
    $\hH=(\constQ^{2,2}[H^\NTSZ]/\constQ^{3,2d+4}_{\ho}[H^\NTSZ])^2$, 
    $\Dc = (\hH/\tharm)\NTSZ\NTSZ^\tp$ 
    be the harmonic time, anharmonic action, modified anharmonic action, and characteristic diffusion matrix of $\HC$ from Definition~\ref{def:characteristicScales}.
    Assume $\ho \ge \hbar$.
    Assume an initial state quantum state $\rho(t\liq 0)$ given as a mixture of Gaussian states with covariance matrix $\sigma = (\hbar/2)\NTSZ\NTSZ^\tp$. 
    Finally,  let $\rho(t)$ and $\cstate(t)$ be the corresponding quantum and classical trajectory for the frictionless dynamics specified by $\HC$ and a homogenous (i.e, constant over phase space) diffusion matrix $\D$.
    Then for any tolerable error growth rate $\erate$ satisfying 
    \begin{equation} \label{eq:erate-thresh}
    \erate \geq  \frac{C_d}{\tharm}\sqrt{\frac{\hbar}{\hH}},
    \end{equation}
    the diffusion strength constraint
    \begin{align}\label{eq:d-threshold}
        \D \ge 
        \left(\frac{C_d}{ \tharm\erate}\right)^{\frac23}
        \left(\frac{\hbar}{\hH}\right)^{\frac43}\Dc
    \end{align} 
    guarantees there exists a quantum trajectory $\rhot(t)$ with strictly positive Wigner function $\WW[\rhot(t)]$ such that
    \begin{enumerate}[a.]
        \item 
        $\rhot(t)$ cannot be distinguished from $\rho(t)$ with probability greater than $\erate t$ by \emph{any} quantum measurement; and
        \item 
        $\WW[\rhot(t)]$ cannot be distinguished from $\cstate(t)$ with probability greater than $\erate t$ by \emph{any} classical variable.
    \end{enumerate}
    Above, $C_d$ is the same universal dimensionless constant depending only on $d$ from Theorem~\ref{thm:mainResult}.
\end{restatable}
\begin{proof}
    We can consult the definitions of the anharmonic factors in Eqs.~\eqref{eq:harmErrConstCtt-def}, \eqref{eq:harmErrConstQtt-def}, and \eqref{eq:harmErrConstQf-def} to see that in the special case of linear Lindblad operators ($\LCk = \ell_{k,a}\alpha^a$, $\D_{ab} =\hbar\scD_{ab}=  \hbar\Real \sum_k {\ell_{k,a}^*}{ \ell_{k,b}^{\phantom *}}$) they reduce to 
    just 
    \begin{align}
        \harmErrConstCtt{\HC^\NTSZ}{\ell_{k,a}\alpha^a} 
        &= 
         \constQ^{3,3}[\HC^\NTSZ]
        \\
        \harmErrConstQtt{\HC^\NTSZ}{\ell_{k,a}\alpha^a} 
        &= 
        \constQ^{3,2d+4}_\hbar[\HC^\NTSZ]
        \\
        \harmErrConstQfBare[\ell_{k,a}\alpha^a,\hbar,\nu] &= 
        0
    \end{align}
    The dynamics are taken to be frictionless, so we will apply Theorem~\ref{thm:mainResult} with 
    Eq.~\eqref{eq:zz-def-frictionless} from Definition~\ref{def:eff-inv-diff-strength} for the relative diffusion strength.  The key quantity is
    \begin{align}
        \inf_\alpha 
        \frac{\lambdamin[\scD^\NTSZ(\alpha)]}{\lambdamax[(\nabla^2\HC)^\NTSZ(\alpha)]}
        & \ge 
        \tharm \hbar^{-1} \lambdamin[\D^\NTSZ(\alpha)]
        \\
        \label{johnny1}
        & \ge 
        \tharm \hbar^{-1} \left(\frac{C_d}{\tharm\erate}\right)^{2/3}
        \left(\frac{\hbar}{\hH}\right)^{4/3} \lambdamin[\Dc^\NTSZ]
        \\
        \label{johnny3}
        & \ge 
        \left(\frac{C_d^2\hbar}{\erate^2 \tharm^2\hH}\right)^{1/3} \\
        \label{johnny4}
        & \ge 1 
    \end{align}
    where we have sequentially applied the definitions 
    $\tharm=\constQ^{2,2}[\HC^\NTSZ]$ and $\D=\hbar\scD$,
    the diffusion constraint~\eqref{eq:d-threshold},  the fact that $\lambdamin[\Dc^\NTSZ] = \lambdamin[(\hH/\tharm)\IdM] = \hH/\tharm$, and the error rate constraint~\eqref{eq:erate-thresh}.  This means that the relative diffusion strength $\rds=1$ so that, by Eqs.~\eqref{eq:main-classical} and \eqref{eq:main-quantum} of Theorem~\ref{thm:mainResult} we have that $\Trnorm{\rhot(t)-\rho(t)}$ and $\Lonenorm{\WW[\rhot(t)]-\rhoc(t)}$ are both upper bounded by
    \begin{align}
        C_d \, t\, \hbar^{\frac12} \constQ^{3,2+4d}_\hbar[\HC^\NTSZ] \le \erate \, t
    \end{align}
    because $\hH = (\tharm \constQ^{3,2+4d}_{\ho}[\HC^\NTSZ])^{-2}$ and $\constQ^{q,r}_{\hH}$ is an increasing function of an action $\hH \ge \hbar$.
\end{proof}
This corollary is more general than the result in our companion paper \cite{hernandez2023decoherence1} because here we do not restrict to Hamiltonians of the form $\HC = \PC^2/2m+V(\XC)$, but it is weaker in that special case because it involves an unknown constant $C_d$.

\section{Gaussian derivatives and integrals}\label{sec:gaussian-derivatives-and-integrals}

\subsection{Gaussian derivatives}\label{sec:gaussian-derivatives}

The Gaussian probability distribution with mean $\alpha$ and covariance matrix $\sigma$ is  
\begin{align}
\tauCas(\alpha+\beta)= \frac{ e^{-\beta^\tp \sigma^{-1}\beta/2}}{(2\pi)^d\sqrt{\det\sigma}} =  \frac{1}{(2\pi)^d\sqrt{\det\sigma}}\exp\left(-\frac{1}{2}\beta^a \sigma^{-1}_{ab}\beta^b\right)
\end{align}
Let us consider this a real-valued function of any vector $\beta$ and any invertible matrix $\sigma$, including non-symmetric ones, so that $\sigma_{ab}$ and $\sigma_{ba}$ are independent variables for the purposes of partial derivatives. However, at the end we will evaluate these derivatives on the subspace where $\sigma$ is symmetric. 
Recalling our notation $\partial_c = \partial/\partial\beta^c$ so $\partial_c \beta^a = \delta_c^{\pha a}$, 
we have
\begin{align}
\partial_d \left(\beta^a \sigma^{-1}_{ab}\beta^b\right) \,&=  \sigma^{-1}_{db}\beta^b + \beta^a\sigma^{-1}_{ad} 
\\
 \partial_c \partial_d \left(\beta^a \sigma^{-1}_{ab}\beta^b\right) \,&=  
 \sigma^{-1}_{dc} + \sigma^{-1}_{cd}.
\end{align}
so that
\begin{align}
\partial_a \tauCas(\alpha+\beta) = -\sigma^{-1}_{ab}\beta^b\tauCas(\alpha+\beta) 
\end{align}
when $\sigma$ is symmetric. 
We also deploy the standard \cite{petersen2012matrix}  matrix derivative identities\footnote{Some sources express will express this for a not-necessarily invertible matrix $Z$ using the matrix adjugate $\mathrm{adj}(Z)$. A property of the adjugate is that $Z\, \mathrm{adj}(Z) = (\det Z)\IdM$ so that, when $Z$ is invertible, $\mathrm{adj}(Z) = (\det Z)Z^{-1}.$}
\begin{align}
\frac{\partial \det Z}{\partial y} &= (\det Z) \Tr\left[Z^{-1}\frac{\partial Z}{\partial y}\right],\\
\frac{\partial Z^{-1}}{\partial y} &= -Z^{-1}\frac{\partial Z}{\partial y} Z^{-1}
\end{align}
for an invertible matrix $Z$, so in particular 
\begin{align} 
 \frac{\partial \det Z}{\partial Z^{ab}} &= (\det Z) Z^{-1}_{ba},\\
 \frac{\partial Z^{-1}_{cd}}{\partial Z^{ab}} &= -Z^{-1}_{ca}Z^{-1}_{bd}.
\end{align}
Combining these we get
\begin{align}
\frac{1}{2}\partial_a \partial_b \tauCas(\alpha+\beta)=\frac{1}{2}(\sigma^{-1}_{ac}\beta^c\sigma^{-1}_{bd}\beta^d-\sigma^{-1}_{ab})\tauCas(\alpha+\beta) = \frac{\partial}{\partial \sigma^{ab}} \tauCas(\alpha+\beta),
\end{align}
when evaluated for symmetric $\sigma$.  (As expected, this is singular when $\sigma$ is non-invertible.)  Weyl quantizing both sides with $\WW^{-1}=(2\pi\hbar)^{d}\Op$ gives the corresponding quantum expression $\partial_a \partial_b \tauQas = 2\frac{\partial}{\partial \sigma_{ab}} \tauQas$.

\subsection{Gaussian integrals}\label{sec:gaussian-integrals}

Here we recall the evaluation of some Gaussian integrals, as can be done with Wick's theorem. We define the shorthand:
\begin{align}\begin{split}\label{eq:gaussian-int-cl-2}
		\langle (\beta^\tp A^{} \beta)\rangle_\sigma
		:=& \int \!\dd \beta\, \tauC_{0,\sigma}(\beta)  (\beta^\tp A^{} \beta)\\
		=&\int \!\dd \beta\, \tauCas(\alpha+\beta)  (\beta^\tp A^{} \beta) \\
		=& A^{}_{ab} \int \!\dd \beta\, \tauCas(\alpha+\beta) \beta^a\beta^b\\
		=& A^{}_{ab} \sigma^{ab}  \\
		=& \Tr[\sigma A^{}].
\end{split}\end{align}
for any positive semidefinite matrix $A$. ($\sigma$ is also positive, of course.)  Likewise, for $B$, $C$, and $D$ also positive semidefinite, we have
\begin{align}\begin{split}\label{eq:gaussian-int-cl-4}
		\langle (\beta^\tp A^{} \beta)(\beta^\tp B^{} \beta)\rangle_\sigma
		:=& \int \!\dd \beta\, \tauCas(\alpha+\beta)  (\beta^\tp A^{} \beta)(\beta^\tp B^{} \beta)\\
		= & A^{}_{ab} B^{}_{cd}  \left[\sigma^{ab}\sigma^{cd} + 2\sigma^{ad} \sigma^{bc} \right]\\
		= & \Tr[\sigma A^{}]\Tr[\sigma B^{}] + 2 \Tr[\sigma A^{} \sigma B^{}]
\end{split}\end{align}
and
\begin{align}\begin{split}\label{eq:gaussian-int-cl-6}
		\langle (\beta^\tp A^{} \beta)(\beta^\tp B^{} \beta)(\beta^\tp C^{} \beta)\rangle_\sigma
		:=& \int \!\dd \beta\, \tauCas(\alpha+\beta) (\beta^\tp A^{} \beta)(\beta^\tp B^{} \beta)(\beta^\tp C^{} \beta)
		\\
		=& A^{}_{ab} B^{}_{cd} C^{}_{ef} \Big[\sigma^{ab}\sigma^{cd}\sigma^{ef} +2\left(\sigma^{ab}\sigma^{cf}\sigma^{de} + \sigma^{af}\sigma^{cd}\sigma^{be} + \sigma^{ad}\sigma^{bc}\sigma^{ef}  \right) \\
		&\qquad\qquad\qquad + 4\left(\sigma^{ad}\sigma^{be}\sigma^{cf} + \sigma^{af}\sigma^{bc}\sigma^{de}  \right) \Big]\\
		=& \Tr[\sigma A^{}] \Tr[\sigma B^{}] \Tr[\sigma C^{}] + 2\Tr[\sigma A^{}] \Tr[\sigma B^{} \sigma C^{}] +  2 \Tr[\sigma B^{}] \Tr[\sigma C^{} \sigma A^{}] \\
		&\qquad + 2 \Tr[\sigma C^{}] \Tr[\sigma B^{} \sigma A^{}]
		+ 8 \Tr[\sigma A^{} \sigma B^{} \sigma C^{}] .
\end{split}\end{align}
and
\begin{align}\begin{split}\label{eq:gaussian-int-cl-8}
		\langle (\beta^\tp A^{} \beta)&(\beta^\tp B^{} \beta)(\beta^\tp C^{} \beta)(\beta^\tp D^{} \beta)\rangle_\sigma \\
		:=& \int \!\dd \beta\, \tauCas(\alpha+\beta) (\beta^\tp A^{} \beta)(\beta^\tp B^{} \beta)(\beta^\tp C^{} \beta)(\beta^\tp D^{} \beta)
		\\
		=& A^{}_{ab} B^{}_{cd} C^{}_{ef}  D^{}_{gh} \Big[\sigma^{ab}\sigma^{cd}\sigma^{ef} \sigma^{gh} \\
		&+2\big(
		\sigma^{ab}\sigma^{cd}\sigma^{eg} \sigma^{fh} +
		\sigma^{ab}\sigma^{ce}\sigma^{df} \sigma^{gh} +	\sigma^{ac}\sigma^{bd}\sigma^{ef} \sigma^{gh} +\\
		&\qquad\sigma^{ab}\sigma^{cg}\sigma^{ef} \sigma^{dh} +
		\sigma^{ae}\sigma^{cd}\sigma^{bf} \sigma^{gh} +
		\sigma^{ag}\sigma^{cd}\sigma^{ef} \sigma^{bh}
		\big) \\
		&+4\big(
		\sigma^{ac}\sigma^{bd}\sigma^{eg} \sigma^{fh} +
		\sigma^{ae}\sigma^{cg}\sigma^{bf} \sigma^{dh} +
		\sigma^{ag}\sigma^{ce}\sigma^{df} \sigma^{bh}
		\big) \\
		&+8\big(
		\sigma^{ac}\sigma^{de}\sigma^{fb} \sigma^{gh} +
		\sigma^{ac}\sigma^{dg}\sigma^{ef} \sigma^{gb} +
		\sigma^{ae}\sigma^{cd}\sigma^{fg} \sigma^{hb} +
		\sigma^{ab}\sigma^{ce}\sigma^{fg} \sigma^{hd}
		\big)\\
		&+16\big(
		\sigma^{ac}\sigma^{de}\sigma^{fg} \sigma^{hb} +
		\sigma^{ae}\sigma^{fc}\sigma^{dg} \sigma^{hb} +
		\sigma^{ac}\sigma^{dg}\sigma^{he} \sigma^{fb}
		\big)\Big]\\
		=& \Tr[\sigma A] \Tr[\sigma B] \Tr[\sigma C]  \Tr[\sigma D]  +\\
		&+2\big(
		\Tr[\sigma A] \Tr[\sigma B] \Tr[\sigma C \sigma D] +
		\Tr[\sigma A] \Tr[\sigma D] \Tr[\sigma B \sigma C] +
		\Tr[\sigma C] \Tr[\sigma D] \Tr[\sigma A \sigma B] +\\
		&\qquad \Tr[\sigma A] \Tr[\sigma C] \Tr[\sigma B \sigma D] +
		\Tr[\sigma B] \Tr[\sigma D] \Tr[\sigma A \sigma C] +
		\Tr[\sigma B] \Tr[\sigma C] \Tr[\sigma A \sigma D]
		\big) \\
		&+4\big(
		\Tr[\sigma A \sigma B]\Tr[\sigma C \sigma D]+
		\Tr[\sigma A \sigma C]\Tr[\sigma B \sigma D] +
		\Tr[\sigma A \sigma D] \Tr[\sigma B \sigma C]
		\big) \\
		&+8\big(
		\Tr[\sigma A \sigma B\sigma C]\Tr[\sigma D]+
		\Tr[\sigma A \sigma B\sigma D]\Tr[\sigma C]+
		\Tr[\sigma A \sigma C\sigma D]\Tr[\sigma B]+
		\Tr[\sigma B \sigma C\sigma D]\Tr[\sigma A]
		\big)\\
		&+16\big(
		\Tr[\sigma A \sigma B\sigma C \sigma D] +
		\Tr[\sigma A \sigma C\sigma B \sigma D] +
		\Tr[\sigma A \sigma B\sigma D \sigma C]
		\big)\Big].
\end{split}\end{align}

\bibliographystyle{unsrt}
\bibliography{references}

\end{document}